\documentclass[fleqn]{2023SCGE}
\usepackage[utf8]{inputenc}
\DeclareUnicodeCharacter{0306}{\u{}}
\usepackage{graphicx}
\usepackage{mwe,tikz,ulem}
\usepackage[percent]{overpic}
\usepackage{xspace}
\usepackage[numbers]{natbib}
\usepackage{float}
\usepackage{color}
\usepackage{makeidx}
\usepackage[columns=1]{idxlayout}
\usepackage{textcomp}

\def\simgt{\lower.5ex\hbox{\gtsima}}

\newcommand{\tmop}{\rm}

\newcommand{\ysx}{} % the revision of the 1st round referees' comments. Cyan previously, and editor asked to remove highlight. 

\newcommand{\yisx}{} % Define \yisx as a simple text wrapper

\begin{document}
\ensubject{subject}

%%%%%%%%%%%%%%%%%%%%%%%%%%%%%%%%%%%%%%%%%%%%%%%%%%%%%%%
%%% Authors do not modify the information below
%%% ????????????????
%%% ??????????, ????????????{}, ???????????????????
%Letter to the Editor??Article%??????
\ArticleType{Article}%??Article
\SpecialTopic{SPECIAL TOPIC: }%???????
\Year{ }
\Month{ }
\Vol{ }
\No{ }
\DOI{ }
\ArtNo{ }
%\ReceiveDate{January 11, 2016}
%\AcceptDate{April 6, 2016}
%\OnlineDate{January 1, 2016}
%%%%%%%%%%%%%%%%%%%%%%%%%%%%%%%%%%%%%%%%%%%%%%%%%%%%%%%

%%% title: ????
%%% \title{}{title for citation}
\title{Prospects for Time-Domain and Multi-Messenger
 Science with eXTP}{eXTP WG4}

\author[1]{Shu-Xu Yi}{sxyi@ihep.ac.cn}%
\author[2]{Wen Zhao}{wzhao7@ustc.edu.cn}
\author[3,4]{Ren-Xin Xu}{r.x.xu@pku.edu.cn}
\author[5]{Xue-Feng Wu}{xfwu@pmo.ac.cn}
\author[6]{Giulia Stratta}{}
\author[7]{Simone Dall'Osso}{}
\author[1]{\\Yan-Jun Xu}{}
\author[8, 1]{Andrea Santangelo}{}
\author[9]{Silvia Zane}{}
\author[1]{Shuang-Nan Zhang}{}
\author[1]{Hua Feng}{}{}
\author[10]{\\Huan Yang}{}
\author[10]{Junjie Mao}{}
\author[11]{Junqiang Ge}{}
\author[4]{Lijing Shao}{}
\author[13]{Mi-Xiang Lan}{}%
\author[14]{He Gao}{}%
\author[14]{Lin Lin}{}%
\author[2]{\\Ning Jiang}{}%
\author[15]{Qingwen Wu}{}
\author[12]{Tong Liu}{}
\author[16]{Yun-Wei Yu}{}%
\author[17,18]{Xiang-Yu Wang}{}%
\author[19]{Jin Zhang}{}%
\author[21]{Dafne Guetta}{}%
\author[5]{\\Jin-Jun Geng}{}
\author[5]{Di Xiao}{}
\author[17,18]{Yong-Feng Huang}{}
\author[3]{Yacheng Kang}{}
\author[1,22]{Tian-Yong Cao}{}
\author[1]{Zhen Zhang}{}
\author[23]{\\Zhenwei Lyu}{}
\author[24,25]{Zhen Pan}{}
\author[11]{Yunfeng Chen}{}
\author[38]{Yong Gao}{}
\author[12]{Ang Li}{}
\author[14]{Yu-Cong Fu}{}
\author[26]{Shuo Xiao}{}
\author[22]{\\Wei-Yang Wang}{}
\author[17]{Fayin Wang}{}
\author[17]{Zhenyin Zhao}{}
\author[15]{Weihua Lei}{}
\author[27]{Rong-Feng Shen}{}
\author[28]{Lixin Dai}{}
\author[16]{Guang-Lei Wu}{}
\author[16]{\\Liang-Duan Liu}{}
\author[1]{Bing Li}{}
\author[29]{Xilong Fan}{}
\author[30]{Xing-Jiang Zhu}{}
\author[11]{Youjun Lu}{}
\author[31]{Fan Xu}{}
\author[32]{Kangfa Cheng}{}
\author[20]{\\Da-Bin Lin}{}
\author[33]{Xiao-Hong Zhao}{}
\author[5]{Jun-Jie Wei}{}
\author[17,18]{Bin-Bin Zhang}{}
\author[33]{Ji-Rong Mao}{}
\author[2]{Yongquan Xue}{}
\author[31]{\\Xinwen Shu}{}
\author[11]{Wenjie Zhang}{}
\author[12]{Wei-Li Lin}{}
\author[34,35]{Achille Fiore}{}
\author[3,4]{Zhuo Li}{}
\author[36]{Antonio Martin-Carrillo}{}
\author[36]{\\Joseph Fisher}{}
\author[20]{Fei Xie}{}
\author[5]{Ye Li}{}
\author[37]{Sandro Mereghetti}{}
\author[1]{Shao-Lin Xiong}{}
\author[39]{Yu-Han Yang}{}
\author[39]{\\Eleonora Troja}{}
\author[2]{Zi-Gao Dai}{}
\author[5]{Da-Ming Wei}{}
\author[20]{En-Wei Liang}{}
\author[40]{\\J.E. Horvath}{}
\author[40]{G. R. Cunha Sampaio}{}
\author[40]{L.G. Barão}{}
\author[41]{L.M.de Sá}{}

\address[1]{State Key Laboratory of Particle Astrophysics, Institute of High Energy Physics, Chinese Academy of Sciences, Beijing 100049, China}%
\address[2]{Department of Astronomy, University of Science and Technology of China, Hefei 230026, China}%
\address[3]{Department of Astronomy, School of Physics, Peking University, Beijing 100871, China}
\address[4]{Kavli Institute for Astronomy and Astrophysics, Peking University, Beijing 100871, China}
\address[5]{Purple Mountain Observatory, Chinese Academy of Sciences, Nanjing 210023, China}
\address[6]{INAF, Osservatorio di Astrofisica e Scienza dello Spazio, Via Piero Gobetti 101, I-40129 Bologna, Italy}
\address[7]{INAF – Istituto di Radioastronomia, Via Piero Gobetti 101, I-40129 Bologna, Italy}
\address[8]{Institute for Astronomy and Astrophysics, Department of High-Energy Astrophysics, Sand 1, 72076 Tuebingen, Germany}
\address[9]{Mullard Space Science Laboratory, University College London, Holmbury St Mary, Dorking, Surrey RH5 6NT, UK}
\address[10]{Department of Astronomy, Tsinghua University, Beijing 100084, China}
\address[11]{National Astronomical Observatories, Chinese Academy of Sciences, Beijing 100101, China}
\address[12]{Department of Astronomy, Xiamen University, Xiamen 361005, China}
\address[13]{Center for Theoretical Physics and College of Physics, Jilin University, Changchun, 130012, China}
\address[14]{School of Physics and Astronomy, Beijing Normal University, Beijing 100875, China}
\address[15]{School of Physics, Huazhong University of Science and Technology, Luoyu Road 1037, Hongshan District, Wuhan 430074, China}
\address[16]{Institute of Astrophysics, Central China Normal University, Wuhan 430079, China}
\address[17]{School of Astronomy and Space Science, Nanjing University, Nanjing 210023, China}
\address[18]{Key Laboratory of Modern Astronomy and Astrophysics (Nanjing University), Ministry of Education, Nanjing 210023, China}
\address[19]{School of Physics, Beijing Institute of Technology, Beijing 100081, China}
\address[20]{Guangxi Key Laboratory for Relativistic Astrophysics, School of Physical Science and Technology, Guangxi University, Nanning 530004, China}
\address[21]{Department of Physics, Ariel University, Ariel 40700, Israel}
\address[22]{University of Chinese Academy of Sciences, Chinese Academy of Sciences, Beijing 100049, People’s Republic of China}
\address[23]{Leicester International Institute, Dalian University of Technology, Panjin 124221, China}
\address[24]{Tsung-Dao Lee Institute, Shanghai Jiao-Tong University, Shanghai, 520 Shengrong Road, 201210, China}
\address[25]{School of Physics \& Astronomy, Shanghai Jiao-Tong University, Shanghai, 800 Dongchuan Road, 200240, China}
\address[26]{School of Physics and Electronic Science, Guizhou Normal University, No. 116 Baoshan North Road, Guiyang, Guizhou 550001, China}
\address[27]{School of Physics and Astronomy, Sun Yat-sen University, No. 2 Daxue Road, Tangjiawan, Zhuhai 519082, China}
\address[28]{Department of Physics, Faculty of Science, The University of Hong Kong, Room 415A, Chong Yuet Ming Physics Building, Pokfulam Road, Hong Kong 999077, China}
\address[29]{School of Physics and Technology, Wuhan University, Wuhan, Hubei 430072, China}
\address[30]{Department of Physics, Faculty of Arts and Sciences, Beijing Normal University, Zhuhai 519087, China}
\address[31]{Department of Physics, Anhui Normal University, Wuhu, Anhui 241002, China}
\address[32]{School of Physics and Electronic Information, Guangxi Minzu University, Nanning 530006, China}
\address[33]{Yunnan Observatories, Chinese Academy of Sciences, Kunming 650011, China}
\address[34]{INAF - Osservatorio Astronomico d’Abruzzo, Via Mentore Maggini Snc, 64100 Teramo, Italy}
\address[35]{INAF – Osservatorio Astronomico di Padova, Vicolo dell’Osservatorio 5, I-35122 Padova, Italy}
\address[36]{School of Physics and Centre for Space Research, University College Dublin, Belfield D04 V1W8, Dublin, Ireland}
\address[37]{INAF–Istituto di Astrofisica Spaziale e Fisica Cosmica di Milano, via A. Corti 12, 20133 Milano, Italy}
\address[38]{Max-Planck-Institut f\"ur Gravitationsphysik (Albert-Einstein-Institut), Am M\"uhlenberg 1, D-14476 Potsdam-Golm, Germany}
\address[39]{Department of Physics, University of Rome “Tor Vergata”, via della Ricerca Scientifica 1, I-00133 Rome, Italy}
\address[40]{Astronomy Department, IAG-USP. R. do Matão 1226, 05508-090 São Paulo SP, Brazil}
\address[41]{Universität Heidelberg, Zentrum für Astronomie (ZAH), Institut für Theoretische Astrophysik, Albert Ueberle Str. 2, 69120, Heidelberg, Germany}
\AuthorMark{First Author} % 标记 First Author 为对应作者
\AuthorCitation{First Author et al.} % 引用中显示的作者信息
\abstract{
In this new era of time-domain and multi-messenger astronomy, various new transients and new phenomena are constantly being discovered thanks to the rapid advances
in observations, which provide the excellent opportunity to study the physics in the extreme environments. The enhanced X-ray Timing and Polarimetry mission
(eXTP), planned to be launched in 2030, has several key advantages, including advanced polarimetry, high sensitivity \& large effective area, and wide energy range coverage, which make it a groundbreaking project in high-energy astrophysics. In this article, we briefly introduce the potential time-domain and multi-messenger targets for eXTP, including gravitational-wave (GW) counterparts, gamma-ray bursts (GRBs), magnetars and fast radio bursts (FRBs), tidal disruption events (TDEs), supernovae, high energy neutrinos and TeV active galactic nucleis (AGNs), and so on. We discuss the advantages of future eXTP observations for detecting these sources, their detection capabilities, the abilities to distinguish theoretical models, and their applications in gravity and cosmology.

%is a science mission dedicated at the X-ray aspect of astronomical events,  which holds the potential to address a series of fundamental
%open questions in astrophysics. 

%which consists of three main instruments: the Spectroscopic
%Focusing Array (SFA), the Polarimetry Focusing array (PFA)
%and the Wide-band and Wide-field Camera (W2C),
}
%%% Keywords. ?????
\keywords{eXTP mission, time-domain astronomy, multi-messenger astronomy, high-energy astrophysics}
\PACS{95.85.Nv,98.70.Rz,97.60.Gb,95.55.Ka,14.60.Lm}
\maketitle

%\tableofcontents%?????

%%%%%%%%%%%%%%%%%%%%%%%%%%%%%%%%%%%%%%%%%%%%%%%%%%%%%%%
%%% The main text. ???????
%???????????????????\cref{fig1}
%\twocolumn\onecolumn
%%%%%%%%%%%%%%%%%%%%%%%%%%%%%%%%%%%%%%%%%%%%%%%%%%%%%%%
\begin{multicols}{2}
\section{Introduction}\label{sec:Introduction}
%{\it Contributors: Xue-Feng Wu, Sandro Mereghetti, Jin-Jun Geng, Di Xiao, Shao-Lin Xiong}
%Outlines
% A brief review of the time-domain and multi-messenger science
%The full capabilities of simultaneous multi-wavelength/messenger and polarimetry with excellent sensitivities (wide field-of-view) being realized in the coming decades.   
% GW, neutrino 
%A brief review of previous missions
%The status of different scientific topics
%The introduction of the eXTP mission
%The key scientific goals of eXTP
%The potential breakthrough by eXTP

\noindent Time-domain and multi-messenger astronomy (TDAMM) observations have been unlocking vast discovery potential across virtually all domains of astrophysics by providing unprecedented insights into the dynamic and transient phenomena of the Universe~\citep{2020arXiv200705546C}. The X-ray sky harbors a rich tapestry of phenomena, including (but not limited to) gravitational wave electromagnetic counterparts (GWEM),
gamma-ray bursts (GRBs), magnetar bursts and fast radio bursts ({\ysx FRBs}), tidal disruption events (TDEs), supernovae and fast blue optical transients (FBOTs),  as well as TeV/neutrino-associated outbursts. 
In the past decades, these transients have sparked enormous interest and motivation among the astronomy community. 

The study of time-domain high-energy transients has achieved remarkable success, largely attributed to the pioneering contributions of X-ray satellites. These include both narrow-field X-ray telescopes, such as Chandra, XMM-Newton, and the Neil Gehrels Swift Observatory (Swift/XRT), as well as wide-field X-ray monitoring instruments like Rossi X-ray Timing Explorer (RXTE) and the International Gamma-ray Astrophysics Laboratory (INTEGRAL). These advanced facilities have revolutionized our understanding of astrophysical phenomena, delivering groundbreaking discoveries in fields such as stellar explosions and compact objects. Recent progress in transient X-ray sky exploration has been significantly enhanced by the successful launch and operation of the Einstein Probe (EP) \citep{EP} and the Space-based multi-band astronomical Variable Objects Monitor (SVOM) \citep{SVOM} missions. They are currently discovering a substantial number of X-ray transients, yet the origins of many of these sources still await identification.

Multi-messenger astronomy has entered a golden era since the window of GW observation opened in 2015~\citep{Abbott2016}. In addition to traditional electromagnetic wave signals, now we may gather information of an astrophysical system via other three messengers, namely, GWs, neutrinos, and cosmic rays. Despite great progress and achievements in this field, many fundamental questions remain unresolved, particularly in anticipation of the next generation of multi-messenger observational facilities \citep{Wilkes22}. Between the end of the 2020s and early 2030s, great advances are expected in GW and neutrino astronomy due to planned upgrades of current detectors, such as for instance the A$+$ \citep{Abbott2020LRR23} and A$\#$\footnote{\url{https://dcc.ligo.org/public/0183/T2200287/002/T2200287v2_PO5report.pdf}} configurations 
 of the two Advanced Laser Interferometer Gravitational Wave Observatory (aLIGO) interferometers in the US \citep{Aasi2015}, and the ongoing construction of the next generation neutrino detector IceCube-Gen2~\citep{IceCube_Gen2}, which is a major upgrade of the current IceCube in the South Pole.  
During the second half of the 2030s, next generation ground-based GW detectors such as Einstein Telescope (ET)~\citep{ET_3rdGW} and Cosmic Explorer (CE)~\citep{Cosmic_Explorer} are expected to be in operation, with planned sensitivity about one order of magnitude higher than the current one. At the same time, the Laser Interferometer Space Antenna (LISA), planned to be launched in 2037~\citep{LISA}, as well as Taiji and TianQin, will extend the observations of GW sky down to the mHz domain, allowing us to further explore the Universe in the low-frequency domain of gravitational radiation. In addition, various Pulsar Timing Arrays (PTAs) are likely to open a new window for the GW detection at nHz frequency bands in the near future.
The flagship X-ray observatories deployed in this epoch will greatly match and complement the broader and deeper horizon advanced by these facilities.

The enhanced X-ray Timing and Polarimetry mission (eXTP), planned to be launched early in 2030, is a flagship mission dedicated to the X-ray aspect of astronomical events in the time-domain and multi-messenger era. 

In the new baseline design, the scientific payload of eXTP consists of three main instruments: the Spectroscopic Focusing Array (SFA), the Polarimetry Focusing Array (PFA) and the Wide-band and Wide-field Camera (W2C). Here, we provide a brief introduction to the scientific payloads and a more detailed description of main instruments and mission could be found in the pioneering article~\citep{2016SPIE.9905E..1QZ}.

The SFA consists of five SFA-T (where T denotes Timing) X-ray focusing telescopes covering the energy range $0.5$--$10\, \mathrm{keV}$, featuring a total effective area of $2750\,{\rm cm^2}$ at $1.5\, \mathrm{keV}$ and $1670\,{\rm cm^2}$ at $6\,\mathrm{keV}$. The designed angular resolution of the SFA is $\le 1^\prime$ (HPD) with a $18^{\prime}$ field of view (FoV). The SFA-T are equipped with silicon-drift detectors (SDDs), which combine good spectral resolution ($\sim$ 180~eV at 1.5~keV) with very short dead time and a high time resolution of $10\,{\mu\mathrm {s}}$. They are therefore well-suited for studies of X-ray emitting compact objects at the shortest time scales. The SFA array also includes one unit of the SFA-I (where I signifies Imaging) telescope equipped with pn-CCD detectors (p-n junction charged coupled device), to enhance imaging capabilities, which will detect weak and extended sources or provide  strong upper limits on their flux. The expected FoV of SFA-I is $18^\prime \times 18^\prime$. Therefore, the overall sensitivity of SFA could reach around $3.3\times 10^{-15}\,{\rm ergs\,cm^{-2}\,s^{-1}}$ for an exposure time of $1\, \mathrm{Ms}$. Since it is not excluded that the SFA might in the end include six SFA-T units, simulations presented here have taken this possibility into consideration. 

The PFA features three identical telescopes, with an angular resolution better than $30^{\prime\prime}$ (HPD) in a $9.8^{\prime} \times 9.8^{\prime}$ FoV, and a total effective area of $250\,{\mathrm{ cm^{2}}}$ at $3\, \mathrm{keV}$ (considering the detector efficiency). Polarization measurements are carried out by gas pixel detectors (GPDs) working at 2---- 8\,keV with an expected energy resolution of 20\% at 6\,keV and a time resolution better than $10\,{\mathrm {\mu{s}}}$ \cite{Costa01,Bellazzini03,Bellazzini07,Bellazzini13,eXTP2019}. The instrument reaches an expected minimum detectable polarization (MDP) at $99\%$ confidence level ($\mathrm{MDP}_{99}$) of about $2\%$ in $1\,\mathrm {Ms}$ for a milliCrab-like source.

%The science payload is completed by the W2C, consisting of a coded mask camera that covers around $60^{\circ} \times 60^{\circ}$ of the sky with a sensitivity of $4\times 10^{-7}\,{\rm ergs\, cm^{-2}\,s^{-1}}$ for an exposure time of $1\,\mathrm {s}$ in the energy range $10$--$1000\, \mathrm{keV}$. The instrument has an angular resolution of $20^{\prime}$ and an energy resolution of about $30$\% at 60\,keV.
The W2C is a secondary instrument of the science payload, featuring a coded mask camera with a FoV of approximately {1500} {square degrees} (Full-Width Zero Response, FWZR). The instrument achieves a sensitivity of $4\times 10^{-7}\,{\rm ergs\, cm^{-2}\,s^{-1}}$ (1\,s exposure) across the 10–600\,keV energy range, with an angular resolution of $20^{\prime}$, a source location accuracy of 5', and an energy resolution better than $30\%$ at 60\,keV.

{\ysx The combination of the vast collecting area, the unprecedented polarimetric capabilties and its broad energy area} of eXTP \footnote{see an explicit comparison between the detectors on eXTP and other missions in Table 1 of the mission white paper \citep{zhang2025enhancedxraytimingpolarimetry}} hold the potential to address a series of fundamental open questions in astrophysics, 
including emission mechanisms of X-rays from various high-energy transients, the structure of outflows launched from a variety of central compact objects, the composition and magnetic field configuration of relativistic jets, etc. Specifically, the SFA can detect X-ray counterpart from binary neutron star mergers like GW170817 within $\sim 100$~Mpc, enabling comprehensive monitoring from the pre-merger phase through the burst and into the afterglow phase complemented by the W2C observations.
Meanwhile, the PFA could achieve a minimum detectable polarization of 1.7\% for bright GRB X-ray flares (flux level of $\sim 1$~mJy and duration of a few kiloseconds) with $\sim$~500 s of exposure. Basically, the observational strategies of different sources depend on their characteristic timescale and brightness. For sufficiently "fast" transients like GW counterparts, GRBs, magnetar bursts, FRBs or high-energy neutrinos, we generally need SFA and PFA to do quick follow up in ToO mode within a typical duration of 10 minutes. For longer-duration magnetar outburst, TDEs or SNe, combined ToO and regular monitoring is needed to characterize their complete evolution history, including both the early exotic and long-term X-ray behavior. In addition, W2C can monitor a large sky region for the relatively bright bursts, such as GRBs, magnetar bursts and jetted TDEs. Remarkably, the simultaneous observation of {\ysx SFA}, PFA and W2C of eXTP would unprecedentedly reveal the temporal, spectral, polarimetry and positional properties of these transients. 

This paper outlines the scientific goals of eXTP in the realms of time-domain and multi-messenger astronomy. It begins by examining GW counterparts (Section \ref{sec:GWn}), showcasing the eXTP's ability to investigate the X-ray features of these captivating events. The paper then delves into eXTP's contributions to understanding various astrophysical sources, including GRBs (Section \ref{sec:GRB}), magnetars and related radio transients (Section \ref{sec:FRB}), 
TDEs (Section \ref{sec:TDE}), supernovae (Section \ref{sec:SNe}), and other high-energy outbursts linked to TeV and neutrino emissions (Section \ref{sec:Neutrino}). Each section provides detailed information on how eXTP improves our understanding of these phenomena. A concise summary is offered in Section \ref{sec:sum}, encapsulating the broader impact of the mission on the field. {\ysx There are other types of targets that are also related to time-domain science but not covered in this white paper, as we have categorized them into other white papers in the series: Type I X-ray bursts are included in the white paper of ``dense matter" \citep{li2025densematterneutronstars}; transients related with black hole-star binaries are included in the white paper of ``strong gravity" \citep{bu2025probingstronggravityregion}; transients related with neutron star-stellar binaries are included in the white paper of ``strong magnetism" \citep{ge2025physicsstrongmagnetismextp}; transients related with stellar flares are included in the white paper of ``observatory science" \citep{zhou2025observatoryscienceextp}.}

%{\snz In some places no explicit arguments are made if eXTP observations can indeed address the specific scientific questions, e.g., the required eXTP observations, such as how long with which instrument in what mode, regular observations or ToO? It will be better if some quantitative statements are made about the feasibility to achieve each specific goal with eXTP. I will try to pick up some of these places.} 
%\textcolor{blue}{Reply: Some discussions are added in the second to last paragraph of Introduction.}

\section{Gravitational Wave Counterparts}\label{sec:GWn}

\subsection{Stellar-Mass Compact Binary Mergers}
%{\it Contributors: Shu-Xu Yi, Tian-Yong Cao, Ya-Cheng Kang, Li-Jing Shao, Zhen Zhang, Xilong Fan, Shao-Lin Xiong}

Ground-based laser interferometry gravitational wave detectors, such as currently operating aLIGO, aVirgo \citep{Acernese2015}, and KAGRA \citep{Akutsu2021}, as well as planned third-generation detectors like CE and ET that may run concurrently with eXTP~\cite{Kalogera:2021bya}, have a sensitivity frequency range of approximately 1-10000 Hz. Within this range, the primary astronomical sources of gravitational waves are compact binary mergers of stellar mass. For example, binary black hole (BBH) mergers, binary neutron star (BNS) mergers, and black hole-neutron star (BHNS) mergers were predominantly detected in the first four observing runs of the LIGO-Virgo Collaboration (LVC) and LIGO-Virgo-KAGRA (LVK). It has been observed that gravitational wave signals detected so far predominantly come from binary black hole mergers. However, it is generally believed that binary black hole mergers do not produce any electromagnetic signals \citep{2016arXiv160402537Z}, unless scenarios where the black holes are charged \citep{2016ApJ...826...82L,2016ApJ...827L..31Z} or the mergers take place in gaseous environments such as the accretion disks of Active Galactic Nuclei (AGNs) \citep{2020PhRvL.124y1102G}. %modifications to gravity are considered .

One of the most extensively studied GW sources that produced electromagnetic radiation is the BNS merger event GW170817 \citep{2017ApJ...848L..13A}. %Current stellar evolution theories predict that the merger rate density of BNSs is greater than that of binary black holes. As the third-generation GW detectors significantly increase the observable volume of the universe, the detection rate of BNSs, and/or neutron star-black hole mergers is expected to surpass that of binary black hole mergers.
{\ysx Current stellar evolution theory \citep{MandelBroekgaardenRatesCompact2022} and observations \citep{LIGOGWTC3population2023} both constrain the BNS merger 
rate density with uncertainties over a few orders of magnitude. As the third-generation GW detectors significantly increase the observable volume of the Universe, this leaves room for the detection rate of BNSs, and/or BHNS mergers to become comparable to or greater than that of BBH mergers.}\footnote{One can perform the estimation using the online GW detection simulator \texttt{GW-Universe Toolbox}\citep{2022A&A...663A.155Y}:\url{http://gw-universe.ihep.ac.cn}} More importantly, third-generation GW detector networks can achieve a 90\% credible region of localization uncertainty ($\Delta \Omega_{90 \%}$) as small as $\mathcal{O}\left(0.01\right)\,\mathrm{deg}^2$ for the best-localized BNS merger events \citep{2022PhRvD.105d3010L, 2024PhRvD.110d3001L}; moreover, with a signal-to-noise ratio threshold of 12, more than $60\%$ of the detected BNSs are expected to have $\Delta \Omega_{90 \%} \lesssim 100\,\mathrm{deg}^2$ \citep{2022ApJ...941..208I}. 

\begin{figure*}
\centering
\includegraphics[width=0.9\textwidth]{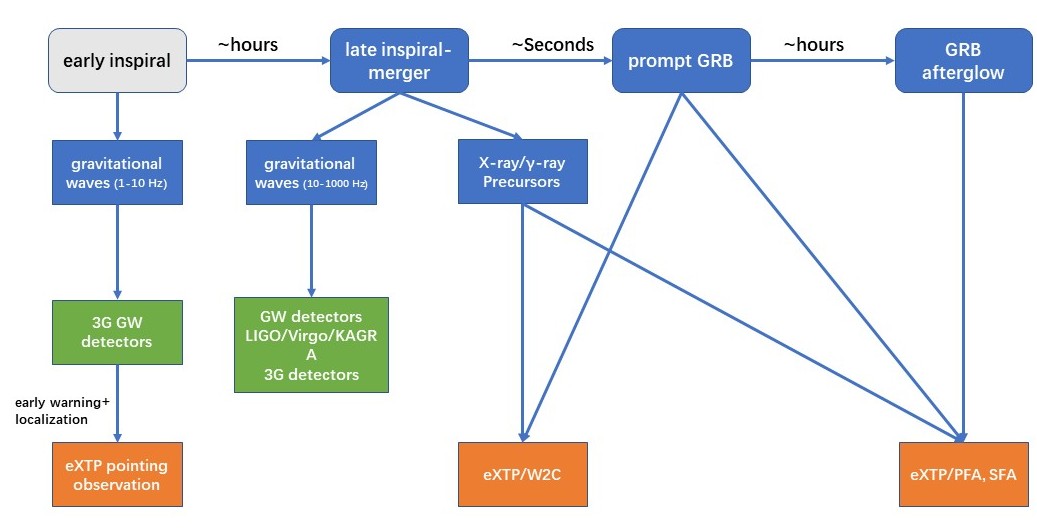}
\caption{Diagram shows the expected signals from different states of the binary mergers and the equipments which are expected to detect them.}
\label{fig:MSF0}
\end{figure*}

Generally, the mergers of BNSs and neutron star black hole binaries are accompanied by a rich spectrum of electromagnetic signals, which have the potential to be observed with equipment onboard the eXTP. In Figure \ref{fig:MSF0}, we outline the signals from different stages of these binary mergers and specify which equipment is expected to detect those signals. In particular, a sky localization of $\lesssim 100\,\mathrm{deg}^2$, as achievable by GW detectors, is less than one-tenth of the FoV of our eXTP/W2C, making it sufficiently precise for identifying and localizing potential gamma or X-ray counterparts. Subsequently, eXTP can provide a further refined localization and characterization of the source. In the following of this subsection, we will give a detailed description of each type of signal.

After the BNS merger (within 0.1-2 seconds), a large amount of ejected material rapidly accretes onto the newly formed compact object, potentially generating relativistic jets via the Blandford-Znajek (BZ) mechanism \citep{1977MNRAS.179..433B} or neutrino annihilation process \citep{2017NewAR..79....1L}, resulting in high-energy gamma-ray emissions that constitute the prompt emission of Type I GRBs (Type-I is defined as merger-originated GRB). As the gamma-ray counterpart to GW170817, GRB170817A serves as an example that confirms this scenario. Events GRB 211211A and GRB 230307A are also believed to originate from binary neutron star mergers and may potentially be kHz GW counterparts. As detailed below, future eXTP observations with W2C or SPA and PFA could allow us to detect similar multi-messenger events.

Before quantitatively analyzing the expected joint detection rates of BNS mergers with eXTP/W2C and different GW detectors, one needs to obtain the BNS population in our Universe. The number density per unit time for BNS mergers in the universe can be estimated as \citep{2015ApJ...812...33S},
\begin{align}\label{eq:bns_tot_number}
    \dot{N}_{\mathrm{BNS}} \approx \int_{0}^{z_{\mathrm{cut}}} \frac{\dot{\rho}_{0, \mathrm{BNS}} f(z)}{1+z} \frac{d V(z)}{d z} d z\,,
\end{align}
where $z_{\mathrm{cut}}$ is the redshift cutoff and we set it to 4; $\dot{\rho}_{0, \mathrm{BNS}}$ is the local BNS merger rate and we adopt ${\dot{\rho}_{0, \mathrm{BNS}} \simeq 106_{-84}^{+190}\, \mathrm{Gpc}^{-3}\,\mathrm{yr}^{-1}}$ from \cite{LIGOGWTC3population2023}; $f(z)$ is the
dimensionless redshift distribution factor, which depends on the delay-time distribution model of BNS mergers superposed on the star formation rate \citep{Zhu:2021ram}. Given that there are various types of merger population models in literature and their predicted results vary widely among different models, especially for the high-redshift regime, we adopt the log-normal delay model here for quick and simple simulations \citep{2019ApJ...881L..40S, Zhu:2021ram, 2023SSPMA..53j0014K}. The comoving volume element $\frac{d V(z)}{d z}$ in Equation~(\ref{eq:bns_tot_number}) can be expressed as,
\begin{align}\label{eq:com_v_element}
    \frac{d V}{d z}=\frac{c}{H_{0}} \frac{4 \pi D_{\mathrm{L}}^{2}}{(1+z)^{2} \sqrt{\Omega_{\Lambda}+\Omega_{\mathrm{m}}(1+z)^{3}}}\,,
\end{align}
where $D_{\mathrm{L}}$ is the luminosity distance,
\begin{align}\label{eq:luminosity_distance}
    D_{\mathrm{L}}(z)=(1+z) \frac{c}{H_{0}} \int_{0}^{z} \frac{d z'}{\sqrt{\Omega_{\Lambda}+\Omega_{\mathrm{m}}(1+z')^{3}}}\,.
\end{align}

Based on Equation~(\ref{eq:bns_tot_number}), we follow Zhu et al.~\cite{Zhu:2021ram} to randomly simulate the BNS merger population in our Universe and explore the joint observations with diffent GW detectors and eXTP/W2C. Note that not all of the BNS mergers can successfully produce relativistic jets and GRB events. Given that the local event rate density of short GRBs is typically lower than $\dot{\rho}_{0, \mathrm{BNS}}$ by a factor of a few \citep{2018NatCo...9..447Z, Zhang_2018}, we assume that in our simulation, half of all GW-triggered events can successfully launch a relativistic jet with a Gaussian structured profile, producing GRB\,170817A-like short GRBs (see e.g. Ref.~\cite{2019ApJ...881L..40S}). These GRB events are believed to be triggered when the $\gamma$-band flux of each BNS merger is larger than the effective sensitivity limit of eXTP/W2C. Considering the around $ 1.1$\,sr field of view (FoV) for eXTP/W2C, we expect that there are $\simeq 6_{-5}^{+11}$ yearly triggered short GRB events with different viewing angles (see e.g. Ref.~\cite{Kang:2022nmz}). Given the sensitivity curve of GW detectors, similar estimations can also be carried out for GW detections (see e.g. Ref.~\cite{Liu:2021dcr}). After the $\sim$ 2030s, the next generation ground-based GW detectors are expected to start their observation, which will have overlapping observational time with the eXTP mission. The currently proposed next-generation GW detector plans include LIGO Voyager (a
possible upgrade upon LIGO A+)~\cite{2020CQGra..37p5003A}, the ET in Europe \citep{ET_3rdGW, 2020JCAP...03..050M, 2023JCAP...07..068B}, and the
CE in the USA \citep{Cosmic_Explorer, Kalogera:2021bya}. As a crude order-of-magnitude estimate, we assume that these next-generation GW detectors can offer early-warning alerts of BNS mergers for eXTP/W2C with accurate localization and timing precision, that is, without FoV discounts. Therefore, we list their expected joint detection results with eXTP/W2C in Table~\ref{tab:bns_joint_detection}. 

%---------------------------------------------------------------------
\begin{table*}
    \renewcommand\arraystretch{2}
    \centering
    \caption{Yearly joint detection numbers of BNS mergers with different GW detectors and eXTP/W2C. The signal-to-noise ratio threshold for GW detection is set to be 12. $\theta_{\mathrm{obs}}$ represents the polar viewing angle of each BNS merger event. In our simulation, we assume that half of all GW-triggered events can successfully launch a relativistic jet with a Gaussian structured profile, producing GRB\,170817A-like short GRBs. Note that the percentages in brackets show the ratio to the total BNS mergers per year in our Universe. In the last column, we also list the maximum detectable redshifts for joint observations. More descriptions of the GW detecting strategy and detectors' configuration can be found in e.g. Refs.~\cite{Liu:2021dcr, Zhu:2021ram}.} 
    \setlength{\tabcolsep}{1.2cm}{\begin{tabular}{c c c c}\\
    \vspace{-4.5em}\\
    \toprule
    \toprule
    \vspace{-3em}\\
    & $\theta_{\mathrm{obs}} \lesssim 5^{\circ}$ 
    & $\theta_{\mathrm{obs}} > 5^{\circ}$ 
    & $z_{\max}$\\
    \midrule
    LIGO Voyager       
    & $2_{-2}^{+4}$ (0.02\textperthousand)      
    & $4_{-3}^{+7}$ (0.04\textperthousand)        
    & 0.38\\
    \cline{1-4}
    ET       
    & $41_{-33}^{+74}$ (0.43\textperthousand)  
    & $17_{-14}^{+31}$ (0.18\textperthousand)     
    & 0.87\\
    \cline{1-4}
    CE        
    & $52_{-41}^{+94}$ (0.55\textperthousand)   
    & $17_{-14}^{+31}$ (0.18\textperthousand)        
    & 0.93\vspace{0em}\\ 
    \bottomrule
    \end{tabular}}
    \label{tab:bns_joint_detection}
\end{table*}
%---------------------------------------------------------------------

Simultaneously capturing the prompt emission of gamma-ray bursts alongside gravitational waves has significant scientific implications. Firstly, the time difference in arrival between the gamma-ray burst and the gravitational wave can be utilized to constrain the propagation speed of gravitational waves \citep{2017PhRvL.119p1102C,2017ApJ...848L..13A,2024RAA....24h5004R}. The hope for more joint detection cases, especially on-axis cases (which have a smaller intrinsic time delay), is to enhance the constraints on gravitational wave speed to $\delta c/c\sim10^{-17}$ for 2nd generation GW detections ($\delta c/c\sim10^{-18}$ with 3rd generation GW detections)\citep{2024RAA....24h5004R}. {\yisx With the advantage from the mask code camera,} the W2C has much better localization ability ($5'$) than most current GRB detectors ({\yisx {\it Fermi}/GBM}, as a comparison, has the localziation ability of about 5-15$^\circ$), which can offer a much smaller sky map for the multi-wavelength follow-up. This is expected to result in more/earlier multiwavelength observations on BNS/BH-NS mergers, which is crucial for understanding GRB physics/nuclear astrophysics~\cite{Liu:2022mcd,Kang:2022nmz,2025ApJ...983L..34L}.  The high localization accuracy is critical for breaking parameter degeneracies in merging binary systems inferred from gravitational wave (GW) data, enabling more precise estimations of the chirp mass and luminosity distance. To quantitatively demonstrate these improvements, we perform a Bayesian parameter inference analysis using the \texttt{IMRPhenomPv2\_NRTidal} waveform model with the following setup:  
\begin{itemize}

\item Source parameters: Component masses $m_1 = 1.4\,M_\odot$, $m_2 = 1.4\,M_\odot$; dimensionless spins $a_1 = 0.05$, $a_2 = 0.03$; tidal deformabilities $\Lambda_1 = 800$, $\Lambda_2 = 900$; luminosity distance $D_L = 1\,\mathrm{Gpc}$; and orbital inclination $\theta_{jn} = 0$.  \\
\item Waveform settings: Reference frequency $50\,\mathrm{Hz}$, signal duration $6\,\mathrm{s}$, and sampling frequency $1024\,\mathrm{Hz}$.  \\
\item Priors: For GW-only analysis, uniform priors are adopted for sky localization (RA, Dec), orbital inclination ($\theta_{jn}$), and luminosity distance (\texttt{PowerLaw} $\alpha=2$, $D_L \in [500, 2000]\,\mathrm{Mpc}$). When incorporating eXTP/W2C localization, RA and Dec are restricted to Gaussians with $\sigma = 5^\prime$, while $\theta_{jn}$ is restricted to $[0, 5^\circ]$.  \\
\end{itemize}
The simulations show that combining GW data with simultaneous eXTP/W2C detection of the prompt emission reduces the uncertainty of the luminosity distance from $\sim1\,\mathrm{Gpc}^{+247\,\mathrm{Mpc}}_{-180\,\mathrm{Mpc}}$ to $\sim1\,\mathrm{Gpc}^{+27\,\mathrm{Mpc}}_{-25\,\mathrm{Mpc}}$, significantly enhancing the precision of parameter estimation. These are beneficial for studying the Hubble constant, the luminosity and opening angle of SGRBs \cite{2017PhRvL.119r1102F,2010ApJ...725..496N}, and for studying compact objects in the mass gap. 

Detecting prompt radiation associated with gravitational waves is also beneficial in confirming the astrophysical origins of gravitational wave sources, helping to lower the detection threshold for gravitational waves and thus expanding the detection range \cite{2024MNRAS.534.2715D}. The prompt emissions corresponding to gravitational waves also serve as an important intermediate link connecting their multi-band afterglows~\cite{Liu:2021dcr}. The low time delay between gravitational waves and gamma-ray bursts allows for better identification of prompt radiation compared to subsequent multi-band afterglows, while the joint localization provided by the gravitational wave detector network and gamma-ray burst detector network further narrows the source position uncertainty, greatly facilitating follow-up multi-band searches~\cite{Zhu:2021ram}.

After a binary neutron star merger occurs (from 1 hour to 1000 days), the external shock produced through the interactions of relativistic jets with the interstellar medium continues to produce X-ray emissions, contributing to the X-ray afterglow of Type I gamma-ray bursts. When the remnant mass from a binary neutron star merger is less than the Tolman-Oppenheimer-Volkoff (TOV) mass, a long-lived magnetar may emerge~\cite{YuYW2018}. The stellar wind from this magnetar continues to inject energy into the external shock, resulting in distinct characteristics of the X-ray afterglow light curves~\cite{Dai_2004,YuYW2010,YuYW2013}.  In contrast to the prompt emissions from highly relativistic jets, these signals exhibit higher isotropy, thus having a greater likelihood of detection. According to simple estimations, the higher sensitivity of SFA can detect signals within $\sim$100 Mpc for similar cases to GW170817 (whose peak flux is $\sim2\times10^{-14}$ erg/cm$^2$/s within 0.5-10 keV \citep{2020MNRAS.498.5643T}). Distinguishing these signals' spectral and light curve information not only aids in understanding the physics of the interaction between jets and the burst environment but also provides information about the remnants of the merger, thereby constraining aspects such as the state of matter in neutron stars. Furthermore, if the continuous gravitational waves emitted from the remnant are also detectable, joint X-ray - GW analysis could greatly improve the constraint on the ellipticity and frequency of the remnant over that of the X-ray analysis alone \cite{2023MNRAS.522.4294Y}.

%Theoretical models predict {\todo Figures}; the eXTP PFA instrument possesses a minimum polarization resolution of {\todo XXX}, which can be employed to limit the jet’s opening angle and inclination angle. If we assume that the direction of the jet is aligned with the orbital angular momentum of the progenitor binary, it will further assist in constraining information about the progenitor stars through gravitational wave observations. Particularly when studying progenitor stars via gravitational waves, the mass, luminosity distance, and orbital inclination angle information are degenerate. Thus, additional inclination angle information can effectively reduce uncertainties in mass and luminosity distance {\todo Figures}.

Given its high angular resolution, the SFA will assist in the precise localization of gravitational wave sources, thereby aiding in the determination of their host galaxies. Identifying host galaxies and conducting optical follow-up observations can yield independent redshift measurements. Together with independent measurements of the distances of gravitational wave events, this can provide a more accurate constraint on the Hubble constant \citep{2018Natur.562..545C} and serve as independent arbitration for the Hubble tension.

Furthermore, studies of GRB 230307A using LEIA (EP/WXT pathfinder) and GECAM have indicated \citep{Sun2024} that early X-ray analysis reveals critical properties of gamma-ray bursts. Therefore, capturing early X-ray emissions from binary neutron star mergers and potentially exploring their pre-merger X-ray precursor emissions is of great significance. During the late phase of operation of eXTP in the 2030s, we would expect a 3rd generation gravitational wave detector networks, which can capture signals from minutes to even hours before the mergers and facilitate early warnings \citep{2018PhRvD..97f4031Z}. As the time of merger approaches, the spatial location of binary systems becomes increasingly precise \citep{2024PhRvD.110d3001L}, and due to the high maneuverability of the eXTP satellites (30 deg/5 min), there is the potential for receiving valuable early detection X-ray signals even before the binary neutron star merger occurs.

Theoretical models predict that the X-ray polarization degree (PD) of GRBs, time-integrated over the prompt emission phases, are encoded with the geometry of the jet (more specifically, the ratio between the jet viewing angle $\theta_v$ and jet's opening angle $\theta_j$, see \cite{2024ApJ...970...10L}). With typical fluence, the MPD of PFA is sufficient to resolve the PD of the prompt X-ray emission of typical Type-I GRBs (see details in Section 3), and has the potential to be used to constrain on the viewing angle of the jet (although we can see from Section 3 that the constraints depend on the dynamic models of the bursts). If we assume that the direction of the jet is aligned with the orbital angular momentum of the progenitor binary, it will further assist in constraining information about the progenitor stars through gravitational wave observations, such as the mass and luminosity distance \citep{2019PhRvX...9c1028C}. 

Additionally, when the merger remnant is a magnetar, flares resulting from magnetic reconnection activities of the magnetar may also serve as detectable gamma-ray counterparts. \cite{2023ApJ...955....4Y} predicts that this counterpart will exhibit time delays, energy cutoff phenomena, and time-dependent features. The W2C is also expected to detect such electromagnetic counterparts within $\sim100\,\rm{Mpc}$, thus providing direct evidence regarding the nature of the remnants.

In the case of black hole-neutron star (BH-NS) mergers, theoretical predictions suggest that similar to binary neutron star (BNS) mergers, these events may also produce prompt gamma-ray bursts and X-ray afterglows. Observationally, the presence of gamma-ray bursts in connection with such mergers has not yet been firmly established. However, models indicate that the accretion of the neutron star material onto the black hole can initiate relativistic jets, leading to the emission of gamma-rays. The mass ratio between the neutron star and black hole, the spin of black hole, along with the state of matter within the neutron star, plays a crucial role in determining whether the neutron star can be torn apart by tidal forces before reaching the innermost stable circular orbit (ISCO) of the black hole. If the neutron star's mass is below a certain threshold relative to the black hole's mass, it may be completely disrupted, and its material will be accreted by the black hole, potentially contributing to the formation of a disk that produces X-ray emissions. As such, constraints on BH-NS mergers, especially the limits on early X-ray signals from eXTP/SFA, can provide additional insights into the mass ratios of these binary systems, as well as insights into the state of matter in neutron stars. Understanding these parameters is crucial for refining models of neutron star physics and the gravitational wave signals anticipated from such mergers. 

{\ysx Overall, further constraints on the properties of merging black holes and neutron stars will help narrow down the emerging complex landscape of supernova products \citep[e.g.,][]{SukhboldCoreCollapseSupernovae2016,BurrowsPhysicalCorrelations2024}. In tandem with other capabilities such as X-ray spectroscopy of supernova remnants \citep{ZhouObservatoryScienceeXTP2025}, eXTP can participate in connecting the formation of NSs or BHs to the properties of the stellar progenitor, greatly improving the accuracy of theoretical population models.} {\ysx It is worth noting that the degree of optimism regarding the achievement of the above scientific goals largely depends on the actual number of GW counterparts detection. Most of this uncertainty arises from the current estimation of the merger rate density of BNS as function of redshift. }

\subsection{Extreme mass-ratio inspirals %(mHz sources)
}
%{\it Contributors: Zhenwei Lyu, Zhen Pan, Junjie Mao, Huan Yang, Yunfeng Chen}
%Yunfeng and Zhenwei: EMRI population summary

\begin{figure}[H]
\centering
\includegraphics[width=0.8\columnwidth]{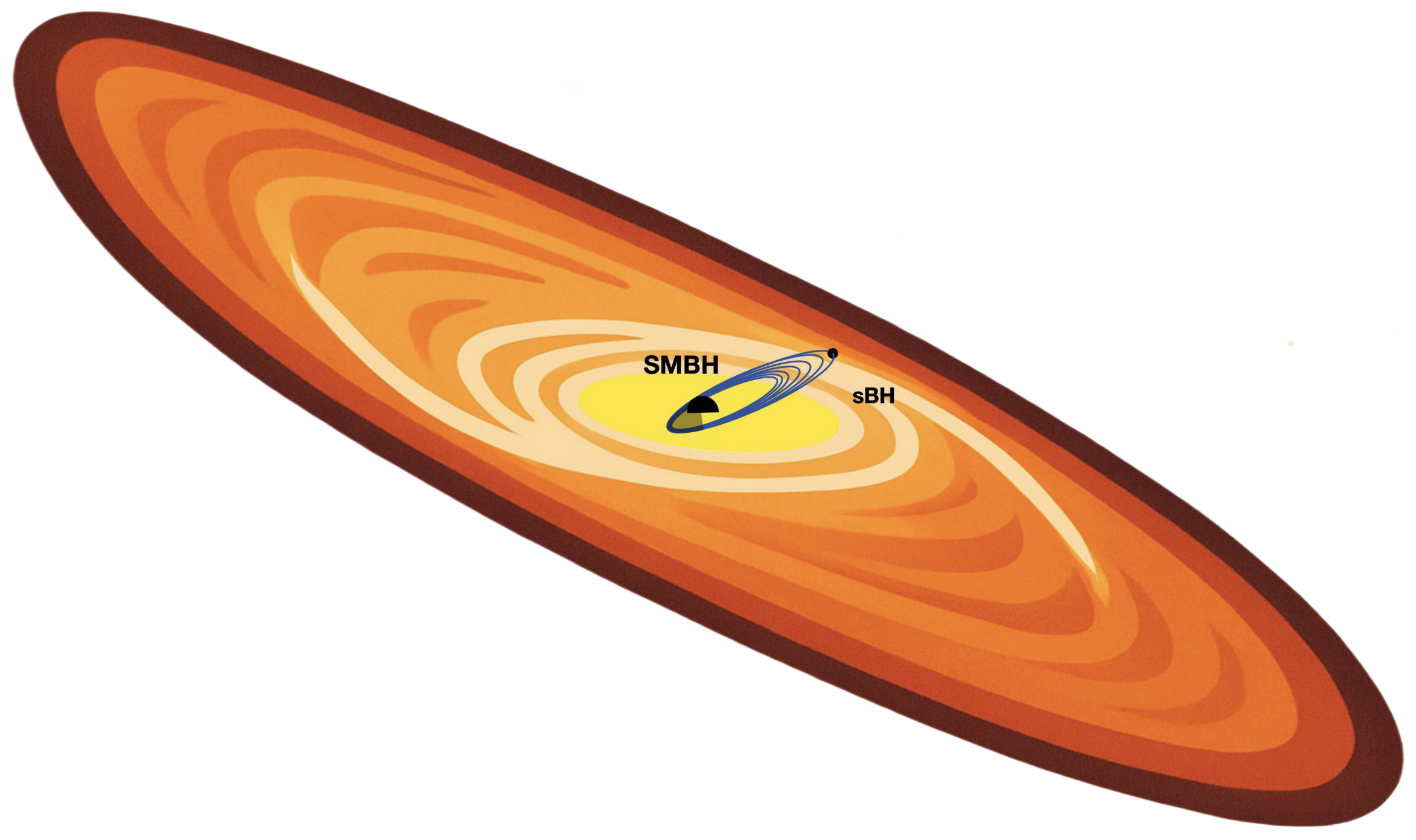}
\caption{Cartoon depiction of a wet EMRI within an AGN accretion disk.}
\label{fig:CartoonForWetEMRI}
\end{figure}

\renewcommand{\arraystretch}{1.2}
\begin{table*}[th]
\caption{EMRI rates in the redshift range $0<z<4.5$, assuming a sBH mass of $\mu=10 M_\odot$. The table includes total EMRI rates, expected LISA detection rates (for SNR$\geq 20$), and the fraction of events where the host AGN galaxy is resolvable. The wet EMRI models are based on the framework established in previous studies~\cite{Pan:2021ksp,Pan:2021oob}, adopting a conservative AGN fraction of $f_{\rm AGN}=1\%$ across the universe. For reference, the final row shows dry EMRI rates, assuming $N_p=10$ plunges per EMRI.}
\centering
\setlength{\tabcolsep}{5pt}
\begin{tabular}{c c c | c  c  c}
\hline\hline
Wet EMRIs & AGN disk  & $T_{\rm disk}$ [yr] &  Total rates [yr$^{-1}$] &  detection rates [yr$^{-1}$] & resolvable hosts [yr$^{-1}$]\\
\hline
& $\alpha$-disk & $10^6$ & 2700 & 120 & 30\\
&  & $10^7$ & 1100 & 40 & 9\\
&  & $10^8$ & 390 & 13 & 3 \\ \cline{2-6} 
& $\beta$-disk & $10^6$ & 3500 & 150 & 32 \\ %\textcolor{blue}{?}
&  & $10^7$ & 1200 & 50 & 10\\
&  & $10^8$ & 390 & 13 & 3\\
\hline
Dry EMRIs & $N_p$ &   &  Total rates [yr$^{-1}$] & detectable rates [yr$^{-1}$]  & \\
\hline
& 10 &  & 480 & 45 & -\\
\hline
\end{tabular}
\label{tbl:wetEMRIs}
\end{table*}

Extreme mass-ratio inspirals (EMRIs) represent one of the key extragalactic sources targeted by space-based gravitational wave observatories operating in the millihertz range, such as LISA, Taiji, and TianQin. EMRIs can be classified into two main types based on their formation channels: those influenced by the presence of accretion disks surrounding supermassive black holes (SMBH) in the mass range $4.0 < \log_{10} (M_{\rm BH}/M_{\odot}) < 7.0$, referred to as ``wet EMRIs'' \cite{Pan:2021ksp}, and those formed through gravitational interactions in dense nuclear star clusters, known as ``dry EMRIs''. Figure~\ref{fig:CartoonForWetEMRI} provides a cartoon illustration of a wet EMRI embedded in a warped AGN accretion disk.

Wet EMRIs are naturally accompanied by electromagnetic radiation from the AGN, with several multi-messenger applications with GW and X-ray observations, as discussed in Sec.~\ref{sec:spincal} and Sec.~\ref{sec:tran}.
The abundance and evolution of such sources are governed by multiple astrophysical factors, including the fraction of SMBHs found in AGNs, the structural properties of accretion disks, and the initial distribution of stellar-mass black holes (sBHs) within surrounding stellar environments. Observational data suggest that approximately $1\% \sim 10\%$ of the SMBHs are hosted in AGNs, with the fraction varying as a function of the redshift.

A reliable estimate of the wet EMRI population requires an accurate SMBH mass function. Recent studies based on optical observations of TDEs have refined our understanding of the local SMBH mass distribution, indicating a nearly flat behavior in logarithmic space~\cite{yao2023tidal}:
\begin{align}\label{eq:mass_function}
    \frac{dN_\bullet}{d\,\log M_\bullet} = 0.005 \left(\frac{M_\bullet}{3\times 10^6\, M_\odot}\right)^{\,\beta} {\,\rm Mpc}^{-3}\,,
\end{align}
where the power-law index is approximately $\beta\approx 0$. This updated mass function provides an improved characterization of SMBHs within the critical mass range of $10^4\sim 10^7 M_\odot$ that is relevant for EMRI formation.

Using the updated mass function, the expected population and detection rates of wet EMRIs have been assessed, with the results presented in Table~\ref{tbl:wetEMRIs}~\cite{Lyu:2024gnk}. The total EMRI rate across all SMBHs is determined by integrating the differential EMRI rate over SMBH mass $M_\bullet$ and redshift $z$, where the differential rate takes the form:
\begin{equation}
\frac{d^2 \mathcal{R}_{\text{wet}}}{dM_{\bullet} dz} = \frac{f_{\text{AGN}}}{1+z} \frac{dN_{\bullet}}{d M_{\bullet}} \frac{dV_c(z)}{dz} C_{\text{cusp}}(M_\bullet,z)\, \Gamma_{\text{wet}}(M_{\bullet}) \,,
\end{equation}
where the $1/(1+z)$ factor accounts for cosmological redshift effects, $V_c(z)$ represents the comoving volume element, and $C_{\text{cusp}}(M_\bullet,z)$ describes the fraction of SMBHs embedded in stellar cusps \cite{Babak:2017tow}. The average wet EMRI rate per AGN, $\Gamma_{\text{wet}}(M_{\bullet})$, depends on key astrophysical parameters such as the accretion disk lifetime $T_{\text{disk}}$, the rate at which sBHs are captured, and their migration timescales \cite{Pan:2021ksp,Pan:2021lyw,Pan:2021oob}.

Table~\ref{tbl:wetEMRIs} presents the total wet EMRI rates for sBHs with a mass of $\mu=10 M_\odot$ over the redshift range $0<z<4.5$, including the expected detection rates by LISA (assuming a signal-to-noise ratio of $\mathrm{SNR}\ge 20$) and the resolvable host AGN galaxies. To identify resolvable host galaxies, the sky localization volume of EMRIs is estimated using Fisher analysis, expressed as:
\begin{equation}
    V_{\text{sky}} = \frac{1}{2} r^3(z)\, \sigma(\ln{d_L})\, \Delta \Omega_s\,,
\end{equation}
where $r(z)$ represents the comoving distance to the source as a function of redshift, and $\sigma(\ln{d_L})$ denotes the relative uncertainty in distance measurement. For LISA-detected EMRIs, Fisher analysis suggests that $V_{\text{sky}}$ can be restricted to $\mathcal{O}(10^2)\, \rm Mpc^3$ at moderate redshifts ($z<0.3$), depending on the SNR and the inclusion of higher order waveform harmonics. This highlights the ability of LISA to achieve high localization precision~\cite{Lyu:2024gnk}.

These estimates are based on the methodology established in previous studies~\cite{Pan:2021ksp,Pan:2021oob,Lyu:2024gnk}, under the conservative assumption that the AGN fraction remains fixed at $f_{\rm AGN}=1\%$ throughout the universe. Across different models presented in Table~\ref{tbl:wetEMRIs}, approximately $20\%–25\%$ of detectable wet EMRIs have resolvable host galaxies, corresponding to a detection rate of 3–30 bright sirens per year. For comparison, the last row of the table includes dry EMRI rates, assuming an average of $N_p=10$ plunges per EMRI event~\cite{Babak_2017}.

\subsubsection{SMBH spin calibration}\label{sec:spincal}
%Junjie: spin calibration

GW wet EMRI observations are expected to precisely (better than 0.1~\%) constrain the spin of the SMBH \cite{Babak_2017}, which offers the opportunity to calibrate traditional electromagnetic wave SMBH spin measurement via the reflection spectroscopy \cite{Tanaka1995,Garcia2014}. Reflection spectroscopy utilizes the distinctive asymmetric profile of the 6.4 keV Fe line, characterized by a sharp blueshifted peak resulting from relativistic beaming effects and an extended redshifted wing caused by gravitational redshift as matter approaches the innermost stable circular orbit (ISCO) \cite{Reynolds_2021}. 

Wet EMRIs observed with LISA, Taiji and TianQin are in the redshift of $\sim0.1-0.3$ \cite{Lyu:2024gnk}. Their active type I SMBH host, which are not X-ray transients, typically have $2-10$ keV observed flux $\lesssim 10^{-12} {\rm~erg~s^{-1}~cm^{-2}}$ \cite{Lyu:2024gnk}. To measure the spin of these low X-ray flux targets using reflection spectroscopy, {\ysx a large effective area at the $3-8$~keV band is essential to well constrain the broad iron line profile. 
For comparison, the effective area of eXTP/SFA  \cite{eXTP2019} is nearly $\sim$2.5 times larger than that of XMM-Newton/EPIC (pn and MOS combined) and $\sim1.8$ times larger than that of NewATHENA/WFI.}

\begin{figure}[H]
\centering
\includegraphics[width=1.0\columnwidth]{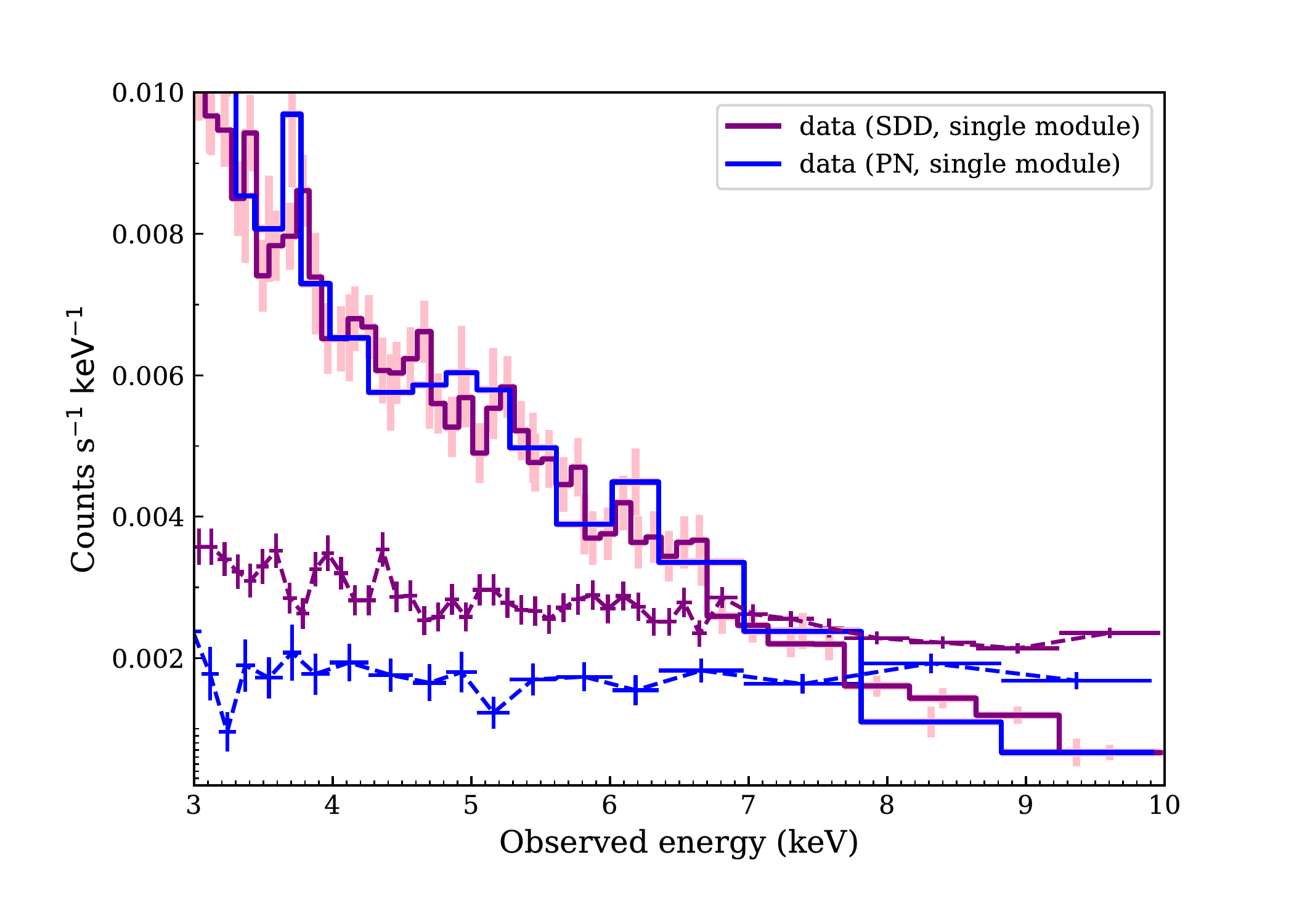}
\caption{Simulated SFA spectra of wet EMRI candidates with instrument background. The exposure time is set to 100 ks. The purple and blue solid curves are observed with single-module Silicon Drift Detector (SDD) and PN CCD detectors, respectively. Dashed curves are the corresponding background levels. }
\label{fig:emri_spec_bkg}
\end{figure}

Although coordinated eXTP and GW observations are not required, a three-stage eXTP observing strategy is necessary. First, use snapshot observations to determine the 2-10 keV flux of wet EMRI targets. Given the large scatter of empirical order-of-magnitude flux estimation \cite{Vasudevan2009}, snapshot observations can effectively pin down the flux level. Assuming a simple absorbed power law spectrum with $N_{\rm H} = 7\times10^{21}~{\rm cm^{-2}}$, $\Gamma = 1.8$, and the observed 2-10 keV flux of $5\times10^{-13}~{\rm erg~s^{-1}~cm^{-2}}$, the 5 and 10 ks eXTP/SFA snapshots were simulated. The 2-10 keV flux can be constrained better than 15~\% and 10~\% ($1~\sigma$ confidence level), respectively. In this case, background data start to dominate source data above 7~keV (Figure~\ref{fig:emri_spec_bkg}). For targets with even lower 2-10 keV flux, background data start to dominate at even lower energies, which makes spin measurement with reflection spectroscopy extremely challenging. 

The second stage of the observing strategy is to observe brighter X-ray ($>5\times10^{-13}~{\rm erg~s^{-1}~cm^{-2}}$) wet EMRI targets with moderate exposure. The main goal of this stage is to select targets with reflection features. We assume a simple absorbed {\tt relxill}\cite{Garcia2014,Dauser2014} spectrum with $N_{\rm H} = 7\times10^{21}~{\rm cm^{-2}}$, $\Gamma = 1.8$, high-energy cutoff $E_{\rm cut} = 300$~keV, the reflection fraction $R_{\rm s} = 1$, the inclination angle of the disk $i = 45^{\circ}$, the inner edge of the accretion disk at the ISCO radius of the spacetime, the emissivity profile of the disk is described by a simple power law with emissivity index $q = 3$, the iron abundance of the disk is the solar iron abundance ($A_{\rm Fe} = 1$), and the ionization parameter of the disk is $\log\xi = 2$ with $\xi$ in units ${\rm erg~s^{-1}~cm^{-2}}$, and the observed $2-10$ keV flux of $5.1\times10^{-13}~{\rm ergs^{-1}cm^{-2}}$. With 100 ks of eXTP/SFA observation, one can verify whether the spectrum has reflection features (i.e., the reflection scale is larger than zero at the $1~\sigma$ confidence level). However, be careful that, at this stage, the SMBH spin cannot be well constrained. 

The third stage of the observing strategy is to observe brighter X-ray ($>5\times10^{-13}~{\rm ergs^{-1}cm^{-2}}$) wet EMRI targets with prominent reflection features. This stage requires deep exposure, which is possible since the active type I AGN hosting the wet EMRI events are not transient sources. With the same model assumption mentioned above, a 900~ks deep exposure will yield a spin measurement better than 10~\% accuracy ($1~\sigma$ confidence level). Given the same flux level and exposure, the smaller the SMBH spin, the larger the uncertainties (e.g., for $a=0.7$, the 1~$\sigma$ uncertainty is $\sim20$~\%). 

Using pn charge-coupled device (CCD) detectors instead of Silicon Drift Detector (SDD) detectors can lower the background so that the background starts to dominate above $\sim7.8$ keV (Figure~\ref{fig:emri_spec_bkg}). This will slightly reduce the total exposure request. %The effective area curves of both types of detectors are almost identical.

\subsubsection{QPE and QPO origin}\label{sec:tran}
%Huan and Zhen: mHz QPEs/QPOs

In a gas rich environment, wet EMRIs are expected to produce quasi-periodic EM emissions, 
including the intensively discussed quasi-periodic eruptions (QPEs) and quasi-periodic oscillations (QPOs).

The currently discovered QPEs are in soft X-ray band with recurrence times longer than a few hours and are from nearby galaxies with redshift $z<0.1$. QPEs are preferentially
found in dwarf galaxies which host a SMBH in the low mass end $\sim (10^5-10^7) M_\odot$
and peak luminosity $\sim (10^{42}-10^{43})$ ergs s$^{-1}$. %\cite{Sun2013,Miniutti2019,Giustini2020,Arcodia2021,Arcodia2022,Chakraborty2021,Evans2023,Guolo2024,Arcodia2024,nicholl2024}.  
Although several models have been proposed to explain the physical origin of QPEs,
 more and more modeling analyses and observational evidence favor the EMRI+accretion disk model \cite{Franchini2023,Linial2023,Tagawa:2023fpb},
 in which QPEs are produced as a result of quasiperiodic collisions between the EMRI and the accretion disk.
 The timing analyzes of the QPE timing data show that there are two populations of EMRI that source QPEs; a low-eccentricity population that is consistent with the prediction of the wet channel and a high-eccentricity population that is compatible with the dry channel \cite{Zhou2024a,Zhou2024b,Zhou2024c}.
 The accretion disk may also be of two different origins: a disk formed from a recent TDE \cite{Linial2023,nicholl2024} or a (warped) AGN disk \cite{Tagawa:2023fpb,Lyu:2024gnk}. 
 Depending on the disk origin,  QPEs are classified as two types: type I QPEs that are associated with a TDE, and type II QPEs that
 are associated with an AGN and without a TDE \cite{Lyu:2024gnk}. 
 In any case,  as the sBH crosses the accretion disk with a relatively velocity $v_{\rm rel}$ higher than the local sound speed,
the gas within the accretion radius $r_{\rm acc}:= 2Gm/v_{\rm rel}^2$ will be shocked and heated up, 
and the total orbital energy loss of the sBH per collision
can be estimated as \cite{Zhou2024a}
\begin{equation}\label{eq:delta_E}
    \begin{aligned}
    \delta E_{\rm sBH} 
    &   
    = 4\pi \ln\Lambda  \frac{G^2m^2}{v_{\rm rel}^2}\frac{\Sigma}{\sin(\iota_{\rm sd})}\ ,\\ 
    &\approx 2\times10^{46} {\rm ergs} \left(\frac{\ln\Lambda}{10}\right) \Sigma_5 m_{30}^2 r_{100} \left(\frac{\sin\iota_{\rm sd}}{0.1}\right)^{-3} \ ,
\end{aligned}
\end{equation} 
where $\ln\Lambda$ is the Coulomb logarithm, $\Sigma$ is the disk surface density, $\Sigma_5:=\Sigma/(10^{5}$ g cm$^{-2}$),
$r_{100}=r/100 M_\bullet$, $m$ is the sBH mass, $m_{30}:=m/30 M_\odot$, $\iota_{\rm sd}$ is the angle of inclination between the EMRI orbital plane and the disk plane.
{\ysx The duration of QPEs is less straightforward to estimate from first principles, which depends on detailed dynamical and radiation processes of the shocked gas. 
From existing QPE sources, an empirical correlation between the QPE duration and the recurrence time ($\approx$ half the EMRI orbital period) has been found \cite{Hernandez-Garcia:2025ruv,Chakraborty:2025ntn,Arcodia:2025lwp}, $t_{\rm dur}\sim   t_{\rm recur}^n$ with the power law index $n\approx 1$. }

 As the EMRI migrates inward and enters the LISA sensitivity band with orbital radius $r\sim 10M_\bullet$ and orbital frequency $f_{\rm obt}\sim $ mHz,
the energy budget per collision is more uncertain due to the large uncertainties in the disk model.
In the commonly used $\alpha$-disk and the $\beta$-disk models \cite{Kocsis2011}, the disk surface densities are formulated as
\begin{equation*}
    \Sigma(r) = 1.7\times 10^5 \ {\rm g\ cm}^{-2} \alpha_{0.01}^{-1} \dot M_{\bullet,0.1}^{-1} r_{100}^{3/2}\ , 
\end{equation*}
and 
\begin{equation*}
    \Sigma(r) =  5.0\times 10^5 \ {\rm g\ cm}^{-2} \alpha_{0.1}^{-4/5} \dot M_{\bullet,0.1}^{3/5} M_{\bullet,6}^{1/5} r_{100}^{-3/5}\ , 
\end{equation*}  
respectively. In the case of $\beta$-disk model,
the energy budget per collision scales as $\delta E_{\rm sBH} \propto r ^{2/5} \propto f_{\rm obt}^{-4/15}$, i.e. the energy budget of mHz QPEs will be a few times lower than the currently observed low-frequency ($f_{\rm obt}\sim 10^{-5}$ Hz) QPEs. {\ysx If the $t_{\rm dur}-t_{\rm recur}$ correlation found in existing low-frequency QPEs also applies to mHz QPEs, 
the duration of mHz QPEs will be $\sim$ a few minutes and the peak luminosity could be comparable to or even higher than that of low-frequency QPEs depending on the radiation efficiency. }
In this case, mHz EMRIs will be an ideal multi-messenger  target for spaceborne GW detectors and X-ray telescopes. 
If the $\alpha$-disk model is a better description, the energy budget
scales as $\delta E_{\rm sBH} \propto r ^{5/2} \propto f_{\rm obt}^{-5/3}$, i.e. the mHz QPEs will be $\mathcal{O}(10^3)$ times weaker than their low-frequency counterparts.
In this case, the weak mHz QPEs are buried in the much stronger variability of disk emissions, therefore, are hard to identify directly from time domain observations but could be identified in the frequency domain as a mHz peak in the power spectrum density,
i.e., mHz quasi-periodic oscillations (QPOs) .

Recent studies \cite{Masterson:2025dph,Kejriwal:2024bna} have tentatively attributed some less abrupt mHz or sub-mHz QPOs in X-ray fluxes of AGNs to wet EMRIs 
that are  embedded in the disk or 
misaligned EMRIs interacting with possibly less dense disks.
This proposal  could be tested with multi-messenger observations.
If the mHz QPO source hosts an EMRI that is detectable by spacebore GW detectors,
the  EMRI frequency evolution on $O(1)$ year is expected to be accurately extracted from the GW signal.
The similar frequency evolution track should be found in X-ray observations of the mHz QPOs if they are sourced by the same EMRI.

Assuming a simple absorbed power-law spectrum with $N_{\rm H} = 5\times10^{20} {\rm cm^{-2}}$, $\Gamma = 1.8$, and the observed 0.3-10 keV flux of $5\times10^{-14} {\rm erg~s^{-1}~cm^{-2}}$, 10 ks eXTP/SFA observations were simulated. Both the 0.3-10 keV flux and the power law photon index can be constrained better than 10 \% ($1~\sigma$ confidence level). This observed flux corresponds to a QPE event at $z=0.26$ with a 0.3 - 10 keV luminosity of $10^{43}~{\rm erg~s^{-1}}$ with a negligible line of sight absorption.

To summarize, mHz EMRIs sourcing either abrupt QPEs or less abrupt QPOs are ideal targets for multi-messenger observations.
Complementary advantages of X-ray observations by eXTP in localizing the source
and of GW observations by spaceborne GW detectors in extracting the source frequency evolution enable a unique identification of the host galaxy.
{\ysx This identification can be firmly established by combining the information of source sky localization and the frequency information of both EMRI GW signals and QPEs.}
Several astrophysical applications follow \cite{Lyu:2024gnk}.

\subsection{Supermassive binary black holes %(nHz GW counterparts)
}\label{sec:GW_SMBH}
%[GE Junqiang:]
%{\it Contributors: Jun-Qiang Ge, Yunfeng Chen, Xing-Jiang Zhu, Youjun Lu}

The massive binary black holes (SMBBHs), as natural products of the hierarchical galaxy mergers in the $\Lambda$ cold dark matter ($\Lambda$CDM) cosmological frame \citep{1980Natur.287..307B, Yu02}, are important astrophysical sources of gravitational wave emissions in the nanohertz (nHz) frequency band ($10^{-9}-10^{-7}$ Hz) when the two black holes (BHs) orbit each other at subparsec separations, and having the total mass of SMBBHs ranging from $10^7$ to $10^{10} M_{\odot}$ \citep{2011MNRAS.414.3251L, 2012MNRAS.420..860S}. For SMBBH mergers with total masses at [$10^5$, $10^7$] $M_{\odot}$ radiate GWs in the frequency range of $10^{-4}-1$ Hz, which can be detected by space-based GW detectors \citep{nHz_TangY2018}, e.g., LISA, Taiji, and TianQin. 

Recently, several PTAs have reported evidence of a stochastic nanohertz GW background (GWB) with Hellings-Downs correlation at the confidence level $2 - 4 \sigma$ \citep{CPTA23hd, NG23hd, EPTA23hd, PPTA23hd, MPTA24hd}. With the accumulation of PTA observations over time, the discovery of the GWB, probabliy originated from SMBBHs, will eventually be realized. Correspondingly, individual SMBBHs will also be discovered \citep{CYL23cgws, CYL24pta}.

\subsubsection{Detectability of individual SMBBHs by GWs}
%[CHEN Yufeng:]

In the nHz frequency band, the detection of GWs from individual SMBBHs is anticipated as the next major milestone following the discovery of the stochastic GWB. 
% , which is believed to be the incoherent combination of the GWs of SMBBHs across the universe. 
The required PTA observational time baseline to achieve this breakthrough, along with the expected SMBBH detection rates, can be assessed by comparing the cosmic population of SMBBHs with the detection capabilities of specific PTA configurations.
% The former (i.e., the cosmic distribution of SMBBHs) can be evaluated given some astrophysical ingredients, including the galaxy stellar mass function (GSMF), the merger rate per galaxy (MRPG), the SMBH-host galaxy mass scaling relation, and the dynamical evolution of SMBBHs inside galaxy merger remnants (see for instance, \citep{CYL24pta}).
%\citep{BBR80, Yu02, CYL20bbh, CYL23cgws, CYL24pta}. 
% The latter (i.e., the PTA detection capability) can be approximately quantified by the PTA's sensitivity curve \citep{Moore15sns4pta, GLY22nf}, which depends on the number of monitored pulsars ($N_{\rm p}$), the timing precision ($\sigma_{\rm t}$), the monitoring cadence ($\Delta t$) and the observational time span ($T_{\rm obs}$).
Current analyses suggest that the first detections of individual SMBBHs are likely to be made by PTAs with superior timing precision such as CPTA and MPTA, rather than those with longer observational time baselines \citep{CYL23cgws, CYL24pta}.
To illustrate the near-term detection potential, Table~\ref{tab:BBHsByCPTA} presents the detection probability and expected number of detections for a mock CPTA configuration, evaluated under two representative SMBBH population models.
% In the table, each number outside the bracket represents the detection probability of individual SMBBHs, while the number inside the bracket represents the corresponding number of detections expected. For the reference model, the SMBH-host galaxy mass scaling relation has a significant redshift evolution (i.e., $\propto (1+z)^{2.07}$) so that the recently detected GWB in the nHz band can be entirely explained by cosmic SMBBHs; while for the empirical model, the SMBH--host galaxy scaling relation has no redshift evolution so that those cosmic SMBBHs can only contribute to $\sim 25\%$ of the detected GWB. 
% The mock CPTA configuration is assumed to have $N_{\rm p}=50$ pulsars being monitored with an average timing precision of $\sigma_{\rm t}=100{\rm \,ns}$ at a cadence of $\Delta t=0.04{\rm\,yr}$. Three choices of $T_{\rm obs}$ and two choices of $\rho_{\rm th}$ (the threshold $S/N$ for individual SMBBH detection) are considered. Clearly, the CPTA (or MPTA) is expected to detect individual SMBBHs in a few years. 
% Besides, the detection breakthrough is expected to be accompanied by a rapid accumulation of detection cases.
The results indicate that CPTA (or MPTA) could achieve the first individual SMBBH detections within a few years. Furthermore, once achieved, this breakthrough is expected to trigger a rapid increase in detected cases as observations continue.

While nHz observations target the SMBBHs at the high mass end ($\sim 10^8$--$10^{10}M_\odot$), spaceborne interferometers like LISA, Taiji and TianQin will probe the mHz band to detect lower-mass systems ($\sim 10^5$--$10^7M_\odot$). The event rate for these mHz SMBBHs is expected to reach several to tens per year \citep{Katz20}, and most sources are expected to exhibit an extreme signal-to-noise ratio, for example, $\rho\gtrsim 100$ \citep{Seoane17lisa}. Crucially, these systems will likely display distinct multimessenger signatures: GWs from the final inspiral and merger can precede potential electromagnetic counterparts by weeks to years \citep{MP05, Haiman09}, due to (1) the binary's decoupling from the circumbinary disk when the GW radiation dominates the decay of the orbit, and (2) the post-merger viscous refilling of the inner disk cavity before accretion-driven emission resumes.

\begin{table*}
\caption{Detection probabilities and corresponding expected numbers of detectable individual SMBBHs (in parentheses) for CPTA under different SMBBH population models, with varying PTA total observational time spans $T_{\rm obs}$ and detection thresholds $\rho_{\rm th}$. Two SMBBH population models are considered: Model A is calibrated such that the detected GWB signal is entirely attributable to SMBBHs; while Model B is not calibrated and SMBBHs only contribute $\sim 25\%$ of the detected GWB signal. To characterize the CPTA's detectability, we assume a mock CPTA configuration with the following parameters: number of pulsars $N_{\rm psr}=50$, average timing precision $\sigma_{\rm t} = 100{\rm\,ns}$, total observational time span $T_{\rm obs}$ (varying) and observational cadence $1/\Delta t = 25{\rm\,yr^{-1}}$.} 
\label{tab:BBHsByCPTA}
\centering
\begin{tabular}{cccccc}
\hline
\hline
$T_{\rm obs}({\rm yr})$ & \multicolumn{2}{c}{Model A} & & \multicolumn{2}{c}{Model B} \\
 \cline{2-3} \cline{5-6}
 & $\rho_{\rm th}=3$ & $\rho_{\rm th}=5$ &&  $\rho_{\rm th}=3$ & $\rho_{\rm th}=5$ \\
\hline
3.4 & 84.7\%(1.85) & 64.0\%(1.01) && 8.5\%(0.088) & 5.0\%(0.052) \\
5.0 & 100\%(10.1) & 99.5\%(5.59) && 41.5\%(0.539) & 26.8\%(0.311) \\
10  & 100\%(34.2) & 100\%(28.3) && 100\%(10.4) & 99.9\%(6.34) \\
\hline
\end{tabular}
\end{table*}

\subsubsection{The electromagnetic observations of SMBBHs}
%[GE Junqiang:]

SMBBHs as nHz/mHz GW emitters also have electromagnetic emissions; however, due to the small separations of these SMBBHs, it is hard to spatially resolve them directly by current telescopes. 
Efforts to find SMBBH systems have been ongoing for several decades, and several hundreds of candidates for SMBBH have been found based on different indirect signatures of electromagnetic signals ranging from $\gamma$-ray to radio bands \citep{2020RAA....20..160W, 2023arXiv231016896D}, and the combination of some signatures would be beneficial in identifying candidates for SMBBH \citep{2024A&A...687A..57G}. By now, none of these SMBBH candidates have been confirmed. 

Studying SMBBHs in the X-ray bands is an important area of research, as X-ray observations can reveal significant interactions between black holes and their environments, as well as accretion processes \citep{2018ApJ...865..140D, 2021ApJ...922..175N, 2022ApJ...928..137G, 2025ApJ...979..155P}. Signatures of the X-ray emission from SMBBHs and the related physical procedures can be mainly divided into three aspects: X-ray periodic variability, double broad K$\alpha$ iron lines, and spectral abnormalities that appear in the continuum.

\subsubsection{Identification of SMBBHs by eXTP}
%[GE Junqiang:]

One indirect signature for SMBBH candidates is the X-ray periodic variability detected by the time-domain observations of the SFA and PFA instruments,
%({\snz Why is SFA not used? SFA has much larger effective area than PFA. Does W2C have the sensitivity to do this? I doubt it. Any way, some quantitative statements should be made on the feasibility of achieving this goal.}\textcolor{purple}{Reply: The multi-epoch observations of SFA takes the dominant role for identifying SMBBHs, we emphasized this point at the end of current and the next paragraphs. As to using W2C for detecting AGNs, yes, a sensitivity of $4\times 10^{-7}\,{\rm ergs\, \mathrm{cm}^{-2}\,s^{-1}}$ is hard to detect AGNs. We hence deleted the descriptions about using W2C for detecting SMBBHs.}), 
which is different from the general variable AGNs in X-ray emission 
\citep{2001A&A...367..470G, 2001MNRAS.324..653L}. There are several mechanisms that can give rise to X-ray periodicities, including 1) the Doppler boosting effect of the two co-rotating BHs \citep{2012MNRAS.420..860S, 2022MNRAS.510.5929C, 2024MNRAS.52710168P}, 2) in the case that the two BHs are surrounded by a gaseous disk, periodic inflows triggered across the gas that excavated by the two BHs \citep{1996ApJ...467L..77A, 2007PASJ...59..427H, 2011MNRAS.415.3033R}, and 3) periodic outbursts produced by the tidally induced mass flows from the accretion disk into BHs, e.g., as in the case of OJ 287 \citep{Sillanpaa1988}.
The periodic variability can be observed more efficiently by the SFA instrument (because of a larger effective area) than the PFA instrument.
Additionally, using the SFA instrument for energy spectral monitoring would provide comprehensive evidence of SMBBHs by focusing on variations of both the continuum and broad Fe K$\alpha$ line profiles.
On the other hand, observations with the PFA instrument would possibly provide further information on polarization angles and help identify different X-ray emission structures, e.g., the corona, the two mini disks, the inner hot accretion flow, or the jet emission \citep{2024A&A...688L..27V}. 

Broad Fe K$\alpha$ lines centered at 6.4 keV appear to be common in AGNs as observed by the XMM-Newton and Rossi X-ray Timing Explorer (RXTE) X-ray telescopes \citep{2007MNRAS.382..194N, 2010A&A...524A..50D}. For a SMBBH system, when each BH accreting with a mini disk, the Doppler and relativistic effects make the observed Fe K$\alpha$ lines quite different from the case of single BHs; they are composed of two broad Fe K$\alpha$ lines from the two mini disks. Since the accretion rates, inclination angles of the two disks, the orbital plane of SMBBHs, and spins of the two BHs can all be different, a population of composite broad Fe K$\alpha$ lines that emitted from the SMBBH system are available for characterizing SMBBHs \citep{2001A&A...377...17Y, 2003ApJ...596L..31T, 2012MNRAS.420..860S, 2014AdSpR..54.1448J, 2025AdSpR..75.1441J}. Once the secondary BH opened an empty gap in the accretion disk of the primary BH, as the gap persists into the innermost disc and moves closer to the primary BH, the ripple effect would appear in the broad Fe K$\alpha$ line by observing a pair of dips in the broad line profiles, which could be detectable and suitable for energy spectral monitoring \citep{2013MNRAS.432.1468M}. The energy spectra at 0.5-10 keV band by eXTP/SFA instrument enables the analysis of the broad Fe K$\alpha$ line for AGNs at different redshifts. Multi-epoch observations of the broad Fe K$\alpha$ line profile would be crucial for identifying rotation feature and ripple effect of the two BHs. 

Considering that the accretion disks around the SMBBHs are different from the single SMBH case, the X-ray energy spectra of SMBBHs would become harder or softer than that of single accreting SMBHs, hence searching for peculiar X-ray spectral indices or abnormal X-ray-to-optical flux ratios becomes an indicator of SMBBH candidates \citep{2014ApJ...785..115R, 2015MNRAS.447L..80F}. Observations of those AGNs with periodic optical light curves might show different X-ray spectral characteristics than single SMBH cases as observed by the Chandra telescope at 2-10 keV \citep{2020ApJ...900..148S}.  
On the other hand, compared to normal AGN spectral energy distributions (SEDs), the SMBBH system may exhibit excess soft X-ray emission in the 0.5-10 keV range due to Compton cooling in shocks generated as streams from the circumbinary
disk hit the disks around each SMBH \citep{2014ApJ...785..115R}.
Currently, no significant abnormalities of X-ray spectra for SMBBH candidates have been found \citep{2014ApJ...785..115R, 2020ApJ...900..148S, 2020MNRAS.495.4061H, 2022ApJ...936..162S}.
The eXTP/SFA instrument with 0.5-10 keV range has natural advantage in detecting the above spectra-related signatures.

\subsubsection{Joint detection of SMBBHs by GW detectors and eXTP}
%[GE Junqiang:]

The joint observations of SMBBHs with eXTP and nHz/mHz GW detectors will provide complementary information on SMBBH masses, distances, and orbital parameters, enabling the multi-messenger astrophysics. Although current PTAs are insufficient to resolve individual SMBBH signals, future improvements could enable the detection of tens to hundreds of individual SMBBHs \citep{2016MNRAS.463L...6S, CYL24pta} (see Table \ref{tab:BBHsByCPTA} for predicted numbers). Future space-based GW detectors, such as LISA, Taiji, and TianQin, are expected to detect a large number of mergers of individual SMBBHs.

For eXTP, the detection number of SMBBHs can be roughly estimated by the quasar detection rate of XMM-Newton. By cross-matching the 4XMM catalog \citep{2020A&A...641A.136W} and the Quasar catalog in the Sixteenth Data Release of the Sloan Digital Sky Survey (SDSS-DR16) \citep{2020ApJS..250....8L}, the detection rate of quasars by XMM-Newton is $\sim 20$ per square degree. Considering that the field of view and flux limit of eXTP are comparable with XMM-Newton, it would detect $\sim 2,000$ quasars if surveying $100$ square degree. In this case, among these quasars, there would be $\sim 2$ SMBBH candidates, if the X-ray observation has similar detection rate ($\sim 0.1\%$) of SMBBH candidates as the optical one \citep{2015MNRAS.453.1562G, 2016MNRAS.463.2145C}.

For a given SMBBH candidate, joint detection by eXTP would significantly improve the precision of source localization by narrowing down the possible location to galaxies exhibiting central X-ray emissions with possible/quasi-periodic variability.
The localization precision for PTAs would be roughly $10 ~{\rm deg}^2$ at SNR$\sim 10$, and $1~{\rm deg}^2$ at SNR$\sim 20$ \citep{2011MNRAS.414.3251L, 2016MNRAS.455.1665B}, with the detailed precision values depending on different PTA detectors \citep{2025ApJ...978..104G}. For eXTP, the designed angular resolution $R\sim 40$ arcsec$^2$, given a detected nHz GW emitter with SNR$\sim 10$, since there is roughly $\sim 0.2$ SMBBH candidate per $10 ~{\rm deg}^2$, the confirmation of the SMBBH system is possible for the eXTP/SFA observation.
In the case of space-based GW detectors, the typical 90\% credible area of localizing a GW source by the LISA-Taiji joint networks is $\mathcal{O}({10^{-2}})~ {\rm deg}^2$ \citep{2021PhRvD.103f4057H, 2024FunRe...4.1072C}. The coalescence events of SMBBHs with total mass $\ge 10^6 M_{\odot}$ are expected to be up to 30 averaged over a ten years period \citep{LISA}, as estimated based on the observed galaxy merger rate \citep{2011ApJ...742..103L} and observed galaxy density at $z\le 2$ \citep{2013A&A...556A..55I}.
On average, there are approximately $20$ quasars per $1~ {\rm deg}^2$. To search for EM counterparts of nHz GW emitters, this would require around 45 observations for the SFA instrument (with a FoV of $\sim 0.022~ {\rm deg}^2$) and roughly 36 observations for the PFA instrument (FoV$\sim 0.028~ {\rm deg}^2$). In contrast, for the LISA-Taiji joint network, the sky localization precision is comparable to the FoV of eXTP, making the search for EM counterparts significantly more efficient than in the PTA case. With just a single observation, we can search for SMBBH signatures and confirm the system through the corresponding GW emissions.

The joint detection of multi-messenger observations can help measure the properties of host galaxies, if we can get a galaxy sample locating in the sky area of GW emitter, we can reveal the evolution history of these galaxies and confirm whether the PTA-detected GWB originates from SMBBHs. Based on the statistical properties of SMBBHs, one can reveal the history and dynamics of galaxy mergers, and enhance tests of general relativity by breaking potential degeneracies between localization parameters and polarization modes.

\subsection{Other classes of GW counterparts}
%{\it Contributors: Li-Jing Shao, Yong-feng Huang, Yong Gao, Ang Li,  Shu-Xu Yi}
\subsubsection{Core-collapse of massive stars}
The Gravitational waves from core-collapse of massive stars, produced via the neutrino-driven mechanism %\cite{Janka2012, Bethe1990, BetheApj1985} 
or magnetorotational mechanism, %\cite{Janka2012, Scheidegger2008, Kotake2012PTEP, Mezzacappa2014}, 
represent prime targets for current detectors like LIGO and next-generation observatories such as ET and CE \citep{LIGO2024arXiv,2023JCAP...07..068B, Cosmic_Explorer}. Continuous advancements in waveform extraction algorithms %\citep{Mezzacappa2024arXiv, Yuan2024arXiv, Yuan2024MNRAS, Hu2022ApJ, Heng2009CQG, Suvorova2019PhRvD, Chan2020PhRvD} 
enhance our detection capabilities for these transient signals. Recent studies indicate that the GW signal generated by Galactic core-collapse supernova events could be detectable with current methodologies \cite{Yuan2024arXiv, Yuan2024MNRAS, Chan2020PhRvD, Mitra2024MNRAS, Powell2024PhRvD}. The combined analysis of gravitational wave signatures with electromagnetic counterparts promises unprecedented insights into the explosion mechanisms and stellar evolution processes of supernovae.

\subsubsection{Close-in planetary systems involving strange stars}

%\textbf{[HUANG Yong-Feng:]}

Due to the extremely high density at the center, the internal composition and structure of the so-called neutron stars are still largely uncertain and fiercely debated \citep{2021Innov...200152G, 2024FrASS..1109463Z}. In fact, strange quark matter (SQM), which is composed of almost equal numbers of up, down and strange quarks, may be absolutely stable and thus could be the true ground state of hadronic matter \citep{1984PhRvD..30..272W}. It is called the SQM hypothesis. If this hypothesis is true, then the observed pulsars may actually be strange quark stars (often shortened as strange stars), but not neutron stars \citep{1986ApJ...310..261A}. 

In the 1 -- 2 $M_\odot$ mass range, a strange star has a radius very similar to that of a neutron star. Additionally, it could also be covered by a normal matter crust. All these features make it very difficult to discriminate between strange stars and neutron stars via observations. In previous studies, extensive attempts have been made to try to tell a strange star from a neutron star. For example, the cooling rate of SQM may be higher than that of hadronic matter, leading to a relatively lower surface temperature for strange stars \citep{1991PhRvL..66.2425P}. It is also noted that SQM has a larger bulk viscosity so that strange stars can spin more rapidly, with the spin period being less than 1 ms \citep{1989Natur.338..234K}. Further more, it has also been argued that the GW emission from a binary strange star merger should be different from that of a binary neutron star merger \citep{2010PhRvD..81b4012B}. However, in all these aspects, the difference between strange stars and neutron stars is generally too small to be tested through current observations.  

Interestingly, in the framework of the SQM hypothesis, the existence of a hydrostatically stable sequence of SQM objects has been predicted, ranging from 1 -- 2 $M_\odot$ strange stars to smaller strange-quark dwarfs and even much smaller strange-quark planets \citep{1995PhRvL..74.3519G}. While the difference between normal strange stars and neutron stars is small, a strange-quark dwarf may differ from a white dwarf markedly. More importantly, the difference between a strange quark planet and a normal matter planet will be even larger since the former will have a radius of only tens or hundreds of meters. Due to their extreme compactness, strange quark planets can spiral very close to their host pulsar without being tidally disrupted \citep{2017ApJ...848..115H}. We thus can try to identify strange-quark objects by searching for close-in pulsar planets \citep{2020ApJ...890...41K}.

While GW astronomy has opened a novel window to the universe, it will also shed new light on the research of strange quark stars. Like inspiraling binary neutron stars or black holes, when a close-in strange-quark planet finally merges with its host strange star, strong GWs will also be produced \citep{2015ApJ...804...21G, 2024MNRAS.531.3905Z}. Thus, merging strange star-strange planet systems would serve as new sources of GW bursts. This kind of merger events occurring in our local universe can be detected by GW detectors, such as the Advanced LIGO and the future Einstein Telescope. This effect provides a unique probe to SQM objects and is, we hope, a powerful tool for testing the SQM hypothesis.

Strong electromagnetic emissions would be generated during the merging process of a close-in strange-quark planetary system. The strange quark planet may have a normal matter crust, which would be stripped off by the tidal force from the host strange star as they become close enough. This will produce a miniature tidal disruption event, which should be associated with an X-ray transient. In the subsequent in-spiralling stage, the bare SQM core of the planet will interact with the strong magnetic field of the strange star, producing fierce radio emission. When the SQM core finally merges and collides with the strange star, an X-ray burst may be produced that should be somewhat polarized due to the strong magnetic field of the host. eXTP may be able to detect the X-ray transient associated with the miniature tidal disruption event and the X-ray burst induced by the final collision. The polarization features will help measure the magnetic field of the strange star and pin down the nature of these events. 

The event rate of these mergers is difficult to estimate, since we do not know what fraction of pulsars are strange stars. Assuming all the pulsars are actually strange objects, Geng et al. argued that at least 10 such events may occur in our Galaxy every year \citep{2015ApJ...804...21G}. But if only a small fraction of the pulsars are strange stars, the event rate will be much lower correspondingly. Thus, it is not easy for eXTP to discover these mergers independently in a blind survey. A more feasible strategy is to closely cooperate with gravitational wave detectors such as aLIGO, aVirgo, and KAGRA, which could provide an alarm when a gravitational wave event involving a planet-mass object is triggered. eXTP can hopefully find the electromagnetic counterpart (especially the afterglow) and help clarify its origin.

\subsubsection{Millisecond magnetars: precession modulated X-ray profile}

%\textbf{[Yong Gao]}

Millisecond magnetars are believed to form through various astrophysical processes, such as core-collapse supernovae %~\cite{Kasen:2009tg,Woosley:2009tu,Dessart:2012vc} 
and the merger of binary neutron stars%~\cite{Rosswog:2003tn,Dai:2006hj,Kiuchi:2023obe}
. These newborn millisecond magnetars can rapidly convert their immense rotational energy into electromagnetic and/or gravitational waves, driven by their strong magnetic fields and rapid spin~\cite{DallOsso:2021xbv}, as well as the possible fallback-accretion-induced dipole field decay \cite{2023RAA....23b5001C}.

The presence of an ultrastrong magnetic field can induce asymmetric deformations of the magnetar by exerting anisotropic stresses. The resulting magnetically-induced ellipticity is given by 
\begin{equation}
\epsilon_{\mathrm{B}}\sim \frac{B^2 R^4}{M^2} = 1.93 \times 10^{-4}  R_{10}^4 M_{1.4}^{-2} B^2_{16}\,,
\end{equation}
where $B$ is the magnetic field, with $B_{16}=B/10^{16}\,\rm G$, $M$ and $R$ are the mass and radius of the magnetar, respectively, with $M_{1.4}=M/1.4\,M_{\odot}$, $R_{10}=R/10\,\rm km$. If the rotation axis is not perfectly aligned with the magnetic (symmetry) axis, the magnetar will undergo free body precision, with period on the order of
\begin{equation}
    P_{\rm fp} \sim P/\epsilon_{\rm B} \sim  5.2 \times 10^{4} P_{1} R_{10}^{-4} M_{1.4}^{2} B^{-2}_{16}\,\rm {ms}\,.
\end{equation}
Here, $P$ is the spin period and $P_{1}=P/1\,\rm ms$. 

During free precession, the magnetic inclination angle varies periodically, resulting in periodic modulations in the observed X-ray light curves and potentially in X-ray polarimetry~\cite{Gao:2022hzd}.
Observational evidence of free precession has been reported in galactic magnetars, inferred from hard X-ray pulse profiles~\cite{Makishima:2014dua} and radio polarization measurements~\cite{Desvignes:2024vle}. More recently, X-ray polarization measurements of Her X-1 with the Imaging X-ray Polarimetry Explorer (IXPE) have suggested that the 35-day periodicity is driven by the free precession of the neutron star~\cite{Heyl:2023neg}. Possible signatures of precessing magnetars have also been identified in the X-ray afterglows of gamma-ray bursts~\cite{Zhang:2024eqa,Suvorov:2020eji,Suvorov:2021uyn}, where quasi-periodic oscillations observed during the X-ray plateau phase are interpreted as periodic modulations in the luminosity of magnetic dipole radiation, driven by quasi-periodic variations in the magnetic inclination angle during free precession.

Additionally, due to large time-varying mass quadrupole and fast spin, a millisecond magnetar may act as a powerful source of transient quasi-continuous gravitational wave emission (i.e., a long transient) in the $\rm kHz$ band~\cite{Zimmermann:1980ba,Gao:2020dwy,Gao:2020zcd,Jones:2000ud,Jones:2001yg,DallOsso:2008kll,DallOsso:2014hpa,DallOsso:2018dos}, making it a promising target for multi-messenger observations.
With its capability for X-ray polarimetry and high-time-resolution X-ray photometry, eXTP will provide an unprecedented opportunity to study early magnetar spin-down and free precession. These observations will not only improve our understanding of magnetar interior structure, but may also help confirm the long-standing conjecture that magnetars can be formed from binary neutron star mergers or core-collapse supernovae. %{\color{blue} [LS: maybe mention that it is called ``transient continuous gravitational waves'']  Yong: added}

\subsubsection{Accreting X-ray neutron stars}

%\textbf{[Lijing Shao: feel free to change texts ;-) ]}

Similarly, an accreting neutron star in a binary system may also develop an asymmetry in neutron star's mass distribution, the so-called neutron star mountain. Depending on the spin period and the accretion rate, such a neutron star could result in emission of continuous gravitational waves in the $10^1$ Hz to $10^3$ Hz band. LVK Collaboration have 
used data from the Third Observing Run
to conduct dedicated searches for continuous gravitational waves from twenty accreting millisecond X-ray pulsars~\cite{LIGOScientific:2021ozr}. 
These neutron stars may also have triaxial deformation and produce interesting modulation signals~\cite{Gao:2020zcd}.
Although no confident detection has been made yet, meaningful upper limits were cast on the maximum gravitational wave strain and were interpreted to constraints on the neutron star ellipticity (thus informing the accretion physics of the binary) and $r$-mode amplitude (thus informing the equation of state of neutron stars). Because of the possible wandering in the neutron star's frequency evolution, which increases the computational cost of continuous gravitational wave searches, observations in the X-ray band from the eXTP satellite would provide complementary information on the spin and accretion (e.g. from X-ray timing or profile observations), and thus be used to reduce the computational cost significantly. 

\subsubsection{X-ray precursor in the late stage of inspiralling binary NS}

%\textbf{[Yong Gao]}

The detection of GW170817 marked the dawn of a new era in multi-messenger astronomy, firmly linking gravitational wave events to post-merger sGRBs and kilonovae. Beyond these well-established electromagnetic counterparts, another yet-to-be-confirmed signal is precursor emission. 

Theoretical models propose two primary mechanisms for producing electromagnetic precursors. One class of models attributes precursor emission to magnetospheric interactions between the neutron star and its compact companion~\cite{Hansen:2000am}. In this scenario, pair production processes may be reignited as the two objects spiral closer together. Analytical studies~\cite{Lai:2012qe,Piro:2012rq} suggest that unipolar induction can efficiently dissipate electromagnetic energy, while force-free numerical simulations~\cite{Most:2020ami} indicate that differential motion - resulting from a misaligned stellar magnetosphere or stellar rotation \cite{Goldreich:1969sb,Zhang:2025ait} - can significantly enhance the emission. The accumulated magnetic twist in the flux tube may lead to powerful flares.

Another class of models focuses on the internal fluid dynamics of neutron stars driven by tidal interactions~\cite{Dall'Osso2013ADS}. The resonant excitation of various oscillation modes, such as interfacial modes~\cite{Tsang:2011ad,Sullivan:2022fsk}, torsional modes~\cite{Suvorov:2022ldw,Zhou:2023dcf}, and g-modes~\cite{Kuan:2021sin,miao2024resolving}, could lead to crustal fracturing, injecting substantial energy into the magnetosphere and potentially producing detectable high-energy emission.

On the observational side, many sGRBs have exhibited precursor emission~\cite{Troja:2010zm}, providing tantalizing hints that these theoretical mechanisms may be at play. In particular, the precursor of GRB 211211A displayed QPOs at $\sim22$ Hz \cite{Xiao:2022quv}, which is suggested to be the oscillations {\ysx within the tidal-yield crusts or interiors of magnetars in the magnetar superflare model~\cite{Zhang:2022qtd}, or the magnetoelastic or crustal oscillations in other models~\cite{Zhang:2022qtd,IsraelEt.al.2005,samuelsson2007neutron,gabler2011magneto,Link_2016,Neill2022MNRAS.tmp.1632N,Suvorov2022arXiv220511112S}}.
%%-> Reply: We revised the sentence and added closely related references.
%%若是只保留一篇，必需保留Zhang:2022qtd，其他文献中代表性列举QPO的详细机制。

%in the tidal-yield magnetar crust in the magnetar superflare model~\cite{Zhang:2022qtd}, {\bf or Alfvén waves propagating through the magnetosphere~\cite{Michel:1982fj}}.% (cite Zhen Zhang et al. Tidally-induced Magnetar Super Flare at the Eve of Coalescence with Its Compact Companion. Yong: added the references). 

Whether these precursor signals can produce enough X-ray photons to be detected by eXTP remains uncertain because precise modeling is still needed to refine predictions. Nevertheless, the search for precursor emission with eXTP offers a unique opportunity to probe the neutron star equation of state through asteroseismology and to investigate the complex electromagnetic dynamics of the magnetosphere in a dynamical spacetime \cite{Goldreich:1969sb,Zhang:2025ait}.

%{\color{blue} [LS: time delay  / luminosity / eXTP is not a survey telescope, need space-borne GW early warning?] }

%\subsubsection{Neutron star oscillation}

%\textbf{[Zhiqiang Miao, Ang Li]}

% ({\snz I don't see how eXTP is useful on this. Any way, some quantitative statements should be made on the feasibility of achieving this goal with eXTP.})

%\subsubsection{core-collapse supernova gravitational waves }
%\textcolor{blue}{XF:  Section 6 is on supernova.  Do we need this type of gw?}

\section{Gamma-Ray Bursts}\label{sec:GRB}

\noindent GRBs rank among the most energetic transient events in the Universe, releasing tremendous amounts of energy in the gamma-ray band on very short timescales \citep{Zhang_2018}. Phenomenologically, they are classified into two main categories based on their duration: ``long'' ($\gtrsim 2$~s) and ``short'' ($\lesssim 2$~s) bursts \citep{1993ApJ...413L.101K}. This duration-based division broadly corresponds to two distinct progenitor scenarios. Specifically, long-duration GRBs (often referred to as Type~II \citep{2009ApJ...703.1696Z}) are generally associated with the collapse of massive stars (the collapsar model), as evidenced by their coincidence with core-collapse supernovae \citep{Woosley_2006ARAA, Hjorth2012grb..book..169H}. In contrast, short-duration GRBs (often referred to as Type~I) are linked to mergers of compact objects, such as binary neutron stars or neutron star-black hole systems, with a prime example being GRB,170817A, which occurred in association with the double neutron-star merger event GW,170817 \citep{Abbott_2017PhRvL, Goldstein_2017ApJ}.

Despite this broad classification, many critical aspects of GRB physics remain unsettled. For instance, the precise nature of the GRB central engine—whether it is predominantly a black hole \citep{2017NewAR..79....1L} or a magnetar \cite{DaiZG1998a,DaiZG1998b}—has not yet been definitively established \citep{Nakar_2007PhR,Zhang_2018}. {\ysx For long GRBs, this is connected to the structure of their massive, rapidly-rotating stellar progenitors \citep{YoonSingleStarProgenitors2006}, the evolution of which remains uncertain and might connect long GRBs to BBH mergers \citep{WuAreLongGammaRayBursts2024}, potentially through chemically homogeneously evolving stars \citep{MarchantANewRouteTowards2016,deMinkTheChemicallyHomogeneous2016}.} Models of magnetars or black hole accretion engines interpret the critical mass as the upper and lower limits of the $M_{\rm TOV}$ (Tolman-Oppenheimer-Volkoff) equation, leading to significant discrepancies in equation of state models based on GRB research~\citep{li2020neutron}. On the other hand, competing theoretical models predict differing jet compositions and energy dissipation mechanisms, such as baryonic versus magnetically dominated outflows \citep{2001ApJ...552L..35Z,Lyutikov_2003}. The mechanisms responsible for launching relativistic jets and determining their composition are also under active investigation. Moreover, the origin of the prompt gamma-ray emission—whether from internal shocks, magnetic reconnection, or photospheric dissipation—remains an open question \citep{Kumar_2015PhR,Zhang_2018,Meng18}. Recent observational campaigns, including those conducted by GECAM, EP and SVOM, have provided valuable insights into the spectral and temporal evolution of prompt emission \citep{2024ApJ...975L..27Y,2024arXiv240416425L,2023arXiv230705689S,2024SCPMA..6789511Z,an2023insight}, including the power-law evolution of MeV spectral line from 37 MeV to 6 MeV \citep{2024SCPMA..6789511Z}.

In the afterglow phase, synchrotron radiation is generally accepted as the primary emission mechanism \citep{Sari_1998ApJ, Meszaros_1997ApJ}. However, the detailed afterglow profile can be significantly influenced by long-term central-engine activity, jet angular structure, and the nature of the circumburst medium \citep{Granot_2002ApJ, Lazzati_2007}. Multi-wavelength observations, from X-ray to radio, have been instrumental in constraining afterglow models and understanding energy injection processes \citep{Panaitescu_2002ApJ}. The detection of very high-energy gamma-ray emission from GRBs, such as those observed by \text{H.E.S.S.}, \text{MAGIC} and LHAASO, has further complicated the picture, suggesting the possibility of inverse Compton scattering or hadronic processes contributing to the afterglow \citep{MAGIC_2019Nature, Abdalla_2019Nature,2023SciA....9J2778C,2023Sci...380.1390L}.

Polarization studies are pivotal as they provide profound insights into jet structure and composition, the radiation mechanisms, and the magnetic field (MF) configuration in the radiation region \citep{Coburn_2003Nature, Toma_2009ApJ}. 
However, the current understanding of polarization in GRBs remains preliminary and fragmented, primarily due to the limited capabilities of existing observational facilities. In this section, we will investigate how the enhanced capabilities of the \text{eXTP} mission can advance our understanding of GRB phenomena by enabling detailed investigations of the temporal evolution of polarization signatures from the prompt emission to the afterglow phase.

%In this section, we investigate the polarization properties across different GRB emission phases. Polarization studies serve as a powerful diagnostic tool for probing magnetic field geometries, jet composition, and radiative mechanisms \citep{Coburn_2003Nature, Toma_2009ApJ}. 
%\textbf{In particular}, eXTP's GRB observations will provide a valuable opportunity to constrain the neutron star equation of state. Models of magnetars or black hole accretion engines interpret the critical mass as the upper and lower limits of the MTOV (Tolman-Oppenheimer-Volkoff) equation, leading to significant discrepancies in equation of state models based on GRB research~\citep{li2020neutron}. This issue requires urgent resolution. eXTP is expected to deepen our understanding of MTOV and, in combination with multi-messenger observations, including SVOM and LIGO, will further constrain the neutron star equation of state (\textbf{Ang Li}).  
%By examining how polarization signatures evolve from the prompt emission to the afterglow phase—observations that can be significantly enhanced by the capabilities of \textit{eXTP}—we aim to shed new light on the physical processes that govern GRB behavior and evolution..

\subsection{Prompt X-ray emission}\label{Sec:GRB Prompt X-ray emission}
%Kangfa Cheng, Mi-Xiang Lan, Jirong Mao 

%Polarization can serve as a probe of the jet structure and composition, the radiation mechanisms, and the magnetic field (MF) configuration in the radiation region.
The polarization of GRB prompt emission has been widely studied, including instantaneous and time-averaged polarization. Some polarization measurements in the prompt emission from the Gamma-Ray Burst Polarimeter (GAP), POLAR, and Cadmium Zinc Telluride Imager (CZTI) have been reported \citep{Yonetoku2012, Kole2020, Chattopadhyay2022}. These measurements have imposed some constraints on the GRB models. However, most of these measurements have a confidence level lower than $5\sigma$. The high-precision measurements in future polarization instruments can better constrain the theoretical models. The PFA on board eXTP is expected to achieve better polarization measurement results in the prompt X-ray emission.

Despite the wealth of observational data accumulated on prompt emission, the underlying radiation mechanisms of GRBs remain elusive.
This persistent ambiguity primarily stems from the absence of theoretical frameworks capable of comprehensively reconciling all observational signatures.
Extensive statistical analyses of temporal and spectral properties have nevertheless delineated two categories about the prompt emission mechanism.
One invokes the non-thermal emission mechanism,
owing to the non-thermal characteristic of the Band component observed in most of GRBs \citep{Band_D-1993-Matteson_J-ApJ.413.281B}.
In this scenario, previous works have shown that the synchrotron
or synchrotron-self-Compton radiation emitted by accelerated electrons is the promising mechanism  \citep[e.g., ][]{Tavani_M-1996-ApJ.466.768T,Racusin_JL-2008-Karpov_SV-Natur.455.183R,Geng18a}.
Another mechanism is a Comptonized quasi-thermal emission from the outflow photosphere
\citep[e.g., ][]{Thompson_C-1994-MNRAS.270.480T},
according to the quasi-thermal components detected in the spectrum of some GRBs \citep[e.g., ][]{Abdo_AA-2009-Ackermann_M-ApJ.706L.138A}.
The involved internal energy dissipation model of the jet are,
e.g., the fireball internal shock model \citep[e.g., ][]{Rees_MJ-1994-Meszaros_P-ApJ.430L.93R},
magnetic reconnection \citep[e.g., ][]{Zhang_B-2005-Kobayashi_S-ApJ.628.315Z},
dissipative photospheric model \citep[e.g., ][]{Rees_MJ-2005-Meszaros_P-ApJ.628.847R},
and an internal-collision-induced magnetic reconnection and turbulence model \citep{Zhang_B-2011-Yan_H-ApJ.726.90Z}.
Distinguishing between these two models in observations is of great significance for our understanding of the radiation mechanism of GRB prompt emission and X-ray flares. 

The polarization measurements can help us effectively distinguish these two models. In general, the synchrotron radiation can produce relatively high polarization degrees (PDs) in the range of $\sim 10\%- 50\%$ (time-averaged) with a large-scale ordered MF in the prompt X-ray emission (see Fig. 2 and 4 in \cite{Sui+Lan+2024}), while the photospheric emission produces a PD lower than $\sim 10\%$ (see Fig. 2 in \cite{Lundman+etal+2014}). Therefore, a high PD measurement suggests the synchrotron model, whereas a low PD measurement indicates the photospheric model.

The MF configuration in the radiation regions of GRB prompt emission is critical but unclear. The small-scale random field and the large-scale ordered field are two candidate MF models. These two MF models usually have different origins. The small-scale random field is generally produced by the Weibel instability in shock \citep{Gruzinov+Waxman+1999, Medvedev+Loeb+1999} or kinetic turbulence \citep{2011ApJ...731...26M,2013ApJ...776...17M}, while the large-scale ordered MF originates from the central objects of GRBs \citep{Spruit+2001}. The time-averaged synchrotron PDs of prompt X-ray emission in the large-scale ordered field ($ \sim 10\%- 50\%$) usually higher than in the small-scale random field $\lesssim 20\%$. Moreover, the ordered MF is generally assumed to have two possible configurations in the prompt emission regions: an ordered toroidal field and an ordered alinged field (aligned in the jet cross section). The ordered toroidal field may originates from a black hole through the Blandford-Znajek mechanism, while the ordered aligned field may originates from a magnetar \citep{Spruit+2001, Lan+etal+2021}. The polarization of these two MF models in GRB prompt emission has been studied in \citep{Lan+etal+2021, Sui+Lan+2024}. The synchrotron PD and the polarization angles (PAs) evolution of the two models are different \citep{Wang+Lan+2023, Cheng+etal+2024b}. A mixed MF model consisting of the ordered component and the random component has also been proposed in GRB prompt emission. The polarization of this mixed MF model has been studied in \citep{Lan+etal+2021, Tuo+etal+2024}. The future polarization measurements may distinguish these different MF models.

The jet structure of GRBs remains uncertain. Currently, there are two main competing jet structures: a uniform top-hat jet and a structured jet. The energy and velocity of the structured jet usually follow an angular distribution. The polarization properties of these two jet models show significant differences under various MF configurations and viewing angles \citep{Gill+etal+2020, Gill+etal+2021}. For an ordered MF, the highest synchrotron PDs of these two jet models can both reach $\sim 50\%$ (time-averaged) in the prompt X-ray emission. However, in the uniform top-hat jet, the synchrotron PDs drop sharply when the line of sight (LOS) slightly deviates from the jet edge, whereas in the structured jet, the synchrotron PDs decrease much more gradually when the LOS deviates from its core. The polarization measurements of GRB prompt emission may distinguish these two jet models. 

\begin{figure*}[!t]
  \centering
  \includegraphics[width=5in]{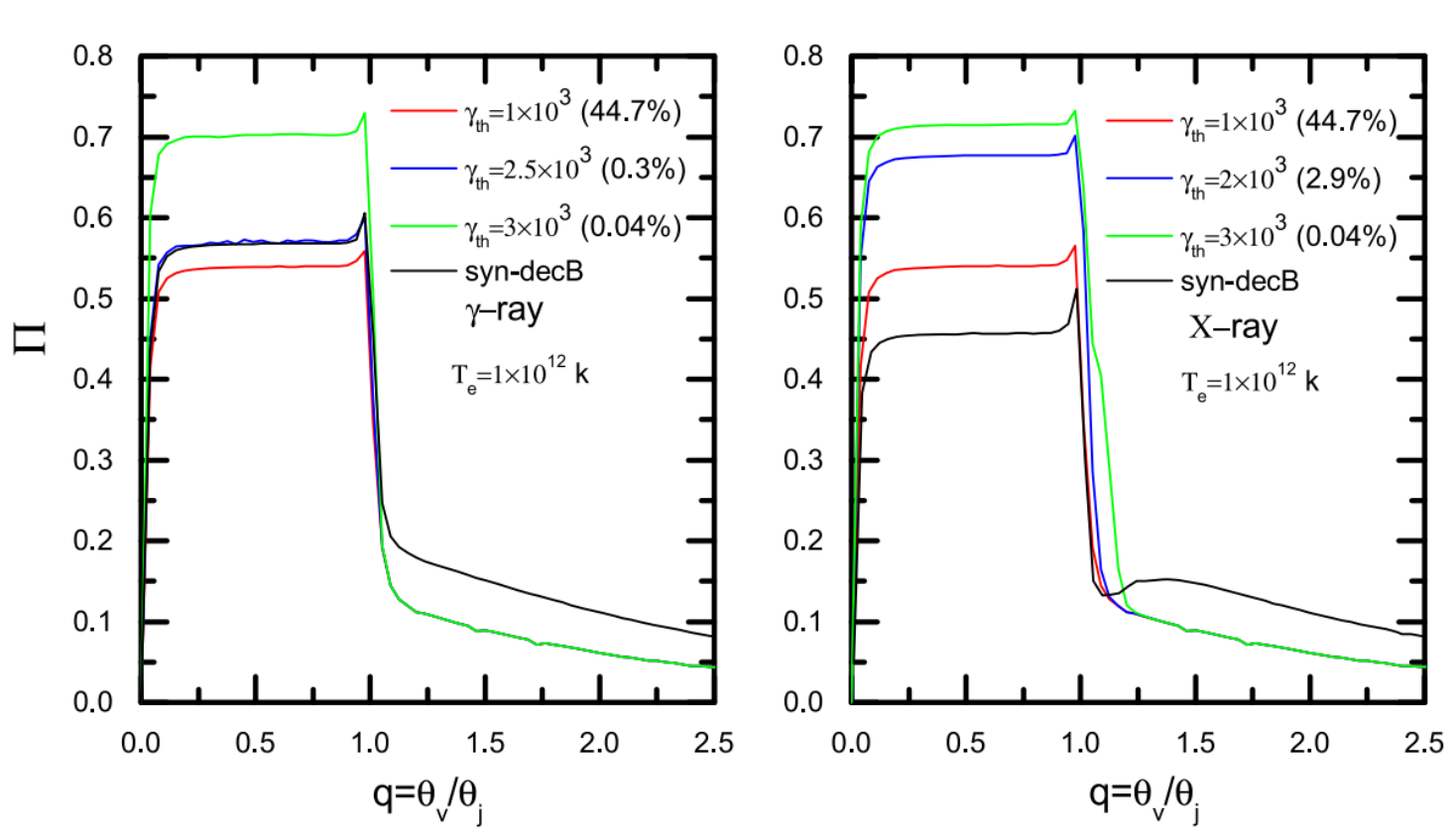}
   \caption{Time-averaged PDs with different normalized viewing angles (q) for various conjunctive Lorentz factors ($\gamma_{\rm{th}}$) in the gamma-ray and X-ray bands. “syn-decB” represents the pure nonthermal electrons case. The energy fractions of the nonthermal electrons for different $\gamma_{\rm{th}}$ are presented in the parentheses. This figure is from \citep{Cheng+etal+2024a}}
\label{gama-XrayPD}
\end{figure*}

Furthermore, polarization measurements can constrain the electron composition in GRB jets. The synchrotron radiation is generally produced by relativistic nonthermal electrons. The relativistic nonthermal electrons originate from the accelerations of internal shocks  \citep{Rees+Meszaros+1994, Paczynski+Xu+1994} or the dissipation of MF energy \citep{Thompson+1994, Zhang+Yan+2011}. However, recent particle-in-cell simulations of baryon-dominated relativistic shocks and Poynting-flux-dominated outflow have indicated that the resulting electron distribution is a combination of relativistic thermal and nonthermal components \citep{Spitkovsky+2008a, Guo+etal+2015}. 
Cheng et al. (2024a) \citep{Cheng+etal+2024a} considered this hybrid electron distribution and calculated the polarization in GRB prompt emission. The results have shown that the time-averaged PDs can be higher than $60\%$ in the gamma-ray and X-ray bands when the electron energy is dominated by relativistic thermal electrons (see Figure \ref{gama-XrayPD}). The high synchrotron PDs ($\gtrsim 60\%$) generally cannot be produced by the pure nonthermal electrons with typical power-law slopes \citep{Cheng+etal+2020}. This polarization feature combining the GRB spectra can help us distinguish the electron composition in the jets of GRBs and provide new insights into the particle acceleration mechanisms of the relativistic outflow. High-precision polarization measurements may verify these theoretical results.
Also, these thermal electrons could be diagnosed using afterglow spectra from synergistic observations in X-rays like eXTP and other optical telescopes \citep{Gao24}.

Studies have been done on the polarizations of prompt X-ray emission \citep{Lundman2018,PL2022,LLW_2024,LL_2025}. The predicted time-integrated PD ranges from $20\%$ to $50\%$ in X-ray band for the synchrotron model \citep{LLW_2024,LL_2025}. However, the predicted PD in X-ray band would vary between the photosphere models \citep{Lundman2018,PL2022}. The duration of the prompt X-ray emission is longer than that in gamma-ray band. For a short burst with a duration less than 2 s in gamma-ray band, its duration would be about 20 s in X-ray band \citep{LLW_2024}. Without GW early warning, it is almost impossible to detect such short duration X-ray emission by eXTP. However, for some ultra-long bursts in gamma-ray band  with duration longer than 1000 s, the corresponding X-ray emission would last for a longer time, which might have the chance to be detected by eXTP. 

\subsection{X-ray flashes}\label{sec:XRF}
%[Fan Xu, Yong-Feng Huang:]

X-ray flashes (XRFs) are a subclass of GRBs typically characterized by a very soft spectrum. They were first discovered by \textit{BeppoSAX} \citep{Heise..2001} and were extensively studied by several missions, including \textit{HETE}-2 and \textit{Swift}. XRFs are typically defined based on the fluence ratio between the X-ray ($S_{X}$) and gamma-ray ($S_{\gamma}$) bands. Comparing with classical GRBs, XRFs exhibit higher $S_X/S_{\gamma}$ ratios and
much lower spectral peak energies, e.g., XRF 060218 is a long, smoothly-evolving burst with peak energy $E_{peak} =$ 4.9 keV\citep{Soderberg2006Nat}, and XRF 020903 with the highest probability value being 3.7 keV\citep{Sakamoto2004ApJ}. However, XRFs and GRBs have similar low-energy photon indices, high-energy photon indices, and durations \citep{Sakamoto..2008}. These properties suggest that XRFs are the low-energy extension of the GRB population \citep{Barraud..2005}.

The origin of XRFs remains under debate. Two main scenarios have been proposed. The first scenario attributes the observed spectral softness to an off-axis viewing geometry \citep{Xu..2023a}. Alternatively, XRFs may be intrinsically faint GRBs observed on-axis, either due to a lower Lorentz factor or the reduced radiation efficiency \citep{Huang..2002}. Notably, these two scenarios are also commonly invoked to explain orphan afterglows \citep{Xu..2023b}.

Since most energy is released in X-ray band, eXTP will be an ideal tool for studying XRFs. The coming X-ray polarization measurement by eXTP may shed new light on the origin of XRFs. The temporal evolution of polarization degree differs significantly between on-axis and off-axis scenarios \citep{Sari..1999,Granot..2002}. Off-axis emission generally leads to a higher polarization degree compared to on-axis cases. 
{\ysx However, the polarization degree is also affected by other factors, including the jet structure, magnetic field configuration, and radiation mechanism. Therefore, it should be used as a complementary diagnostic alongside light curve and spectral evolution, rather than as a standalone criterion for model discrimination. }
The PFA onboard eXTP operating in the soft X-ray band is well-suited for measuring the polarization degree of XRFs. Its high time resolution enables a detailed study of the polarization evolution, which could help distinguish between the two scenarios mentioned above and reveal the nature of XRFs.

\subsection{X-ray plateau phase}
%Mi-Xiang Lan 

\begin{figure*}[!t]
  \centering
  \includegraphics[width=5in]{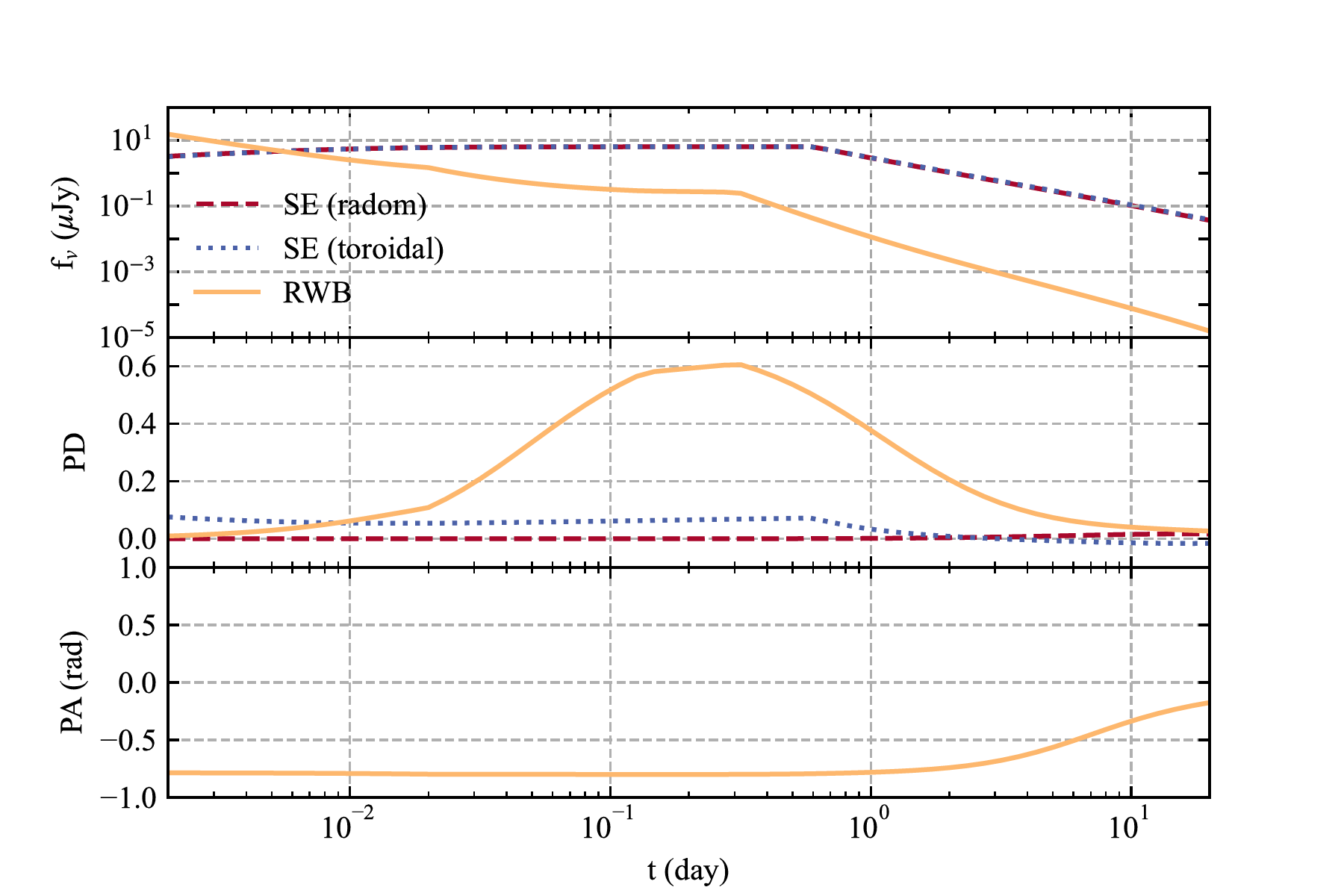}
  \caption{The light curves (upper panel), PD curves (mid panel) and PA curve (lower panel) of the GRB X-ray plateau phase at the energy band of eXTP/PFA (2-8 keV). The brown-solid, red-dashed, and blue-dotted lines correspond to the RWB model, the SE model with a random field in its radiation regions and the SE model with a toroidal field.}
\label{plateaupol}
\end{figure*}

The X-ray plateau, usually lasting 1 day with typical flux of $10^{-12}$-$10^{-10}$\,$\rm erg\cdot cm^{-2} \cdot s^{-1}$, is the shallow decay phase observed in GRB afterglows. About (40-50)$\%$ long GRB and (20-40)$\%$ short GRB had been observed with the plateau phase \citep{2024A&A...692A..73G}. Some interesting correlations involving the plateau phase have been established \citep{2023ApJ...949L..32D}, which may be engaged as useful cosmological probes \citep{2019ApJS..245....1T, 2021ApJ...920..135X}. Literally, there are two popular models to interpret such observational phenomenon, i.e., the relativistic wind bubble (RWB) model \citep{Dai_2004,YuYW2010} and the structured ejecta (SE) model \citep{RM_1998,SM_2000}. Polarization of the two model had been predicted and they are distinguishable with the prospect polarization detection \citep{LWD_2016b}. Depending on the models, there is a polarization degree (PD) bump in the PD curve during the X-ray plateau phase, which is not  expected in the SE model (see Figure. \ref{plateaupol}). Therefore, the PD detection at the X-ray plateau phase would make a clear distinction between the two models \citep{LWD_2016b}. The central engine in the RWB model is a magnetar, while it is a black hole in the SE model. So the polarization detection can provide a unique probe of the central engine at the plateau phase \citep{LWD_2016b}.

\begin{figure*}[!t]
  \centering
  \includegraphics[width=5in]{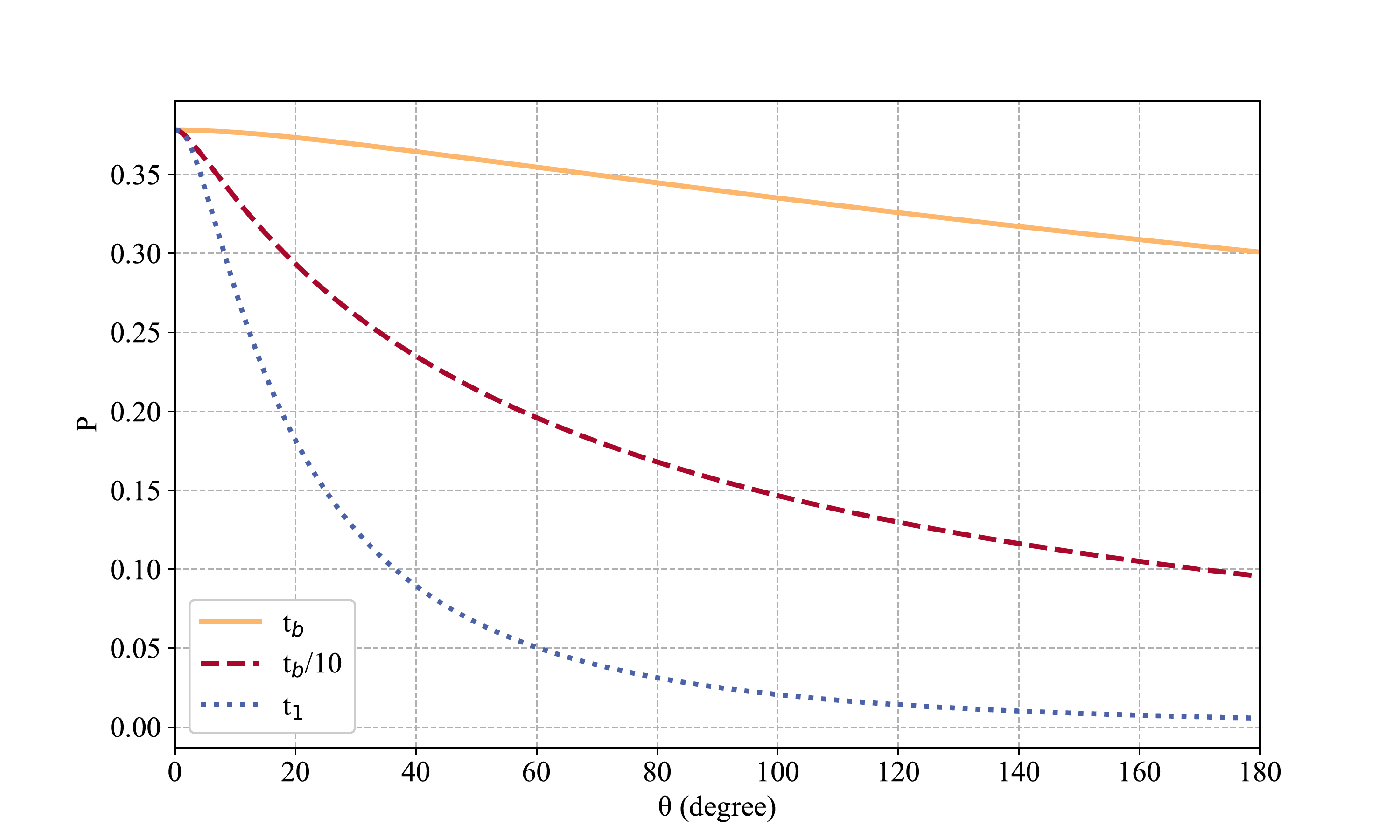}
  \caption{Detection possibility of the GRB X-ray plateau phase for one eXTP/PFA pointing versus the separation angle $\theta$, which is the angle between the eXTP/PFA pointing direction and the light of sight of the target GRB. The brown-solid, red-dashed, and blue-dotted lines refer to the detection possibility of the end, majority, and the initial stage of the plateau phase.}
\label{possibility}
\end{figure*}

Presently, there {\ysx is only one satellite IXPE, which  has the capability of the X-ray polarization detection. However, it is not optimized for the transient targets, like GRBs. After more than three years operation, IXPE only performed one GRB afterglow observation, the GRB 221009A, and collected data after two days of the Fermi trigger. An upper limit of the polarization degree of the afterglow emission of 13.8\% was given \citep{2023ApJ...946L..21N}. } Currently, there have been no polarization detections in the GRB X-ray plateau phase. The PFA on board eXTP would have the sufficient sensitivity (with a minimum detectable polarization degree (MDP) of less than 1.7$\%$ at (1mCrab, $10^6$ s)) and would be able to give a statistical sample of the X-ray polarization at plateau phase. 
The rotation velocity of the eXTP pointing is 3 degree per minute. With the early warning of the detection of the GRB prompt phase, Figure \ref{possibility} shows the detection possibility of the X-ray plateau phase for one eXTP/PFA pointing versus $\theta$, where $\theta$ is the separation  angle between the eXTP/PFA pointing direction and the position of the target GRB. The $t_1$ and $t_b$ refer to the beginning and end times of the plateau phase inferred from the light curve, respectively. For one $\theta$, we have three detection possibilities related to different stage of plateau phase. The case shown by the brown-solid curve refers to the detection possibility of the end of the plateau phase, the red-dashed curve refers to the detection possibility of the majority of the plateau phase, and the blue-dotted curve refers to that of the initial stage. For a GRB with the largest 180 degree separation, it will take 1 hour for eXTP/PFA pointing and there is still about (10-30)$\%$ possibility to detect an X-ray plateau phase. For a GRB with a 90 degree separation, the detection possibility of an X-ray plateau is roughly (16-34)$\%$. Within a relatively close separation (e.g. $\leq$40 degree), it is promising for eXTP/PFA to detect the initial stage of the plateau phase at about 0.01 day, which would be important to distinguish the models of the plateau phase. For example, the detection possibility of the initial stage of a plateau is about 10$\%$ for a GRB with a 40 degree separation. The upper limit of the detection number of the GRB X-ray plateau phase with the eXTP/PFA is about (173-328) per year. The true detection number would depend on the observational strategy of eXTP/PFA. {\ysx In the above estimations, we have compared the plateau flux observed by the Swift satellite with the threshold flux of eXTP/PFA, which shows that the flux of the most ($99\%$) GRB plateaus currently observed \footnote{The percentage was obtained using the most recent plateau data at https:// www.swift.ac.uk.} will be above the flux threshold of eXTP/PFA \citep{2019ApJ...883...97Z,zhang2025enhancedxraytimingpolarimetry}.} 

Depending on {\ysx the intrinsic properties of astrophysical sources,  the polarizations might vary in a relatively large range, so it is hard to make a prediction about the event rate of the polarization detection. }And the detection rate of the polarizations of the GRB X-ray plateau phase by the eXTP/PFA would also depend on the observational strategy. With such a sample, many scientific issues would be studied, for example, the GRB central engine and its activities, the jet/ejecta launching mechanism, and the jet/ejecta composition.

\subsection{X-ray flares}
%Lin Da-Bin, Yacheng Kang

Bright X-ray flares have been identified in about one-third of \textit{Swift} GRBs,
some of which have more than one flare
\citep{Burrows_DN-2005-Romano_P-Sci.309.1833B}.
Phenomenally, these X-ray flares are typically short-lived ($10^2$–$10^5$\,s) and characterized by rapid rise and fall with steep temporal indices, superposing on the background power-law decay component.
In the term of the occurrence time,
the majority of flares happen in the early afterglow
\citep[$<10^3$~s after the burst trigger,][]{Chincarini_G-2007-Moretti_A-ApJ.671.1903C},
while only a few are observed in the very late time \citep[$>10^4$~s, e.g., ][]{Swenson_CA-2010-Maxham_A-ApJ.718L.14S}.
X-ray flares are commonly observed in both long and short GRBs,
despite distinct origins for these two kind of bursts
\citep[e.g.,][]{Campana_S-2006-Tagliaferri_G-A&A.454.113C,Perna_R-2006-Armitage_PJ-ApJ.636L.29P}.

The fluence of X-ray flares ranges from $10^{-9}\,\rm erg\cdot cm^{-1}$ to $10^{-6}\,\rm erg\cdot cm^{-1}$,
clustering around $10^{-7}\,\rm erg\cdot s^{-1}$ 
(see the figure~10 of \citep{Liu_C-2019-Mao_J-ApJ.884.59L}).
Compared with the prompt emission, the fluence of X-ray flares is generally small
with the average fluence approximately 10 times below the average prompt GRB fluence \citep{Falcone_AD-2007-Morris_D-ApJ.671.1921F}.
The 0.3-10~keV isotropic energy of X-ray flares with redshift
approximately follows a log-normal distribution peaked at $\sim 10^{51}$~erg,
with a possible excess at low energies \citep{Chincarini_G-2010-Mao_J-MNRAS.406.2113C}.
Different from the underlying afterglow emission,
X-ray flares are typically harder and show a hard-to-soft evolution within the flare \citep{Margutti_R-2010-Guidorzi_C-MNRAS.406.2149M}.
However, temporal and spectral analyses of X-ray flares reveal many properties analogous to the prompt emission
\citep[e.g.,][]{Chincarini_G-2007-Moretti_A-ApJ.671.1903C,
Falcone_AD-2007-Morris_D-ApJ.671.1921F}.
It is suggested that X-ray flares are the extension of GRB prompt emission, but at later times, at lower energies, and with lower peak photon energies $E_{\rm p}$ of $\nu F_{\nu}$ spectrum.
Together with the prompt emission,
the X-ray flares can provide important clues about the physics of GRB phenomena,
involving the radiation mechanism and the history of central engine activities. See Section~\ref{Sec:GRB Prompt X-ray emission} for more details.
X-ray flares may also signify the interactions between the jet and the circum-burst material or the delayed energy injection \citep{2013ApJ...779...28G, 2018ApJ...852...53R}. 

Traditional X-ray observations struggle to distinguish between different emission models based solely on spectral slopes and light curves  of these X-ray flares \citep{Zhang_B-2006-Fan_YZ-ApJ.642.354Z}. The role of ordered and random magnetic fields in powering these flares remains unclear due to the lack of polarization data \citep{Zhang+Yan+2011, Geng18b}.
Statistic studies have shown that the magnetic field in region of the X-ray flares are one dimension \citep{2013NatPh...9..465W,2016ApJS..224...20Y}. However, the magnetic field in the prompt emission may be three dimension \citep{Lyu2021,Maccary2024}.
Given that the eXTP mission combines high-throughput X-ray spectroscopy, timing, and polarimetry, it enables simultaneous multi-dimensional observations of these X-ray flares.

eXTP observations can help resolving long-standing questions about the origins of X-ray flares and the physics of jets, particularly by decoding magnetic field configurations, discriminating between emission mechanisms, and linking flares to progenitor systems. For instance, if magnetic fields are ordered on large scales, such as the toroidal configurations advected from the central engine, the synchrotron radiation can achieve high polarization degrees \citep[up to $\simeq 70 \%$;][]{Geng18b}; whereas stochastic fields or photospheric emission would result in much lower polarization degrees\citep{Zhang_2018}. The PFA onboard eXTP operates in the 2–8\,keV energy band and can achieve an MDP of $\simeq 10 \%$ for bright flares ($\simeq 10^{-6}\,\mathrm{erg}\cdot\mathrm{cm}^{-2}$). These capabilities enable eXTP to track variations in polarization angle and degree during the rise and decay phases of flares. Therefore, time-resolved polarimetry can test whether polarization evolves during these flares, as expected in some magnetic dissipation scenarios \citep{Zhang+Yan+2011}. 
In addition to the PFA, it is important to note that the SFA onboard eXTP mission offers excellent spectral resolution ($\simeq 150$\,eV at 6\,keV), which will aid in identifying the spectral components (e.g., thermal emission) during X-ray flares.
By correlating polarization with flare luminosity, eXTP can also test whether some X-ray flares arise from interactions with density clumps \citep{2018ApJ...852...53R}. Complementing the PFA and SFA, the W2C will provide real-time GRB alerts and localization with a field of view of $\simeq 1.1$\,sr, ensuring rapid instrument coordination. By correlating polarization signatures with spectral hardening/softening and temporal profiles, eXTP will help disentangle the underlying processes driving X-ray flare emission, offering critical insights into the extreme physics governing GRBs.

\subsection{Normal X-ray afterglow}
%Zhao Xiao-Hong, Mi-Xiang Lan, Jirong Mao

As referred to in section 3.3 and 3.4, the early afterglow phase, which often exhibits complex features such as X-ray plateaus and X-ray flares, typically transitions into a more straightforward phase characterized by an approximately linear decay, known as the normal decay phase \citep{Zhang_B-2006-Fan_YZ-ApJ.642.354Z}. Unlike the intricate behaviors observed in the early afterglow, the normal decay phase is relatively simpler and can be well explained by the standard external forward shock model \citep{Sari_1998ApJ}. In this model, the GRB afterglow is attributed to the self-similar evolution of the forward shock, driven by the interaction of GRB ejecta with the circumburst medium, with synchrotron radiation as the primary emission mechanism.

%Synchrotron 
Polarization arises essentially from the asymmetry of the magnetic field. In the forward shock of a GRB afterglow, the magnetic field is typically generated by plasma instabilities or kinetic turbulence, resulting in a random configuration within the plane normal to the radial direction \citep{Medvedev+Loeb+1999, 2017ApJ...838...78M}. This configuration can produce strong linearly polarized emission near the limb of the emission beam (at an angle of approximately $1/\gamma$, \citep{1999ApJ...513..679G}). However, when the viewing angle is not closely aligned with the jet edge and the jet is ultra-relativistic, the observable region of the afterglow remains largely symmetric. Consequently, polarization vectors from different directions tend to cancel out, leading to a low degree of net polarization (PD), typically less than 10\%, consistent with some afterglow polarization observations in the optical or radio bands \citep{2003A&A...400L...9Covino,2016A&AT...29..205C,2023NatAs...7...80U}. X-ray polarization measurements of a large sample of GRBs, with a wide range of viewing angles, will provide a crucial test of such expectations, which has not been possible to date. 

Nevertheless, 
higher PD values can be achieved even if the viewing angle is off-axis, particularly when the jet undergoes significant deceleration. With an off-axis viewing geometry, the jet deceleration causes the visible region to enlarge. As the jet edge becomes visible, the symmetry is broken, leading to a substantial increase in PD. For an observer situated at the jet edge, the polarization can reach values as high as 20\% \citep{1999ApJ...524L..43S}. Additionally, the magnetic field coherence length can evolve over time. During the early stages of the afterglow, the number of coherent magnetic patches within the visible region of the jet is large, which results in a low measured PD. However, as the jet decelerates, the coherence length of the magnetic field increases, reducing the number of coherent magnetic patches within the visible region. This can result in an increase in PD, which could reach values as high as 10\% \citep{Gruzinov+Waxman+1999}. 

The preceding discussion has focused on homogeneous jets. If the afterglow jet is structured, the evolution of the average magnetic field within its visible region will differ from that of a homogeneous jet, resulting in variations in PD \citep{2004MNRAS.354...86R}. In addition, variations in the density of the circumburst medium can change the jet dynamics, and thus the PD evolution profiles. Therefore, X-ray polarization measurements from eXTP during the normal afterglow phase will be crucial for revealing the magnetic field configuration in the afterglow jet and for probing the viewing angle, jet structure, and the properties of the circumburst environment. 

\begin{figure*}[!t]
  \centering
  \includegraphics[width=5in]{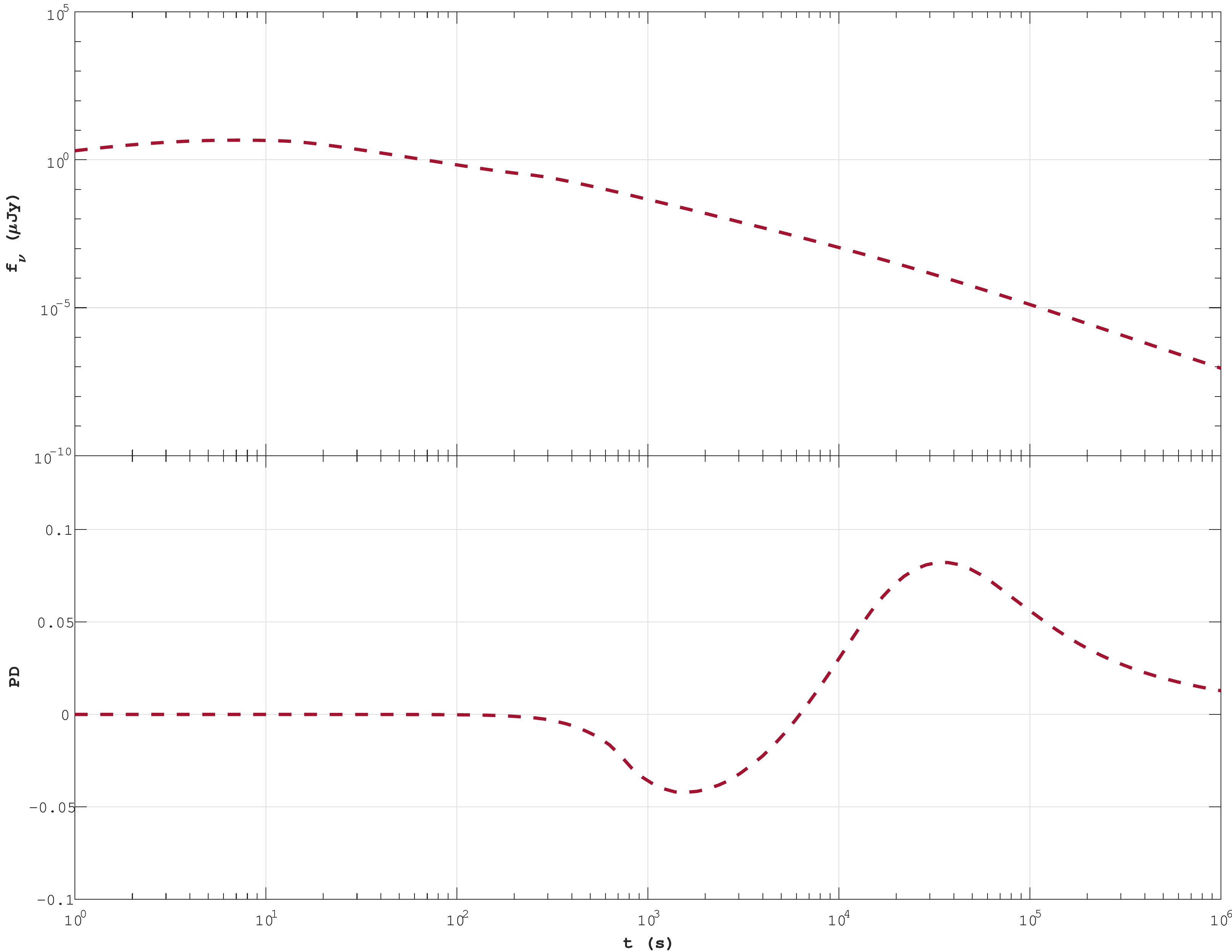}
  \caption{The light curve (upper panel) and PD curve (lower panel) of the normal GRB X-ray afterglow at the energy band of eXTP/PFA (2-8 keV).}
\label{norag}
\end{figure*}

Given these theoretical considerations, the key question is whether eXTP's capabilities are sufficient to detect the predicted polarization signals. We thus make following estimation. The predicted PD ranges from zero to 10$\%$ at both 1000 s and $10^4$ s  \citep{LWD_2023}, which is shown in Fig. \ref{norag}. The MDPs at 2-8 keV for typical afterglow spectrum are all 4.21$\%$ at ($10^{-9}$ $\rm erg~ cm^{-2} ~s^{-1}$, 1000 s), ($10^{-10}$ $\rm erg~ cm^{-2} ~s^{-1}$, $10^4$ s), and ($10^{-11}~\rm{erg~cm^{-2}~s^{-1}}$, $10^5$ s). The proportion of the normal X-ray afterglows above the flux threshold of $10^{-9}$ $\rm erg ~ cm^{-2}~s^{-1}$ at 1000 s is about 8.1$\%$ and drops to about 6.5$\%$ with the flux threshold of $10^{-10}$ $\rm erg~ cm^{-2} s^{-1}$ at $10^4$ s. The reduction of the proportion is due to the decay of afterglows. Therefore, the predicted event rate for polarization detection is less than 13 per year at 1000 s and less than 59 per year at $10^4$ s. The true detection rate would also depend on the observational strategy of eXTP.

{\ysx The IXPE had set an upper limit of $13.8\%$ for the normal X-ray afterglow of GRB 221009A at an exposure time of $\sim$2 days with a flux of $\sim1.5\times10^{-10} $ erg/cm$^2$/s. At the same exposure time with an equivalent flux as GRB 221009A, the MDP of eXTP/PFA would be $0.8\%$. Therefore,  depending on the polarization properties of the source, eXTP/PFA could establish a more restrictive upper limit on the observed PD or give an effective detection. }

%In particular, the MDP at 2-6 keV is about 4\% for an afterglow with a flux of about $10^{-11}~\rm{erg~cm^{-2}~s^{-1}}$ and a power-law spectrum that has the photon index of 2.0 when we have an exposure time of 100 ks.

\subsection{GRB X-ray polarization detection as a probe of quantum gravity}
%Jun-Jie Wei

Many quantum gravity theories propose modifications to the standard dispersion relation
of high-energy particles in an effort to unify the two fundamental theories of modern physics: general relativity and the standard model of particle physics \citep{1998Natur.393..763A,2005LNP...669..351P}. 
These modifications manifest themselves at the Planck energy scale $E_\mathrm{Pl}=\sqrt{\hbar c^{5}/G}\simeq1.22\times10^{19}$ GeV.
In specific cases, particularly in loop quantum gravity, the modifications to
the standard dispersion relation result in a form where the circularly polarized modes
of photons travel with different phase and group velocities
\citep{1999PhRvD..59l4021G,2003PhRvL..90u1601M}.
When these modes travel at different velocities, light propagation experiences vacuum birefringence, 
leading to an energy-dependent rotation of the polarization vector of linearly polarized signals.
Although this vacuum birefringence effect is expected to be negligible at current observable energy levels ($E\ll E_\mathrm{Pl}$),
it can amplify with increasing energy and accumulate over large distances, becoming measurable. 
Therefore, astrophysical measurements that involve high-energy polarimetry and long baselines can 
offer sensitive constraints on the magnitude of the quantum gravity birefringence parameter
\citep{2012PhRvL.109x1104T,2019MNRAS.485.2401W}.

GRBs, the most powerful explosions occurring at cosmological distances, are regarded as exceptional probes 
of quantum gravity. To observe the vacuum birefringence effect, it is crucial to understand the emission 
properties of the GRB source. Additionally, since the polarization degree is directly influenced by 
propagation details, a thorough understanding of the medium's properties is essential. For instance, 
magnetized plasmas can affect the rotation of the linear polarization plane, a phenomenon known as Faraday
rotation. Fortunately, the Faraday rotation angle is pronounced in the low-frequency radio bands but negligible in the optical and higher-energy bands \citep{2007MNRAS.376.1857F}. 
To enhance the sensitivity of birefringence measurements, the development of instruments that can measure
the polarization of photons in the X-ray and gamma-ray bands is necessary. Although gamma-ray polarimeters
are hindered by technical challenges, X-ray polarimetry has significantly advanced with the recent launch
of the Imaging X-ray Polarimetry Explorer (IXPE) \citep{2021AJ....162..208S}, and further improvements are expected with the upcoming eXTP \citep{2016SPIE.9905E..1QZ}.

\section{Magnetars and related radio sources}\label{sec:FRB}

% lin lin and Yu-cong Fu
Magnetars are a group of young slow-rotating neutron stars with an extremely strong magnetic field $\sim10^{14-15}$ G \citep{2017ARA&A..55..261K,2015SSRv..191..315M}. Their extreme variability phenomena in X-ray and radio bands provide us an unique way to study not only the neutron star itself but also the physics under extreme conditions. In the parallel paper WG3, we present how the advanced properties of eXTP can help to unveil the mysterious in strong magnetic fields by detecting for example the timing anomalies in the outbursts, line features in the X-ray spectra and polarization of the X-ray bursts. In this section, we will focus on the emission physics of magnetars in the broad electromagnetic waveband and detailed timing investigation of magnetar X-ray bursts.

In recent years, the understanding of Fast Radio Bursts (FRBs) expanded quickly and the Long-Period radio Transients (LPTs) has been discovered. There is theoretical and observational evidence that magnetars might be connected to these interesting classes of sources. The association of FRB 20200428A and the galactic magnetar SGR J1935+2154 makes the magnetar the only observational confirmed source of FRBs \citep{2020Natur.587...54C,2020Natur.587...59B,2020ApJ...898L..29M,2021NatAs...5..378L}. However, through which physical process the FRB was emitted, the connection between FRB and other magnetar activity and whether magnetars are capable to host all kinds of FRBs are not clear. Therefore, we will also discuss the important role eXTP can play to solve the fundamental question about the origin of FRBs and LPTs.

\subsection{Magnetars as radio pulsars}
%{\it Contributors: Lin Lin, Yu-Cong Fu}

Magnetars are prominent X-ray sources, but only six out of thirty known magnetars have been detected in radio bands. Their transient radio behavior is connected with X-ray outbursts but not necessary to be synchronized. Both periodic radio pulses and bright single-pulse ``FRB-like'' events have been detected from magnetars. The radio emission properties are very diverse among this small sample. Thanks to the high sensitivity and spectral-polarimeter measurement of eXTP, the multi-wavelength synergy observation with eXTP and radio telescopes can provide more clues to understand the emission physics of magnetar. 

The X-ray pulse profile of magnetars are broad and peaked at the hot spots towards us. These hot spots may be polar region or the active position during the outburst \citep{2022ApJ...924L..27Y}. The periodic radio pulses are narrow and pin the polar region as in normal radio pulsars. The coordinated radio and X-ray observations exhibited that the periodic radio pulses in some cases are close to the peak of the X-ray pulse profile, such as XTE J1810-197 and 1E 1547.0-5408\citep{2021ApJ...907....7I,2022MNRAS.510.1996C}. While in Swift J1818.0-1607 and SGR J1935+2154, they are located in the valley of the X-ray pulse profile\citep{2023SciA....9F6198Z,2023MNRAS.523.2401B}. Several bright isolated radio pulses have been detected from SGR J1935+2154 and they do not have any preference of the rotational phase\citep{2023SciA....9F6198Z}. This indicates the radio bursts and periodic pulses have different origin. 

\begin{figure*}[!t]
  \centering
  \includegraphics[width=5in]{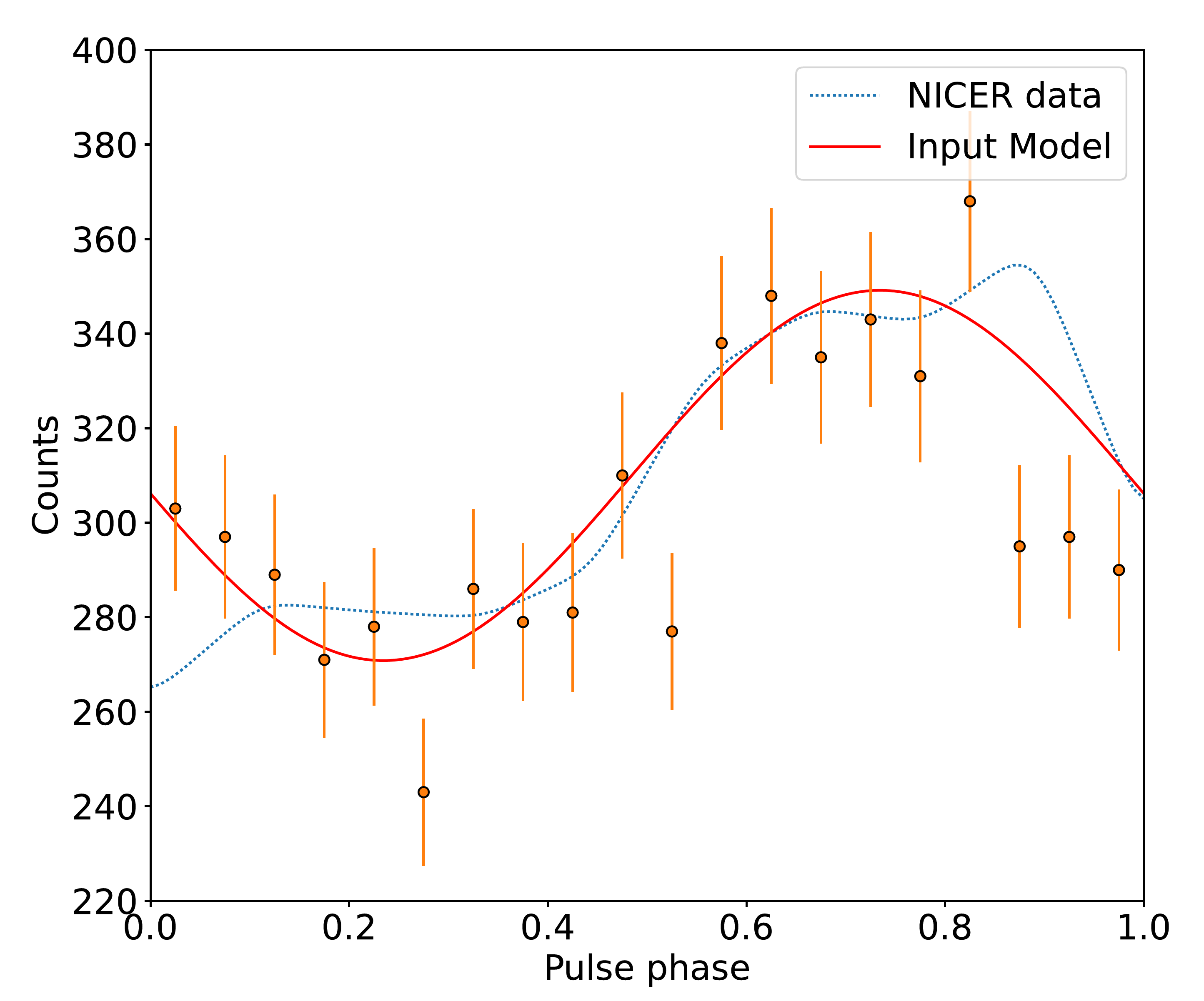}
  \caption{The simulated X-ray pulse profile of magnetar detected with eXTP/SFA. }
\label{magnetarprofile}
\end{figure*}

\begin{figure*}[!t]
  \centering
  \includegraphics[width=5in]{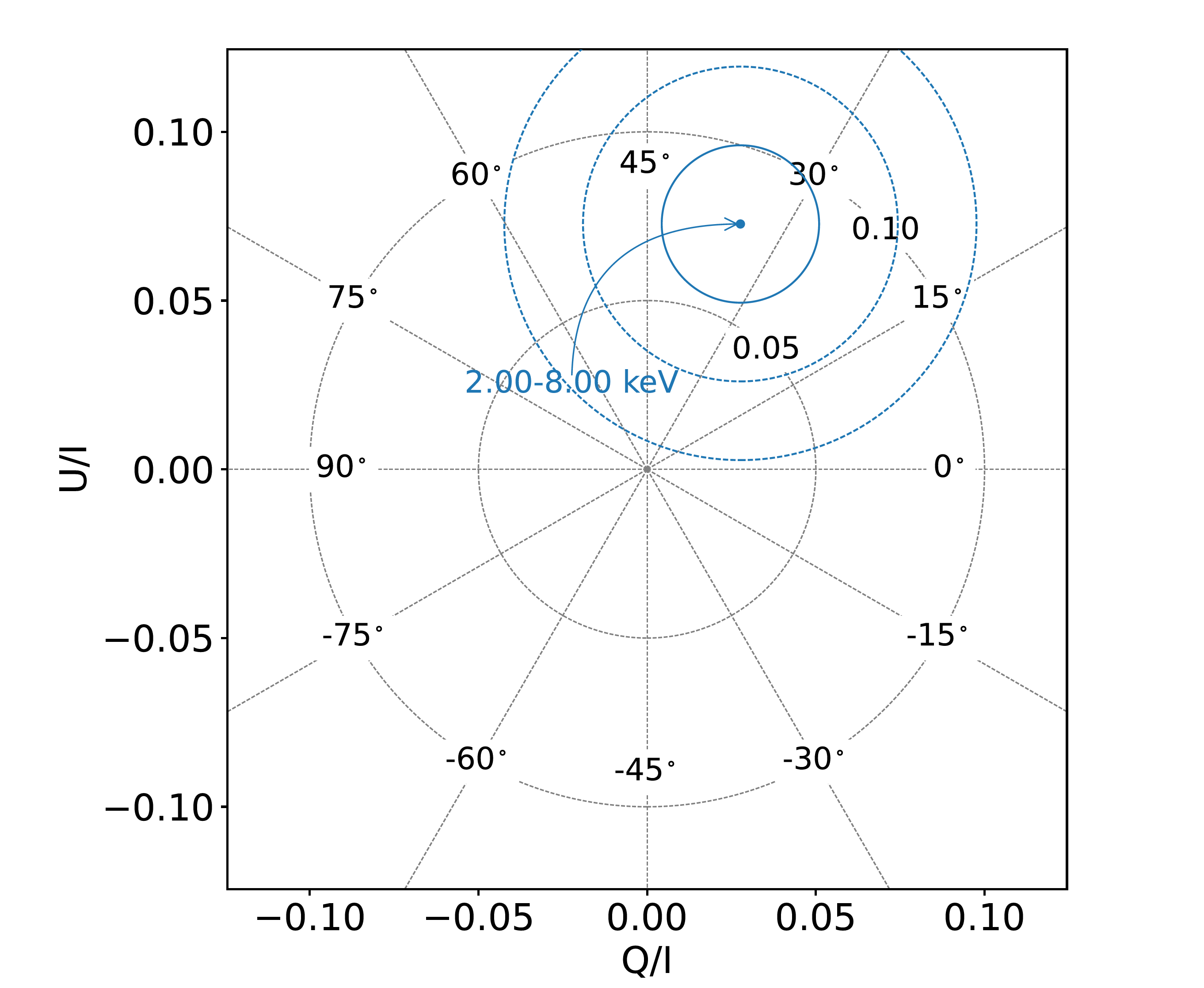}
%  \caption{The simulated X-ray polarization of magnetar outburst detected with eXTP/PFA. }
  \caption{A 300 ks simulated observation with eXTP/PFA can reveal 10\% polarization in an outburst similar to that of SGR J1935+2154 in 2020 October.}
\label{magnetarpolarization}
\end{figure*}

Simulations using the same X-ray parameters of SGR J1935+2154 during the radio pulsar phase in 2020 October\citep{2023NatAs...7..339Y,2023SciA....9F6198Z}, when the X-ray flux had already decayed to $\sim$2.2$\times10^{-11}$ erg s$^{-1}$ cm$^{-2}$, show that the eXTP/SFA will detect a significant X-ray pulse profile (Figure \ref{magnetarprofile}) with 50 ks exposure time. 
Note that the typical outburst flux is about 1-2 orders of magnitude brighter than this level. If the X-ray flux is $\sim$2.2$\times$10$^{-10}$ erg s$^{-1}$ cm$^{-2}$, then the exposure time will be reduced to 10 ks to get a similar significant profile. 
Additionally, eXTP/PFA is capable to constrain the polarization property of the X-ray outburst. 
If we assume a 10\% polarization degree in the X-ray outburst of SGR J1935+2154 in 2020 October, then $\sim300$ ks would be required to detect it at 99\% confidence level using PFA 
%At the same flux level, PFA can detect 10\% polarization at 99\% confidence level with 300 ks exposure,
as shown in  Figure \ref{magnetarpolarization}. 
Note that the only polarimetric observation of a magnetar in outburst obtained up to now, indicates a significant polarization increasing from 15\% at 2-3 keV to more than 50\% at 6-8 keV \citep{2025ApJ...985L..34R,2025ApJ...985L..35S}.
With the future eXTP observations we can provide pulse phase-resolved spectro-polarimetry, which compared to possibly present radio emission, can provide important information on the emission processes and on the magnetic field geometry of the emitting region. 
%during the radio pulses or bursts and to study how the radio emission related to the hot spots of magnetar as well as the magnetic field geometry of the emitting region. 

Both the appearance of the radio pulses in SGR J1935+2154 and the sudden disappearance of the radio emission in 1E 1547.0-5408 are possibly related to   spin-up or spin-down glitches \citep{2023NatAs...7..339Y,2023ApJ...945..153L,2024RAA....24a5016G,2025ApJ...980...99Y}. The connection and the exact time of the glitch are not clear due to the limited monitoring of the source. With eXTP, we can have longer exposure as well as more frequent monitoring of the radio magnetars to nail down the changing point. It is even possible to find out whether the polarization properties change or not after the glitch. This would indicate that the reconfiguration of the local magnetic field may relevant to the switching on/off of  the radio emission.

\subsection{Magnetar X-ray bursts}
%Shuo Xiao
%{\it Contributors: Shuo Xiao, Shao-Lin Xiong}

Timing analysis is a key approach to reveal the physical properties of magnetars, but timing analysis of magnetar X-ray bursts has remained quite limited, with previous research primarily concentrating on characterizing the duration of bursts. For example, bursts from soft gamma-ray repeaters (SGRs) can be generally divided into three types according to their duration and luminosity \citep{2006csxs.book..547W}: short-duration bursts ($\sim$ 0.1 s), intermediate bursts (seconds) and giant flares (tens of seconds to minutes). On the other hand, the duration only describes the global emission properties of a burst, which does not capture the information concerning individual pulses in a burst. Therefore, using thousands of XRBs observations (e.g. \citep{lin2020fermi,li2021hxmt}) of {\it Insight}-HXMT, GECAM and Fermi-GBM, their minimum variation timescale (MVT) \citep{xiao2023minimum,xiao2024log}, spectral lag \citep{xiao2023discovery}, power spectra \citep{2022ApJ...931...56L,roberts2023quasiperiodic,xiao2024individual} and individual pulse characteristics \citep{xiao2024self,yang2024comprehensive} are studied. For example, the first XRB observed by {\it Insight}-HXMT in association with FRB 200428 was found to have an interesting QPO at $\sim$ 40 Hz with $3.4\ \sigma$ \citep{2022ApJ...931...56L}.
%With a high sensitivity (3.3$\times10^{-15}\ \rm{erg/cm^2/s}$ for 5 $\sigma$ at $10^6$ s), 
Thanks to its high sensitivity, shorter dead time ($<\ 5\%$@1Crab \citep{zhang2025enhancedxraytimingpolarimetry}), high temporal resolution (2 $\mu$s) and larger transmission count rate, eXTP will have significant advantages in the discovery of refined structures, QPO searches and spectral lag studies.

The MVT is approximately the rise time of the shortest pulse in a light curve. It   can give insight into the radiation region and Lorentz factors for possible jets, and also allows to distinguish the origin of a burst. For example, GRB 200415A had a magnetar origin, and an MVT of $\sim$ 2 ms \citep{yang2020grb}, which is consistent with magnetar X-ray bursts rather than most GRBs. 
However, limited by the effective area and maximum transmission rates of current instruments (e.g. $\sim$ 30,000, $4\times10^5$ and 375,000 counts  s$^{-1}$ for {\it Insight}-HXMT/HE \citep{2020JHEAp..26...58X}, GECAM \citep{zhang2019energy} and GBM \citep{meegan2009fermi}, respectively), the smallest MVT identified so far is $\sim$ 0.4 ms for XRBs from SGR J1935+2154, it is difficult to identify finer pulse structures with durations of less than sub-milliseconds. 
On the other hand, with its much larger effective area and enhanced transmission rate, eXTP will have the opportunity to discover structures of a hundred microseconds or even more extreme, thus more tightly limiting the radiation region, which is crucial to constrain both pulsar-like and GRB-like models  (see \citep{2020Natur.587...45Z} for a review).

Spectral lag of the low-energy photons with respect to the high-energy ones is a common phenomenon in astronomical sources. It is conventionally defined as positive lag when low-energy photons follow high-energy photons.
The spectral lag of magnetar bursts is usually regarded as negligible in previous studies, such as GRB 200415A with magnetar origin \citep{yang2020grb}. 
The spectral lags of about 61\% (non-zero significance $>1$$\sigma$) bursts from SGR J1935+2154 are linearly dependent on the photon energy, which may be explained by a linear change of the temperature of the blackbody-emitting plasma with time \citep{xiao2023discovery}. Besides, the distribution of the slope can be well fitted with three Gaussians, which may correspond to different origins of the bursts. However, the study of their origin requires more precise spectral lag measurements, which rely on instruments like eXTP with high sensitivity and temporal resolution.

QPOs have been observed in several magnetars, including SGR 1806-20 (e.g. \cite{2005ApJ...628L..53I,huppenkothen2014quasi}), SGR 1900+14 \citep{strohmayer2005discovery}, SGR J1550-5418 \citep{2014ApJ...793..129H}, and SGR J1935+2154 \citep{2022ApJ...931...56L,roberts2023quasiperiodic,xiao2024individual}. Additionally, QPO signals have been detected in an extragalactic magnetar associated with GRB 200415A \citep{castro2021very} and even in a magnetar possibly produced in a compact binary merger, as seen in GRB 211211A \citep{xiao2024peculiar}. Theoretical models suggest that these QPOs originate from oscillatory motions in the crust of magnetars (e.g. \cite{2014ApJ...787..128H,2022ApJ...939L..25Z}). Consequently, their detection provides a potential avenue for probing the neutron star equation of state and internal magnetic field by linking specific QPO frequencies to global seismic modes \citep{huppenkothen2013quasi}. However, limited by the sensitivity or dead time of current instruments, it is difficult to confirm QPOs at both low and high frequencies. Therefore, eXTP provides more opportunities to investigate the interior structure of neutron stars based on searched QPO.

In addition, we will also investigate the temporal properties evolution of magnetar bursts, such as duration, MVT and waiting time, across different active episodes. Understanding how these properties evolve over time can shed light on potential changes in the neutron star’s internal structure and magnetic field configuration. For example, previous studies on SGR J1550–5418 have revealed spectral evolution in bursts during its 2008–2009 active episode, with bursts in October 2008 best described by a blackbody model, while those in early 2009 were better fitted by an optically thin thermal bremsstrahlung model \citep{von2012detection}. However, no significant evolution in burst duration was identified \citep{von2012detection,collazzi2015five}. More recently, \cite{xiao2024log} reported that the MVT of bursts from this magnetar exhibited substantial variations between different phases of activity, suggesting that changes in the magnetospheric configuration following intense bursting episodes may influence burst properties. Therefore, a detailed investigation of the evolution of these properties may offer new insights into the physical conditions governing magnetar bursts and their potential connection to FRBs.

Magnetar bursts also have important applications in the field of astronomical techniques due to the timing characteristics of their smaller MVTs and spectral lags. Besides, SFA onboard eXTP has high time resolution (2 $\mu \rm s$) and absolute time resolution (2 $\mu \rm s$), it not only has advantages and potential in timing analysis but also plays a crucial role in astronomical techniques. For example:\\
(1) Time Delay Localization is a well-established method for localizing transient events by measuring the arrival time differences of signals detected by satellites at different positions (which requires high time resolution and absolute time accuracy) (e.g., \citealp{hurley2011interplanetary, pal2013interplanetary, xiao2021enhanced}). The precise localization of GRBs and SGRs is essential for multimessenger and multiwavelength astronomy, facilitating follow-up observations and the identification of associated transients, neutrinos, and gravitational waves. The absolute time accuracy of 2 $\mu \rm s$ for SFA corresponds to an angular uncertainty of only about 0.007 degrees for time delay localization in low Earth orbit satellites. More importantly, the high time resolution of 2 $\mu \rm s$ for SFA, combined with the Li-CCF method, can further improve localization accuracy \citep{li2004timescale, xiao2021enhanced}.\\
(2) Pulsar and Magnetar Navigation Experiment is crucial for both orbit estimation of Earth satellites and deep-space navigation of spacecrafts (e.g., \citealp{witze2018nasa, zheng2019orbit, luo2023pulsar}), which requires the arrival times of observed events to be corrected to the Solar System barycenter (DE200) (i.e., high time resolution and absolute time accuracy are required). The absolute time accuracy of 2 $\mu \rm s$ for SFA corresponds to a light travel distance of only about 600 meters, which significantly enhances the precision of pulsar-based navigation.\\
(3) Absolute time accuracy calibration is crucial for precise synchronization across space missions. Magnetar bursts, with their sharp light curves and negligible spectral lags, are ideal for precise on-orbit time calibration. Using bursts from SGR J1935+2154, time delays between instruments like GECAM-B and GBM can be determined with uncertainties as small as 0.2 ms \citep{xiao2022ground}. However, current instruments are limited by sensitivity, leading to relatively large uncertainties. eXTP's superior sensitivity and time resolution allow for more accurate and faster time delay calculations, even for short-duration magnetar bursts. This provides a significant advantage for efficient on-orbit time calibration. With its enhanced capabilities, eXTP will improve synchronization across space missions, enabling precise timing for applications such as multi-messenger astronomy and timing analysis.\\
(4) Cosmic Ray Event Identification is another important application. When cosmic rays interact with the detector or the satellite structure, they can produce multiple secondary particles that are recorded nearly simultaneously by multiple detectors (e.g., detectors of SFA, PFA and W2C), necessitating high relative timing accuracy for proper identification. With a time resolution of 2 $\mu \rm s$ for SFA, the false coincidence rate between two detectors with an assumed count rate of 1000 is only about 0.4 per second. Additionally, relative timing accuracy is critical for joint analyses involving multiple detectors.
These capabilities demonstrate that eXTP's high time resolution and absolute time accuracy provide significant technical advantages in high-energy astrophysics and space science applications. {\ysx We would like to note that the topic of  Quantum Electrodynamics (QED) propagation effect in magnetars X-ary bursts have been covered in another white paper in the series: the white paper of ``strong magnetism"\citep{ge2025physicsstrongmagnetismextp}.}

\subsection{Fast Radio Bursts}
%Weiyang Wang
%{\it Contributors: Weiyang Wang, Lin Lin, Yu-cong Fu, Shao-Lin Xiong}

Fast radio bursts (FRBs) are millisecond radio flashes with extremely large bright temperatures \cite{lorimer07}.
Although the detection of FRBs from galactic magnetar SGR J1935+2154 shows that at least a portion of FRBs could come from magnetars, the origin(s) of most (extragalactic) FRBs is still unknown.
FRBs can be classified into two catalogs: repeaters and apparently one-off bursts.
However, despite hundreds of sources have already been discovered, whether all FRBs can repeat and share a common origin is still an open question.
%It is tempting to speculate that some FRB, which includes known repeaters and some sources with longer quiescent periods,
%are repeaters, while some apparently one-off FRBs are truly non-repeaters.
Various ideas have been proposed that most, if not all, FRB sources may be repeaters with widely different rates and geometrically-constrained visibility (e.g. \cite{2024ApJ...973..123D,2025ApJ...982...45B}
%ref
%Dall'Osso, La Placa, Stella, Bakala, Possenti et al. 2024, The Astrophysical Journal, Volume 973, Issue 2, id.123; Beniamini & Kumar 2025, The Astrophysical Journal, Volume 982, Issue 1, id.45
).  However, it is also speculated that a sizable fraction of apparently one-off sources are truly non-repetears. 
Repeating FRBs seem to have longer durations with narrow bands than apparently one-off bursts \cite{2021ApJ...923....1P}.
The apparently one-off FRBs also tend to have higher luminosities than repeaters, leading to possibly distinct origins.
Specifically, non-repeating bursts are likely associated with binary compact star mergers, whereas repeating bursts are thought to originate from starquakes of magnetar \cite{2020RAA....20...56J,2018ApJ...852..140W,2025arXiv250109248W}.

Considering the differences in radio luminosity between repeating and non-repeating bursts, it is possible that their X-ray counterparts also exhibit noticeable differences in luminosity.
We assume that the ratio between X-ray and radio luminosity for all FRBs is the same, which is of order $\sim10^5$ \cite{2021NatAs...5..378L}.
Statistical findings indicate that the luminosity of nearly all non-repeating bursts exceeds $10^{43}$ erg s$^{-1}$, a value which we can tentatively adopt as the lower limit for the luminosity of non-repeating bursts \cite{2020MNRAS.494..665L}, indicating an X-ray luminosity of $10^{47}$ erg s$^{-1}$.
Some X-ray counterparts that are relatively close to us may be observable. Considering the sensitivity of eXTP/W2C, we can detect the associated X-ray burst of FRBs closer than 25 Mpc with $3\sigma$ confidential level.
%The 3$\sigma$ sensitivity of eXTP/W2C to detect a 0.1 s burst with the same spectral shape as the X-ray burst associated with FRB 20200428A \citep{2021NatAs...5..378L} is $\sim10^{-6}$ erg s$^{-1}$ cm$^{-2}$. This indicate that we can detect the X-ray burst with luminosity of $10^{47}$ erg s$^{-1}$ closer than $\sim25$ Mpc.
Until now, no associated high-energy counterparts of extragalactic FRBs have been confirmed.
The identification of such counterparts would have profound implications for the understanding of FRBs' origin(s).
Moreover, FRBs exhibit abundant polarization features in the radio band \cite{2020Natur.586..693L,2022Natur.609..685X}, and those of their X-ray counterparts are also highly anticipated. 
%linlin
If the duration of the X-ray burst is 0.1 s and the flux is over 10$^{-4}$ erg s$^{-1}$ cm$^{-2}$, then the eXTP/PFA can detected 10\% polarization in 95\% confidence level.

In addition, for the nearby extragalactic FRBs (such as FRB 250316A), the pointed observation with eXTP/SFA and eXTP/PFA will be able to capture the persistent or impulsive X-ray emission from the FRB source.
{\ysx The X-ray emissions are most likely generated from the regions near the stellar surface, so that X-ray emissions may experience vacuum polarization effect, leading to a variety of polarization properties.}
For the galactic magnetars, eXTP could do pointed observations to implement spectral-polarimetry measurements of magnetar X-ray bursts coincident with FRBs if they occur, as well as the outburst or persistent x-ray emission from the magnetar before and/or after the FRB.

\subsection{Long-period radio transients}
%{\it Contributors: Fayin Wang, Zhenyin Zhao}
%Fayin Wang \& Zhenyin Zhao

Long-period radio transients (LPTs) are recently discovered as low-frequency radio transients with unknown origin \cite{Hurley-Walker2022,Hurley-Walker2023,Caleb2024,deRuiter2024,Hurley-Walker2024,Men2025,Wang2025,Lee2025,Bloot2025,Dong2025}. The pulses usually last about minutes and have a period of tens of minutes to several hours. The radio emission are bright (with luminosity around $10^{27}-10^{32}$ erg s$^{-1}$), coherent and highly polarized. {\ysx So far, at least a dozen such sources have been reported, and the properties of those with X-ray observations are listed in Table \ref{tab1}. More complete information can be found in the LTP catalog\footnote{LTP catalog: \url{https://lpt.mwa-image-plane.cloud.edu.au/published/galactic_view}.}. }

The high-energy counterpart could provide clues to the origin of LPTs. Initially, no significant X-ray emissions were detected from the LPTs (with the upper limit of $L_{\mathrm{x}}<10^{30-33}$ erg s$^{-1}$, see Table \ref{tab1}). ASKAP J1832-0911 is the first LPT source to detect X-ray transients associated with bright radio emissions \cite{Wang2025}. The peak X-ray luminosity is $\sim 10^{33}$ erg s$^{-1}$ but the spectral model is not well restricted (with a photon index $\Gamma=0.0 \pm 0.5$ or a blackbody temperature $kT\sim 2$ eV and blackbody radius $\lesssim 0.1 \mathrm{~km}$) \cite{Wang2025}.

Many models have been proposed to explain LPTs. Long-period pulsars or magnetars are a possible origin\cite{Ronchi2022,Cooper2024}, supported by the discovery of complex pulse profiles and orthogonal polarization modes (OPMs) from some LPTs \cite{Hurley-Walker2022,Hurley-Walker2023,Wang2025}. {\ysx However, the low spin-down energy ($\dot{E}<10^{29}$ erg s$^{-1}$) is inefficient to explain the detected radio luminosities, suggesting an alternative process for momentum loss.} The long spin periods may be caused by supernova fallback accretion \cite{Ronchi2022,Tong2023}, {\ysx or by the contribution of relativistic wind emission \cite{horvath2025evolution,kirk2009theory}, a phenomenon observed in young pulsars such as Crab and Vela (and their PWNs), and which may remain relevant for a long time in the neutron star's lifetime. Models of wind emission \cite{harding1999magnetar} relate it to a Particle Luminosity parameter $L_{p}$, which implies a torque $\propto \Omega$, in contrast to dipole emission (the 'canonical' mechanism), which is $\propto \Omega^{3}$. It has been argued that both processes can act simultaneously, but that the wind may be the main mechanism of the radio emission observed from LPTs, assuming they are old pulsars (ages > 1 Gyr), such that the dipole term is no longer relevant. The important factor is that wind activity can emit high-energy pulses \cite{petri2016theory}, but only a few sources have X-ray detections, due to their weak intensities} Another type of model based on magnetars believes that the ultra-long period originates from the precession of magnetars \cite{Eksi2022}. The radio emission from LPTs may also be driven by a rotating, magnetized white dwarf (so-called 'white dwarf pulsar')\cite{Katz2022}. Motivated by the discovery of M dwarf companion from ILT J1101+5521\cite{deRuiter2024} and GLEAM-X J 0704-37\cite{Hurley-Walker2024,Rodriguez2025}, the white dwarf/M dwarf (WD-MD) unipolar induction magnetic interaction model is proposed to explain the longer period LPT source \cite{Qu2025}.

In the future, by monitoring known LTPs with eXTP, their origins may be constrained. The soft X-ray $(\lesssim 10 \mathrm{keV})$ emission of isolated magnetars is usually described by a combination of a thermal component with a blackbody with temperature $k T \sim 0.3-0.9$ keV and a power law component with photon index $\Gamma \sim 2-4$ \cite{2018MNRAS.474..961C}. The non-thermal components are from the resonant cyclotron up-scattering of thermal photons in the magnetosphere. The emission is dominated by the extraordinary mode (X-mode) for $\lesssim 1 \mathrm{MeV}$ band \cite{Beloborodov2013}, and the characteristic polarization fraction is $\sim 10 \%-30 \%$ \cite{Fernandez2011}. For the WD–MD close binary system, synchrotron radiation of relativistic electrons from the shocked stellar wind can result in broadband (radio to X-ray) emission \cite{Geng2016}. {\ysx The multiwavelength radiation has been detected for AR Scorpii \cite{Marsh2016} and J1912-4410 \cite{Pelisoli2023}, and it is very different from other LPTs.} The long-term evolution of X-ray emission is found to be modulated with the beat frequency between the spin frequency of WD and orbital frequency for the WD–MD binary system AR Scorpii \cite{Takata2021}, {\ysx although it much depends on the emission geometry and is not always found in binary systems. For the unipolar induction magnetic interaction model, X-rays can also be produced via inverse Compton (IC) scattering or cyclotron radiation \cite{Qu2025}.} The polarization properties and long-term variations of X-ray emission can be tested by eXTP.

\begin{table*}[]
\centering
\label{tab1}
\begin{tabular}{ccccccc}
\hline
LPT name &   Period (min) & Frequency (MHz ) & $L_{\mathrm{radio}}$ (erg s$^{-1}$)  & $L_{\mathrm{x}}$ (erg s$^{-1}$) & Counterpart &  Ref.   \\
\hline
GLEAM-X J162759.5-523504.3 & 18.18 &  88-215 &  $\sim 10^{31}$ & $< 10^{32}$  &  None    &  \cite{Hurley-Walker2022}                    \\
GPM J1839–10 &  21.97  &  $\sim 100-1000$ & $\sim 10^{28}$ & $< 10^{33}$ &  None &   \cite{Hurley-Walker2023,Men2025} \\
ASKAP J1935+2148 & 53.76 &  887.5-1284  & $\sim 10^{29}-10^{30}$ & $< 10^{30}$ & NIR source & \cite{Caleb2024} \\
ASKAP J1755-2527 &    &  750-1000  & $\sim 10^{30}$ & $< 10^{32}$ & None & \cite{Dobie2024} \\
ILT J1101+5521 & 125.5  & 120-168 & $\sim 10^{27}$ & $< 10^{30}$ & M dwarf & \cite{deRuiter2024} \\
GLEAM-X J 0704-37  & 174.9  & 170-1050 & $\sim 10^{29}$ & $< 10^{30}$ & M dwarf & \cite{Hurley-Walker2024} \\ 
ASKAP/DART J1832-0911 & 44.27  & 320-3124 & $\sim 10^{32}$ & $\sim 10^{33}$ & X-ray source & \cite{Li2024,Wang2025} \\ 
ASKAP J1839-0756 & 387  & 816-2100 & $\sim 10^{28}$ & $< 10^{33}$ & None & \cite{Lee2025} \\ 
CHIME/ILT J1634+44  & 14  & 120-1440 & $\sim 10^{29}-10^{30}$ & $< 10^{32}$ & None & \cite{Bloot2025,Dong2025} \\ 
\hline
\end{tabular}
\caption{The radio and X-ray emission of known LPTs.}
\end{table*}

%\section{Cataclysmic Variable}\label{sec:CVs}

\section{Tidal Disruption Events}\label{sec:TDE}
%{\it Contributors: Ning Jiang, Weihua Lei, RongFeng Shen, Lixin Dai, Qingwen Wu, Yongquan Xue, Xinwen Shu, Wenjie Zhang}
\subsection{Overview and X-ray Properties of TDEs} \label{subsec:xtde}
%Ning Jiang

One of the major breakthroughs in time-domain astronomy over the past decade has been the discovery of a rapidly increasing number of tidal disruption events (TDEs, \cite{Gezari2021}). A TDE occurs when an unlucky star wanders into the tidal sphere of a supermassive black hole (SMBH) at the center of a galaxy and is then torn apart and partially accreted. Such a phenomenon occurs rarely, i.e., approximately once every $10^4-10^5$ years~\cite{Stone2016} for an individual galaxy, and produces a burst of electromagnetic radiation over months to years~\citep{Rees1988}. Following the theoretical prediction that the accretion SED peaks in the soft X-ray or extreme ultraviolet (EUV), TDEs were first serendipitously identified as soft X-ray transients in galactic nuclei from archival data since the late 1990s~\cite{Bade1996}. The X-ray spectra of these TDEs are very soft near peak, which can be well modeled by a blackbody with temperature in the range of $kT_{\rm bb}=0.04-0.12$~keV (or, alternatively, with powerlaw indexes of $\Gamma_{X} = 4–5$), followed by a spectral hardening over time~\cite{Saxton2020}.
However, the growth in the number of X-ray TDEs has been slow due to the absence of time-domain X-ray surveys until the recent launch of the eROSITA~\cite{Sazonov2021} and EP~\cite{EP}.  

The field of TDEs has undergone a transformation since the identification of a population of optical TDEs~\citep{vV2011,Gezari2012}, the number of which is steadily increasing thanks to the advent of wide-field optical time-domain surveys, and has gradually opened up the new era of statistical analysis enabled by Zwicky Transient Facility (ZTF~\citep{vV2021}). In the near future, the Wide Field Survey Telescope (WFST~\cite{WFST}) and Legacy Survey of Space and Time (LSST) at the Vera C. Rubin Observatory~\citep{Bricman2020} are expected to lead to an even greater capability of optical TDE detection. Thanks to rapid advances in observations, TDEs are gradually unveiling their immense scientific values.
First of all, TDEs offer a unique means of probing dormant SMBHs~\citep{Yao2023}, including intermediate-mass BHs (IMBHs) and SMBH binaries~\citep{Ricarte2016, Chang2025}. Additionally, the gas and dust echoes of TDEs enable us to investigate the sub-parsec environments of quiescent SMBHs~\citep{Jiang2021}, which are otherwise inaccessible by other methods. TDEs also provide a unique opportunity to investigate the accretion physics of SMBHs and outflows by showcasing the entire life cycle of BH activities~\citep{Dai2021}. Recently, increasing evidence for spatial and temporal coincidences of (candidate) TDEs and IceCube high-energy neutrinos suggests that TDEs could be a potential source of these mysterious neutrinos~\citep{Stein2021,Jiang2023}, and makes TDEs a formal target for multi-messenger astronomy. 

X-ray monitoring has become a standard and crucial component of follow-up campaigns following the detection of optical TDEs. A challenge arises from the observational fact that optical TDEs are predominantly X-ray faint while some exhibit delayed X-ray emission~\citep{Guolo2024}. It raises the central question of TDEs, i.e., what is powering the bright optical flares, and whether X-ray and optical TDEs belong to different populations or can be described in a unified framework. Some propose that UV/optical emission originates from the reprocessing of X-ray photons in an extended envelope or outflow~\cite{Dai2018}, while others suggest that the emission results from shocks generated by the stream-stream collision of stellar debris~\cite{Piran2015}. The delayed X-ray emission, when observed, can be attributed to either a structural change in the reprocessing layer as the accretion rate decreases from super-Eddington to sub-Eddington~\cite{Thomsen2022}, or a delayed onset of accretion following circularization. 

%The high sensitivity of eXTP/SFA will serve as a powerful tool for monitoring the X-ray evolution of TDEs. Continuous and deep X-ray monitoring of optical TDEs will not only help with constraining the X-ray bright fraction and onset time but also reveal various interesting short-term variability behaviors, such as quasi-periodic oscillations or modulations~\cite{Pasham2019,Pasham2024nature}, and hours-timescale large-amplitude X-ray dips~\cite{Yao2024}. They may indicate unique physical processes occurring in the environments of strong gravity, such as Lense-Thirring precession, providing a distinct opportunity to explore the properties and instability of the accretion disk formed by TDEs, as well as the mass and spin of SMBHs. Another intriguing discovery is the detection of QPEs in the long-term late plateau phase of optical emission~\cite{nicholl2024}, providing crucial insights into the origin of this new type of mysterious nuclear transients.

In addition to the typical soft X-ray and optical TDEs, there exists a rare subset of TDEs that are exceptionally X-ray bright (with isotropic X-ray luminosities of $L_{\rm X,iso}\sim10^{47-48}~\rm erg\,s^{-1}$) yet with dramatic short-time variability. These TDEs feature non-thermal X-ray spectra and are believed to be characterized by powerful relativistic jets aligned with our line of sight, rather than accretion. To date, only four such relativistic TDEs have been identified, with Swift~J1644+57~\cite{Burrows2011} and AT~2022cmc~\cite{Andreoni2022} being the most prominent sources discovered in the X-ray and optical bands, respectively. These jetted TDEs represent ideal targets for the eXTP/SFA due to their high X-ray luminosity and rapid variability, offering valuable insights into the physics of jet formation and evolution.

\subsection{Black hole spin}\label{subsec:spin}
%WeiHua Lei

In most BH accretion environments, the disk (and/or jet) angular momentum and BH spin axis are parallel; however, the transient disk of a TDE will generally have some tilt with respect to the SMBH equatorial plane. Since the TDE disk forms very close to the black hole, general relativity (GR) effects must be taken into account. An accretion disk inclined out of the equatorial plane of a spinning BH by an angle (assumed to equal the inclination of the stellar orbit before disruption) will be subject to Lense-Thirring torques with a strong radial dependence.

For a thin disk, it is expected that the Bardeen-Petterson effect~\cite{Bardeen1975} will induce a warp in the disk structure. However, for the thicker disks expected in many TDEs, simulations combining GR and magnetohydrodynamic effects (GRMHD) have shown that the disk precesses as a solid body rotator \cite{Stone2012}. Such a precessing disk might offer a novel way to measure spin in black hole systems, through timing observations. 

Lei, Zhang \& Gao (2013) argued that the jet precession is a possible consequence of this effect, providing an explanation of the quasi-periodic modulation ($\sim 2.7$ days) of Sw J1644+57's X-ray light curve \cite{Lei2013}. Based on this observation, Franchini et al. (2016) concluded that the SMBH should have a moderate spin value of $a_\bullet \sim 0.6$ \cite{Franchini2016}.

QPOs are regularly seen in stellar-mass BHs. Recently, QPOs have also been observed in AGNs and a couple of potential IMBHs. Despite the lack of a specific physical explanation, most models strongly link the origin of QPOs with orbits and/or resonances in the inner accretion disk close to the BH.

Reis et al. (2012) discovered a $\sim 200$-second X-ray QPO from the 2–10 keV power spectra of both the Suzaku and XMM-Newton observations in Sw J1644+57 \cite{Reis2012}. Abramowicz \& Liu (2012) regarded this observed QPO as one of ``3:2 twin peak QPO'', assuming that the second frequency was not observed based on the resonance in two eigen-modes of disk oscillations \cite{Abramowicz2012}. Tchekhovskoy et al. (2014) constrained the SMBH spin by considering three scenarios for this 200-second QPO \cite{Tchekhovskoy2014}: (1) a complete or (2) partial tidal disruption of a lower mass main-sequence star by an SMBH; or (3) a complete disruption of a white dwarf by an IMBH.

\begin{figure*}
%\sidecaption[b]
\centering
\includegraphics[width=1.0\textwidth]{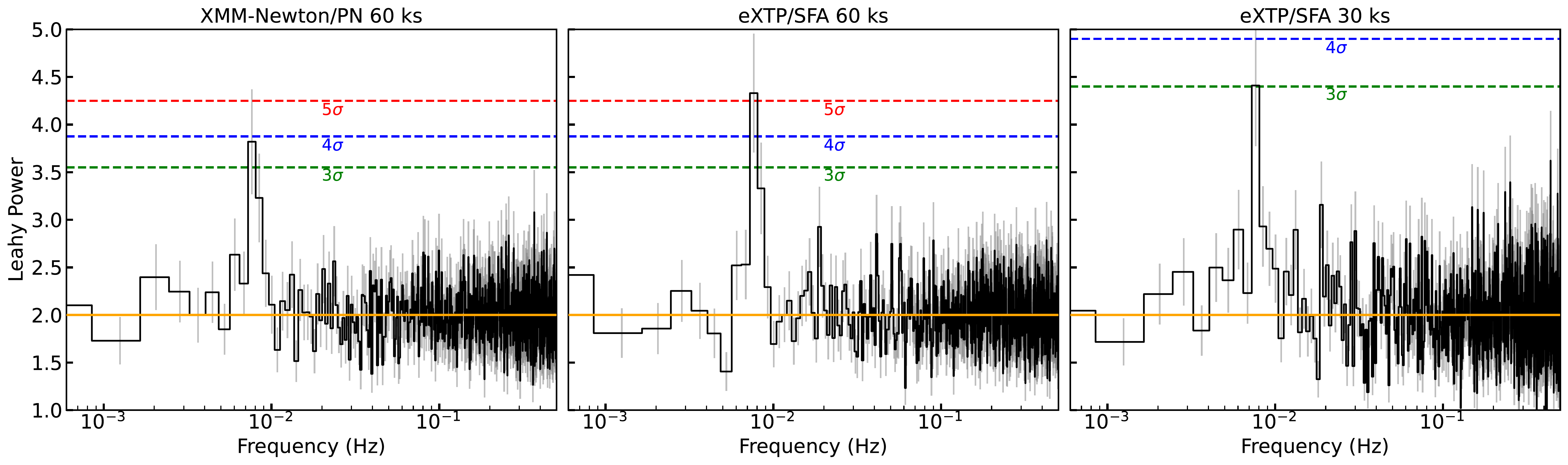}
\caption{
X-ray power spectra for ASASSN-14li, showing a QPO at 7.65 mHz. Left: The average X-ray PDS from six continuous 10,000-second light curves derived from XMM-Newton observational data \cite{Pasham2019}. 
Middle: The average X-ray PDS from six simulated continuous 10,000-second light curves generated for the {\bf eXTP}/SFA. Right: The average X-ray PDS from three simulated continuous 10,000-second light curves generated for the SFA. All simulated light curves were generated based on the XMM-Newton light curve of ASASSN-14li.
}
\label{fig:TDEqposimulation}
\end{figure*}

The unusual features of Sw J1644+57 suggest that this TDE is closely related to the onset of a relativistic jet from an SMBH. The jet is expected to be magnetically dominated (Burrows et al. 2011). Lei \& Zhang (2011) suggested that Blandford-Znajek process is the plausible mechanism to launch the relativistic jet from this source, and they used the available data to constrain the BH spin for Sw J1644+57 and Sw J2058+05 \cite{Lei2011}. Andreoni et al. (2022) adopted the same method to study the BH spin in AT2022cmc \cite{Andreoni2022}. They found that the BHs of these sources have a moderate to high spin, suggesting that BH spin is likely the crucial factor of powering the jet/outflow from these BH systems. 

Due to the high sensitivity of eXTP/SFA, high-cadence X-ray monitoring of TDEs by eXTP will be efficient in discovering quasi-periodic modulations, providing a useful tool for investigating the GR effects near SMBH, and then making it possible to measure the BH spin. The eXTP will have a good chance to detect jettd TDEs. The X-ray polarization observations of such jetted TDEs with eXTP will help to comprehend the connection between jet production and BH spin.

\subsection{X-ray polarizations of TDEs}\label{subsec:tdepol}
%Rongfeng Shen (Feel free to edit)

\begin{table*}[]
    \centering
    \begin{tabular}{c|cccc}
    \hline
         & Stream shock & Inner accretion flow & Corona & Jet\\
         \hline
       Spectrum  & thermal & thermal & non-thermal & non-thermal\\
       Polarization degree (\%)& a few & a few to 10 & $<\sim 1$ & $\sim$ a few tens \\
       Variability & slow & fast & mild & fast \\       
       Time (w.r.t. optical peak) & earlier or concurrent & concurrent & later & unknown\\
       \hline
    \end{tabular}
    \caption{Possible X-ray emitting regions in a TDE and their expected emission properties, which are highly speculative. } \label{tab:tdepol}
    \label{tab:my_label}
\end{table*}

Possible X-ray emitting regions or components in TDEs are stream-collision shocks, inner accretion flow, hot corona, and the jet. The polarization measurement ability of eXTP could be utilized to delineate the physical picture of a TDE.   

Firstly, the stream-stream collision via shocks, i.e., the so-called `circularization' of the debris stream, is often thought to precede the accretion of the stellar debris by the BH. The stream collision has been frequently invoked as the emission site for the optical / UV flares seen in most TDEs discovered so far \citep[e.g.,][\S\ref{subsec:xtde}]{Piran2015}, yet no work has looked into the question that whether this collision could produce detectable X-rays as well. Indeed, a couple of events show not only late X-ray peaks, but also some early X-rays, i.e., concurrently with the optical peak (e.g., AT 2019azh). Motivated by an open-minded style of thinking, we consider this as an open possibility. 

Yet the most probable X-ray emitting sites may be related to the central BH accretion. An bare accretion disk might not emit X-rays that are seen in TDEs, due to (1) the super-Eddington nature of the system, which launches fast wind or outflow that blocks or reprocesses the central emission, and (2) the high mass of the central BH, which causes the surface temperature of an standard Shukura \& Sunyeav thin disk to be too low to emit soft X-rays. It is likely that X-rays are emitted from a low-density, polar region near the BH, as shown in numerical simulations of the super-Eddington stage of a TDE, (e.g., \cite{Dai2018}). The high-inclination region is dominated by the high density outflow and emits in optical and UV bands. 

When the accretion rate diminishes and becomes low enough, a corona of hot or relativistic electrons may form. Non-thermal X-rays could be produced from the corona by Comptonization of the disk seed photons. 

Lastly, a TDE might generate relativistic jets and a few jetted TDE candidates have been detected, predominantly in X-rays (see \S\ref{subsec:xtde}-\ref{subsec:spin}). In those cases, X-rays are thought to be produced by synchrotron radiation within the jet. Similar to the prompt emission of GRBs (\S\ref{Sec:GRB Prompt X-ray emission}), relatively high PDs are expected in such cases, given that some requirements about the line-of-sight alignment and the magnetic property within the jet are satisfied. 

Geometrical asymmetry and magnetic-field configuration are two global factors that affect the PD of an emitting source. Based on their properties, one can naively draw the following order for the four possible X-ray emitting components in terms of PD:

Jet $>$ Inner accretion flow $>$ Stream shock $>$ Corona.

Table \ref{tab:tdepol} lists the four possible X-ray components. Their expected properties such as spectrum, variability and time are also given, mostly in a relative manner. Through detecting a sample of TDEs, the timing and polarization measurement capabilities of eXTP could help to identify or differentiate those components.

\subsection{X-ray spectral and timing studies of TDEs}\label{subsec:tdest}
%Lixin Dai, Ning Jiang

TDEs exhibit a variety of X-ray spectral properties. Blueshifted X-ray absorption features reveal that fast outflows are produced from TDEs ~\cite{Kara2018, Kosec2023}. Some of such outflows can reach speeds faster than $0.1c$, which are likely launched from super-Eddington accretion flows due to the large radiation pressure ~\cite{Jiang2014, Dai2018}. At the same time, narrower absorption lines with smaller blueshifts with $v\sim \rm few \times100 \ \rm km \ s^{-1}$ have also been observed~\cite{Miller2015}, which are possibly produced due to
the absorption in a slow disk wind or the debris stream.

%[\textcolor{blue}{Copied from Jiang Ning}] 

Moreover, TDEs show various interesting long-term and short-term variability behaviors. The former include X-ray brightening after a few months post flare peaks observed in many optical TDEs. The latter include quasi-periodic oscillations or modulations~\cite{Pasham2019,Pasham2024nature}, and hours-timescale large-amplitude X-ray dips~\cite{Yao2024}. They indicate unique physical processes occurring in the environments of strong gravity, such as Lense-Thirring precession, providing a distinct opportunity to explore the properties and instability of the accretion disk formed by TDEs, as well as the mass and spin of SMBHs.
Another intriguing discovery is the detection of QPEs in the long-term late plateau phase of optical emission~\cite{nicholl2024}, providing crucial insights into the origin of this new type of mysterious nuclear transients. 

Besides Sw J1644+57, two other non-relativistic TDEs are found to display QPOs with periods of $\sim$100 s in their soft X-ray emission, including  ASASSN-14li and 3XMM J215022.4–055108 \cite{Pasham2019, Zhang2025}, the latter likely involving an intermediate-mass black hole. With the QPO frequency, it is possible to determine the black hole spin if the black hole mass can be independently estimated using other methods. 
%So far, the number of TDEs with QPO detections is still small, and 
Given the sparse detection rate of X-ray QPOs, it remains unclear whether they are a common phenomenon in TDEs. 
The improved energy resolution and sensitivity provided by eXTP/SFA will allow for better addressing the question. Taking the QPO signal detected in ASASSN-14li as an example and assuming the same X-ray flux, 
we simulated the light curve from eXTP/SFA observations with an exposure of 60 ks. 
The analysis of the power density spectrum suggests that the significance for the QPO detection can be improved from $\sim$4$\sigma$ to 5.2$\sigma$ with the same exposure as XMM-Newton  (Fig.~\ref{fig:TDEqposimulation}). 
Furthermore, to achieve the detection significance as that observed in XMM-Newton, 
only half of exposure is required for eXTP/SFA observations, or eXTP/SFA will allow for probing 
the QPO signal to a fainter flux by a factor of $\approx$2. 
On the other hand, eXTP/SFA will provide the long, uninterrupted exposures that are critical to detecting the X-ray oscillation signal with lower rms amplitudes. 

Last but not least, X-ray reflection and reverberation signatures have been detected in TDEs. For example, a blueshifted, broad Fe K$\alpha$ fluorescent line at 5-9 keV has been observed from the relativistic TDE Swift J1644+57~\cite{Kara2016}. Analysis of the data using Fourier technique also show a lag between the Fe K line and the continuum. Fe K$\alpha$ fluorescent lines are produced by the irradiation of the accretion flows by non-thermal corona emissions. Previously such Fe K$\alpha$ lines are only observed from sub-Eddington accretion sources such as AGNs and X-ray binaries. Modeling efforts~\cite{Thomsen2019, Thomsen2022b} reveal that the Fe line profile and lag as observed in Swift J1644+57 are better explained by a super-Eddington outflow geometry than a thin disk geometry.
Further theoretical studies also show that for super-Eddington accretion flows X-ray spectral and timing analysis can be used to constrain many important properties, such as the black hole mass, corona height, wind geometry and kinematics~\cite{Zhang2024}.

The high sensitivity of eXTP/SFA will serve as a powerful
tool for monitoring the X-ray evolution and variability of TDEs, which will not only help
constrain the X-ray bright fraction and onset time of TDEs but
also reveal important accretion, outflow, jet and other BH physics in TDEs.

\subsection{Tidal disruption of white dwarf and multi-messenger }\label{subsec:tdewd}
%Qingwen Wu

White dwarfs are the final products of the evolution of low-mass stars, which can also be disrupted when they enter into the tidal radius of black hole with mass $\lesssim 10^5M_{\odot}$. Compared the TDEs of main sequence stars by SMBHs, the TDEs of white dwarf involve both compact objects and IMBHs, where the IMBHs have not been conclusively identified. The confirmation of the IMBHs will play a key role in understanding the SMBH formation and evolution, which will fill the big gap between the stellar-mass BHs and supermassive BHs. 

EMRI, as one of most important and promising GW sources, which is the ideal tool to test fundamental theories and nature of massive black hole (MBH). For a typical event of EMRI with a inspiral of 0.5$M_{\odot}$ WD into a $10^5M_{\odot}$ IMBH in a highly eccentric bound orbit with a pericenter distance of $\sim 20 R_g$ will radiate GW frequency of several milliHz, which will be possibly observed by future space GW projects (e.g., LISA, Taiji, and TianQin). The tidal heating effect may be important during the inspiral phase and before the tidal disruption, which will possibly trigger a runaway fusion in the surface layer of hydrogen and produce a soft X-ray flare \citep[e.g.,][]{2017MNRAS.468.2296V,2022ApJ...933..225W}. When the WD enter into the tidal disruption radius, the accretion of IMBH with super Eddington rate will lead to a soft X-ray burst. If the WD stay in an eccentric orbit, the accretion will produce a periodic X-ray outburst, which provide an explanation for the observed X-ray quasi-periodic eruptions (QPEs) at soft X-ray wavebands \citep[e.g., GSN 069,][]{2019Natur.573..381M,2022ApJ...933..225W}. If this is the case, the TDE WD is a good candidate for the multi-messenger object. Assuming the 4 yr observational time for future space GW project, the threshold detection distance of GW is about several tens Mpc for typical IMBHs and WDs at given typical QPE period of several hours. Most of the QPEs for these typical parameters can be detected by the eXTP. The space GW detections can accurately predict the disruption time of a WD and other related physical process, which can serve as an excellent laboratory to test the EMRI, TDE physics as well as the nature of WDs \cite{2018ApJ...856...82H}.

\section{Supernovae}\label{sec:SNe}
%{\it Contributors: Yun-Wei Yu, Guang-Lei Wu, Liang-Duan Liu, Tong Liu, Wei-Li Lin, Achille Fiore, Dafne Guetta}

X-ray observations of supernovae offer unique insights into their underlying mechanisms, though such detections remain observationally scarce compared to optical bands. The most promising X-ray-detectable supernovae are gamma-ray burst (GRB)-associated supernovae, where X-rays originate from either internal dissipation processes in relativistic jets or their interaction with the surrounding environment. Additional X-ray detection prospects include shock breakout events, circumstellar interaction phases, superluminous supernovae (SLSNe), and fast blue optical transients (FBOTs).

\subsection{GRB/FXT-associated supernova}
%Yun-Wei YU

Since the identification of the SN 1998bw/GRB 980425 association \cite{Galama1998,Iwamoto1998}, 
dozens of supernovae connected to GRB have been confirmed observationally \cite{Cano2017,Woosley_2006ARAA}. 
These events are observationally characterized as broad-lined Type Ic supernovae (SNe Ic-BL) with characteristic
 kinetic energies $E_{\rm k} \sim (1–2) \times 10^{52}$ erg. 
 Meanwhile, their optical luminosity is about several times higher than those of normal CCSNe, which may result from the radioactive decays 
 of a relatively high mass of $^{56}$Ni ($\sim0.2-0.5M_{\odot}$) or, alternatively, be aided by a central engine \cite{ZhangYu2022}. A notable example is SN 2006aj, which followed XRF 060218, where a distinct blackbody component was observed in the X-ray spectrum \cite{Campana2006}. Within the framework of the GRB standard model, the XRF emission could be the result of an off-axis observation of a typical relativistic jet, where the jet's direction is far from the line of sight. Alternatively, the XRF may indicate that the jet in this event is intrinsically weaker than those of normal GRBs. In this case, the X-ray emission is produced by a mildly relativistic jet and its cocoon formed during the jet's breakout from the progenitor envelope or, in some cases, by a jet choked within the envelope. In the latter scenario, the XRF can be attributed to the breakout emission from the shock driven by the choked jet interacting with the progenitor material.

The recent discoveries by EP have revealed a new class of supernovae preceded by fast X-ray transients (FXTs; e.g., EP 240414a, EP 241021a, EP 250108a) without gamma-ray counterparts \cite{Sun2024}. In principle, these FXTs could still be explained by the scenarios proposed for the SN 2006aj/XRF 060218 event. However, given their potential differences from GRB-associated events, which require further investigation, we cannot rule out the possibility that these EP-associated supernovae may represent new types of optical transients. Consequently, the origin of the FXTs remains an open question and may not necessarily be linked to a jet. Regardless, these EP discoveries highlight promising prospects for future observations by eXTP, which could help to determine their origins by revealing the polarization of the FXT emission.

\subsection{Shock breakout and interaction}
%Yun-Wei YU

Shock breakout (SBO) emission in core-collapse supernovae occurs when the
radiation-dominated shock wave from the stellar collapse propagates through
the progenitor envelope. The critical transition point occurs when the
optical depth drops below $c / v_s$ \citep{Weaver1976}, where
$v_s$ is the velocity of the shock. This transition triggers a brief but luminous
X-ray/UV flash, with a duration ranging from seconds to hours depending on the
progenitor star's radius ($R_{\ast}$). For example, red supergiants (RSGs)
with extended envelopes ($R_{\ast}\sim 10^{13} \rm cm$) exhibit
shock breakout durations of $\sim 10^3 \text{s} $, while Wolf-Rayet (WR)
stars with compact radii ($R_{\ast}\sim 10^{11} \rm cm$) show
flashes as short as $\sim 10 $ s. The total energy released during the
SBO follows an energy budget given by \citep{Waxman2017}:
\begin{equation}
    E_{\tmop{bo}} \sim 10^{47} R_{\ast, 13}^2 v_{\tmop{bo}, 9} \kappa_{0.34}^{-
   1} \text{ } \tmop{erg},
\end{equation}
where $R_{\ast, 13}$ is the radius of the progenitor in units of $10^{13} \text{ }
\tmop{cm}$, $v_{\tmop{bo}, 9}$ is the breakout velocity in units of $10^9
\text{ } \tmop{cm} $ s$^{-1}$ , and $\kappa_{0.34}$ is the opacity in units of $0.34
\text{ } \tmop{cm}^2 $ g$^{-1}$. The SBO timescale can be estimated by the
crossing time:
\begin{equation}
    t_{\mathrm{bo}}= \frac{c}{\kappa \rho_{\mathrm{bo}}v_{\mathrm{bo}}^2}\sim10^2 \kappa_{0.34}^{-1}\rho_{\mathrm{bo,-9}}^{-1}v_{\mathrm{bo,9}}^{-2}\,\mathrm{s} ,
\end{equation}
where $\rho_{\tmop{bo}, - 9}$ is the density of the breakout point in units of
$10^{- 9} \text{g}  \tmop{cm}^3$. Consequently, the peak luminosity scales
as:
\begin{equation}
    L_{\mathrm{bo}}\sim \frac{E_{\mathrm{bo}}}{t_{\mathrm {bo}}}\sim 10^{45}\rho_{\mathrm{bo,-9}} R_{\ast,13}^2 v_{\mathrm{bo,9}}^{3}\,\mathrm{erg ~s^{-1}}.
\end{equation} 
Thus, detecting SBO emission can provide critical constraints on progenitor
properties such as stellar radius ($R_{\ast}$), explosion asymmetry, and shock
dynamics \citep{Colgate1974, Matzner1999}. The initial
breakout pulse is smeared by the light travel time, causing the observed
luminosity to increase until it reaches the time $t_{\tmop{bo}}$, after which
it remains at a constant plateau of approximately $L_{\tmop{bo}} \cdot
t_{\tmop{bo}} c / R_{\ast}$ until the light-crossing time of the progenitor,
given by $R_{\ast} / c$. Following this plateau, the luminosity decreases as
$L \propto t^{- 4 / 3}$. This bolometric luminosity behavior allows for
independent measurements of $R_{\ast}$ and $t_{\tmop{bo}}$ if the typical
temperature of the SBO falls within the observation window of eXTP. 

% ({\snz Do you mean W2C? If W2C sensitive enough to do this? Any way, some quantitative statements should be made on the feasibility of achieving this goal with eXTP.})

% ({\color{blue}{Reply: The revision has been made to emphasize the advantages of eXTP (large field of view and broad energy coverage).}})

The observed temperature of SBO emission depends on the coupling between gas
and radiation, which determines whether thermal equilibrium is reached. In the
case of red supergiants (RSGs), which are the progenitors of SNe IIP, the
shock velocity is typically around $v_{\tmop{bo}} \sim 10^9 \tmop{cm}$ s$^{-1}$. In this scenario, the radiation emitted is in thermal equilibrium, with
an observed temperature of approximately $\tmop{kT}_{\tmop{bo}} \sim 10 \text{
} \tmop{eV}$, most of which is in the UV bands and expected to be absorbed by
the interstellar medium (ISM). However, the total energy released in the soft
X-ray tail can still reach $\sim 10^{46} \tmop{erg}$ over a duration
of approximately $1000$ s. In the case of more compact progenitors, such as WR stars and some blue BSGs — which are the progenitors of SNe Ib/c — the shock velocity is higher, approximately $v_{\tmop{bo}} \sim 10^{10} \tmop{cm}$ s$^{-1}$.
In these cases, the photons generated from the initial bremsstrahlung and its subsequent Comptonization cannot reach thermal equilibrium with the gas. As a result, the observed temperature is significantly higher than what would be expected in thermal equilibrium, reaching values up to the sub-keV range \citep{Nakar2010, Katz2010}. The emergent spectrum in this non-equilibrium case is a rapidly evolving, complex Comptonized free-free spectrum, with a smearing effect due to light travel time (e.g., \citep{Irwin2024} and references therein). 
Early-time multi-wavelength observations are thus expected to provide valuable insights into the physics of SBO in the non-equilibrium case. For the well-known SN SBO candidate XRO 080109/SN 2008D \citep{Soderberg2008},  the total energy of the radiation in the range of 0.3-10 keV is estimated to be $\sim 2\times10^{46}\,\mathrm{erg}$ with a power-law frequency dependence of the photon number. The light curve and spectrum analysis of SN 2008D reveals features characteristic of an SN Ib/c, indicating a compact progenitor star \citep{Mazzali2008, Modjaz2009}. 
However, the light curve of XRO 080109 shows that the peak was reached after approximately 60 s and declined after about 130 s, which is too long for a compact progenitor star. Therefore, this behavior may be attributed to the SBO occurring in the dense circumstellar medium (CSM) surrounding the progenitor star \citep{Balberg2011}. 

The interaction between high-velocity SN ejecta and circumstellar material (CSM) generates dual shock fronts: a forward shock propagating into the CSM and a reverse shock decelerating the ejecta. These shocks initially produce thermal X-ray emission when the CSM is optically thick  $\tau > c/v_s$ ($v_s$: shock velocity).  While traveling to areas where the optical depth to the CSM edge drops below \( c/v_{\mathrm{s}} \), the thermal radiation can emission escape from the shock and produce thermal SBO emission.  If the CSM is dense and compact surrounding the progenitor, the SBO emission peaks at X-ray bands with a relatively long timescale to the SBO in the progenitor envelope, which relieves the tension between the duration of the SBO and the progenitor type indicated by optical observation in XRO 080109/SN 2008D. In contrast, an extended CSM is responsible for the peak luminosity of certain optical/UV thermal transients \citep{Khatami2024}. The shock continues to interact with the outer CSM, gradually transitioning into a gas-dominated collisionless shock, rather than remaining radiation-dominated. The shock temperature is governed by the equipartition between ions and electrons, reaching up to  $T_{\mathrm{s}}\sim 10^9 v_{\mathrm{s,9}}^2\,\mathrm K$, , which exceeds the equilibrium temperature \citep{Margalit2022}. The hot gas primarily emits X-ray photons through free-free bremsstrahlung, with a luminosity given by 
\begin{equation}\label{eqs:CSMLx}
    \begin{aligned}
        L_{X}\approx3.0\times 10^{39} \bar{\mathrm{g}}_{\mathrm{ff}} C_{n}\left(\frac{\dot{M}}{10^{-5}\,M_{\odot} \mathrm{yr}^{-1}}\right)^2&\left(\frac{v_{\mathrm{wind}}}{10 \,\mathrm {km\,s}^{-1}}\right)^{-2} \\
        \times\left(\frac{t}{10\,\mathrm{days}}\right)^{-1}\,\mathrm{erg~s^{-1}},
    \end{aligned}
\end{equation}
where $\bar{\mathrm{g}}_{\mathrm{ff}}$ is the free-free Gaunt factor, $C_{n}=1$ for the forward shock and $(n-3)(n-4)/4(n-2)$ for the reverse shock, $n$ is the density index of the out envelope of the ejecta, $\dot{M}$ and $v_{\mathrm{wind}}$ are the mass-loos rate and wind velocity of the progenitor, respectively. Equation \ref{eqs:CSMLx} which is highly sensitive to the late evolution of the progenitors. Thus, the X-ray light curve can be used to diagnose the evolutionary state of different progenitor types.

The breakout of
 the SN shock through the dense CSM shell discussed above may also produce a neutrino flux that can account for a significant fraction of
 the observed $>$ 10 TeV neutrino background \cite{Waxman2025}. 
 Therefore we expect a neutrino signal together with the X-ray emission from the shock breakout.
 These neutrinos are emitted in an optically thick region that explains the lack of a high-energy gamma-ray background accompanying the neutrino background.

\subsection{Superluminous supernovae}

SLSNe are a rare subclass of supernovae, with peak luminosities tens to hundreds of times greater than those of typical SNe \citep{2011Natur.474..487Q, 2012Sci...337..927G}. For typical SNe, the radioactive decay of heavy elements like $^{56}$Ni and $^{56}$Co serves as a common power source \citep{1982ApJ...253..785A}. However, this mechanism is insufficient to explain the extreme luminosity evolution observed in most SLSNe. This discrepancy has prompted extensive research into alternative energy sources, although the mechanisms responsible for the extraordinary radiation of SLSNe remain a topic of ongoing debate.

One widely used model posits that a rapidly spinning, highly magnetized neutron star (magnetar) serves as the main power source for the SN luminosity at maximum \citep{2010ApJ...719L.204W,2010ApJ...717..245K}. In this scenario, the rotational energy of the magnetar is transferred to the expanding supernova ejecta via a strong wind. This energy deposition heats the ejecta and drives the extraordinary luminosity observed in SLSNe. The amount of energy that can be released by this mechanism is substantial, as a newly formed magnetar can contain rotational energy on the order of 10$^{52}$ erg, depending on its spin period. This mechanism not only explains the extreme luminosities, but also accounts for the slow rise and decline rates in the light curves of many SLSNe that are hardly interpreted as pair-instability SN candidates \citep{2013Natur.502..346N}.

If an accretion disk forms around the magnetar shortly after the supernova explosion, the compact object may experience either spin-up through accretion torque or spin-down via propeller effects, depending on the disk-magnetosphere interaction \citep{2018ApJ...857...95M,2021ApJ...914L...2L}. This magnetar-disk coupling can substantially modulate the wind energy injection rate, thereby directly shaping the supernova light curve evolution. Such a mechanism offers a critical supplement to the standard magnetar-driven paradigm.

In parallel, SLSNe may derive their energy from ejecta-CSM interaction \citep{1982ApJ...258..790C,2012ApJ...746..121C}. When the expanding ejecta collides with a dense, slowly moving material shell surrounding the progenitor star, the resulting shock system thermalizes the kinetic energy into observable radiation. This energy conversion process becomes particularly efficient when the CSM density is enhanced by intense pre-supernova mass loss from the progenitor.

The high-energy photons produced by these progenitor systems could penetrate the ejecta surface, establishing X-ray observations as a critical diagnostic for distinguishing between powering mechanisms. Spectral-timing analysis in the 0.3-30 keV band enables detection of shock-powered signatures including off-axis jet components, magnetar-driven ionization breakouts, and azimuthally asymmetric CSM distributions. Over the past two decades, multi-messenger campaigns have employed Chandra, NuSTAR, Swift, and XMM-Newton to monitor nearby SLSNe from hours to years post-explosion \citep{2013ApJ...771..136L,2018ApJ...864...45M,2022ApJ...941L..16A}. Notably, the majority of these events remain X-ray quiescent, in a few cases where upper limits reach several $10^{40\sim41}$ erg s$^{-1}$. Even the contested X-ray detection from PTF12dam's location showed contamination-dominated signals, constraining any intrinsic SLSN emission to levels below that of an even weak gamma-ray burst GRB980425 \citep{2018ApJ...864...45M}.

In contrast, SCP06F6 was observed to exhibit an X-ray luminosity of approximately $\sim 10^{45}$ erg s$^{-1}$, 150 days after its discovery \cite{2013ApJ...771..136L}. Such an extreme luminosity may require unique physical conditions not typically associated with most events, such as an unusually dense CSM, an extreme magnetar, or even an unknown energy source. Another known SLSN observed in X-rays is SN 2020tcw \cite{2021TNSAN.139....1M}, with a luminosity of $5.2\times 10^{40}$ erg s$^{-1}$ within the $0.3 - 10$ keV range. Additionally, it is important to note that theoretical models have ruled out detectable X-ray emission from a magnetar engine in the early stages (i.e., within months after maximum), instead favoring a more likely detection at later times ($\gtrsim 3- 30$ years) due to the reduced bound-free opacity of oxygen-rich ejecta \cite{2018MNRAS.481.2407M}, which aligns with expectations for SLSNe. While this late emission is expected to be faint, it may be consistent with the numerous X-ray detection limits recorded for many SLSNe \cite{2024A&A...683A.223S}.

To advance our understanding of SLSNe and their high-energy emission mechanisms, systematic X-ray detections are crucial. WFM onboard eXTP, with large field of view, can promptly detect X-ray counterparts to SLSNe, potentially  revealing new instances of high-energy phenomena. Moreover, the broadband capability (from soft to hard X-ray) of eXTP enables monitoring nearby SLSNe across the X-ray regime and tracking the evolution of their spectral energy distribution. 

% ({\snz To be critical, eXTP's advantages are not its sensitivity; actually Chandra and Newton are more sensitive than eXTP. The readers may wonder how can eXTP do better on this important science?})

% ({\color{blue}{Reply: The revision has been made to emphasize the advantages of eXTP (large field of view and broad energy coverage).}})

\subsection{Fast blue optical transients}\label{sec:FBOT}
%Yun-Wei YU, Liangduan Liu, Bing Li

Fast Blue Optical Transients (FBOTs) exhibit three defining observational characteristics: an exceptionally rapid rise  to maximum brightness ($t_{\rm rise} < 10$ days), intense emission ($L_{\rm peak} > 10^{43}$ ergs s$^{-1}$), and  distinctive blue spectral energy distributions  ($(g-r)< -0.2$ mag) \cite{Drout2014,Pursiainen2018,Ho2021-ZTF}. These extreme properties fundamentally distinguish FBOTs from both normal SNe and SLSNe. While their physical origin remains enigmatic, the proposed mechanisms fall into two main categories: shock-powered models involving dense circumstellar medium \cite{Rest2018,Pellegrino2021} and central-engine-driven models invoking millisecond magnetars or accreting compact objects  \cite{YuYW2015,Pursiainen2018,LiuJF2022}. Theoretically, these X-ray production mechanisms observed in standard SNe and SLSNe could enable FBOT detection in X-rays; however, observational confirmation remains challenging due to the transient nature of these events. To date, only a handful of FBOTs have been detected across the electromagnetic spectrum \cite{Coppejans2020,Bright2022,YaoYH2022}, underscoring the need for improved multi-wavelength monitoring strategies.

%status qpo of X-ray observed from FBOT and their characters 
The most notable example is AT2018cow, which has been observed photometrically and spectroscopically across a broad range of wavelengths, from radio to $\gamma$-ray \cite{Perley2019, Rivera2018, Margutti2019}. \text{This source remains the best-studied FBOT in X-rays.} The X-ray \text{peak} luminosity of AT2018cow, which is of particular interest for eXTP, is on the order of $L_X \sim 10^{43}$\,erg\,s$^{-1}$. Specifically, its light curve shows a variable behavior on timescales of a few tens of hours, also known as X-ray variability. An 224 Hz QPO evidence was found in the average PDS of X-ray \cite{Pasham2022NatAs}.
While the soft X-ray emission decays gradually, the hard X-rays were only detected during the first 10 days and then disappeared abruptly. X-ray emission has also been detected from AT2020xnd and AT2020mrf. AT2020mrf is the first multi-wavelength FBOT identified through an X-ray survey. Its X-ray luminosity in the 0.1-10 keV band was found to be approximately $L_X \sim 2 \times 10^{43}$\,erg\,s$^{-1}$, although the detection by eROSITA was delayed by about 35 days after the first optical observation. This suggests that the X-ray emission of AT2020mrf is likely the highest among the FBOTs \cite{YaoYH2022}. In contrast, the X-ray luminosity of AT2020xnd is relatively lower than that of AT2020mrf but still comparable to that of AT2018cow at around 20-40 days \cite{Bright2022}. EP240414a are the other most prominent FBOTs (or FBOT candidates) with confirmed X-ray observations. It was detected by EP-TXT in very early days at a low luminosity status  because of the sensitivity of the EP-FXT and the outbrusts are rich in X-rays\cite{vanDalen2025ApJ.982L47V}. These examples strongly demonstrate the potential of X-ray observations for studying FBOTs.

In any case, the lack of X-ray emission for the majority of FBOTs may simply be because the optimal time for X-ray observation had already passed when they were discovered through archival data. Timely follow-up observations of FBOTs are clearly essential, and eXTP is undoubtedly expected to play a crucial role in this effort. Such potential X-ray observations could help distinguish between different models of FBOTs and, furthermore, reveal differences in their progenitors compared to those of SLSNe, as their progenitors may include ultra-stripped stars \cite{Tauris2013, Suwa2015, Sawada2022}, electron capture SNe \cite{Moriya2016, Mor2022}, super-Chandrasekhar white dwarfs \cite{Kasliwal2010, Brooks2017, YuYW2019a}, and even binary NSs, binary WDs, or NS-WD binaries \cite{YuYW2013, YuYW2019b, Zenati2019}.

FBOTs often show a combination of thermal (blackbody) and non-thermal (power-law) components in their 
X-ray spectra. In the soft X-ray band (0.3–4 keV), thermal emission observed as a blackbody spectrum 
with temperatures of $\sim$ 1–3 \rm keV, likely originating from shock-heated ejecta interacting with
circumstellar material \cite{Lilong2024ApJ963L13L}. 
For example, the AT2018cow displayed a soft X-ray blackbody component peaking at $\sim 1 \rm  keV$, 
attributed to thermalization of photons trapped in expanding ejecta. The hard X-ray non-thermal 
emission ($> 10 \rm keV$) ofen shonw as an observed power-law component (spectral index $\Gamma \approx 2–3$),
possibly due to synchrotron radiation from relativistic electrons accelerated in shocks or magnetic 
dissipation in magnetar winds \cite{Romansky2024AdSpR.74.4290R}. 
In AT2018cow, the hard X-ray emission was explained by high-energy photons leaking from a striped 
magnetar wind \cite{Lilong2024ApJ963L13L}.

Recent studies suggest FBOTs may exhibit high X-ray polarization (up to 50–60\%), which could reveal the magnetic field geometry of magnetar-Driven models or quantum electrodynamics (QED) effects. The X-mode dominates photon escape in strong fields, leading to high polarization degrees, as seen in magnetar models for FBOTs. A highly ordered toroidal magnetic field around a millisecond magnetar would produce polarized synchrotron emission with periodic rotation of polarization angles, synchronized with the spin of a magnetar (e.g., 225 Hz pulsations in AT2018cow\cite{Pasham2022NatAs.6..249P}) . In extreme magnetic fields ($ >10^{14}$ G), vacuum birefringence may cause energy-dependent polarization angle shifts(e.g., differences between soft and hard X-rays)\cite{LaiDong2023PNAS12016534L} . Studying the X-ray polarization of FBOTs provides valuable insights into their geometry, origin, and underlying mechanisms. Some FBOTs are thought to be associated with the launch of relativistic jets. X-ray polarization observations by the eXTP mission can help determine whether the transient involves jet-like structures by analyzing the polarization direction.

Moreover, X-ray polarization can shed light on the symmetry of the FBOT explosion. A symmetric, spherical explosion would generate minimal polarization, while an asymmetric explosion, or one involving jet-like structures, is likely to produce significant polarization. Monitoring changes in polarization over time can provide additional information on the evolution of the event, revealing whether it evolves symmetrically or exhibits more complex, asymmetric behavior. The future prospects of advanced X-ray observatories like eXTP that combined high sensitivity, broad energy coverage (0.5–30 keV), and polarization resolution will resolve polarization-angle variations to confirm magnetar-driven models, detect QED effects in extreme magnetic fields via energy-resolved polarization, and monitor long-term X-ray evolution to distinguish between magnetar spin-down and black hole accretion. For a comprehensive understanding, multi-wavelength campaigns (e.g., combining X-ray, radio, and optical data) are essential to link central engine activity with ejecta dynamics. Further polarimetric observations in both the radio, optical and X-ray bands are needed to verify whether common phase-dependent polarization properties are also present in those sources, as photons undergo birefringent and dispersive propagation effects which result in frequency-dependent conversions of polarized radiation behaved in radio magnetar XTE J1810-197 \cite{Lower2024NatAs.8.606L}. The QED-driven polarization shifts can provide critical insights in magnetic field diagnostics and central engine confirmation of FBOTs.

\section{High energy neutrinos and TeV AGNs}\label{sec:Neutrino}
%{\it Contributors: Xiangyu Wang, Jin Zhang, Dafne Guetta, Zhuo Li}

High-energy neutrinos and TeV gamma-rays are dual signatures of hadronic processes involving high-energy cosmic rays, i.e., energetic charged particles from the outer space, whose origin is  largely unknown. In astrophysical sources, such as AGNs, the energy in the jet or accretion disk is transformed into the acceleration of protons or heavier nuclei. These particles interact with radiation fields or ambient matter near the source, producing secondary particles like pions, whose decay chains generate both neutrinos and TeV gamma-rays. In particular, blazars, a special type of radio-loud (RL) AGNs, are the main confirmed extragalactic sources of TeV emission. The jets of RL-AGNs are also thought to be accelerators of ultra-high energy cosmic rays (UHECR), the cosmic rays with  energy greater than $10^{18}{\rm eV}$  \citep{2022Univ....8..607R}. Therefore,   neutrino and  TeV gamma-ray emission are crucial messengers for studying some fundamental problems in high-energy astrophysics, such as the  sources of UHECRs,  the composition (i.e., the normal proton-electron composition or Poynting flux dominated composition) and radiation mechanisms of AGN jets, and etc.

\subsection{X-ray counterparts of neutrinos and TeV AGNs}\label{sec: xray_nu} 

\begin{figure*}
%\sidecaption[b]
\centering
\includegraphics[width=1.0\textwidth]{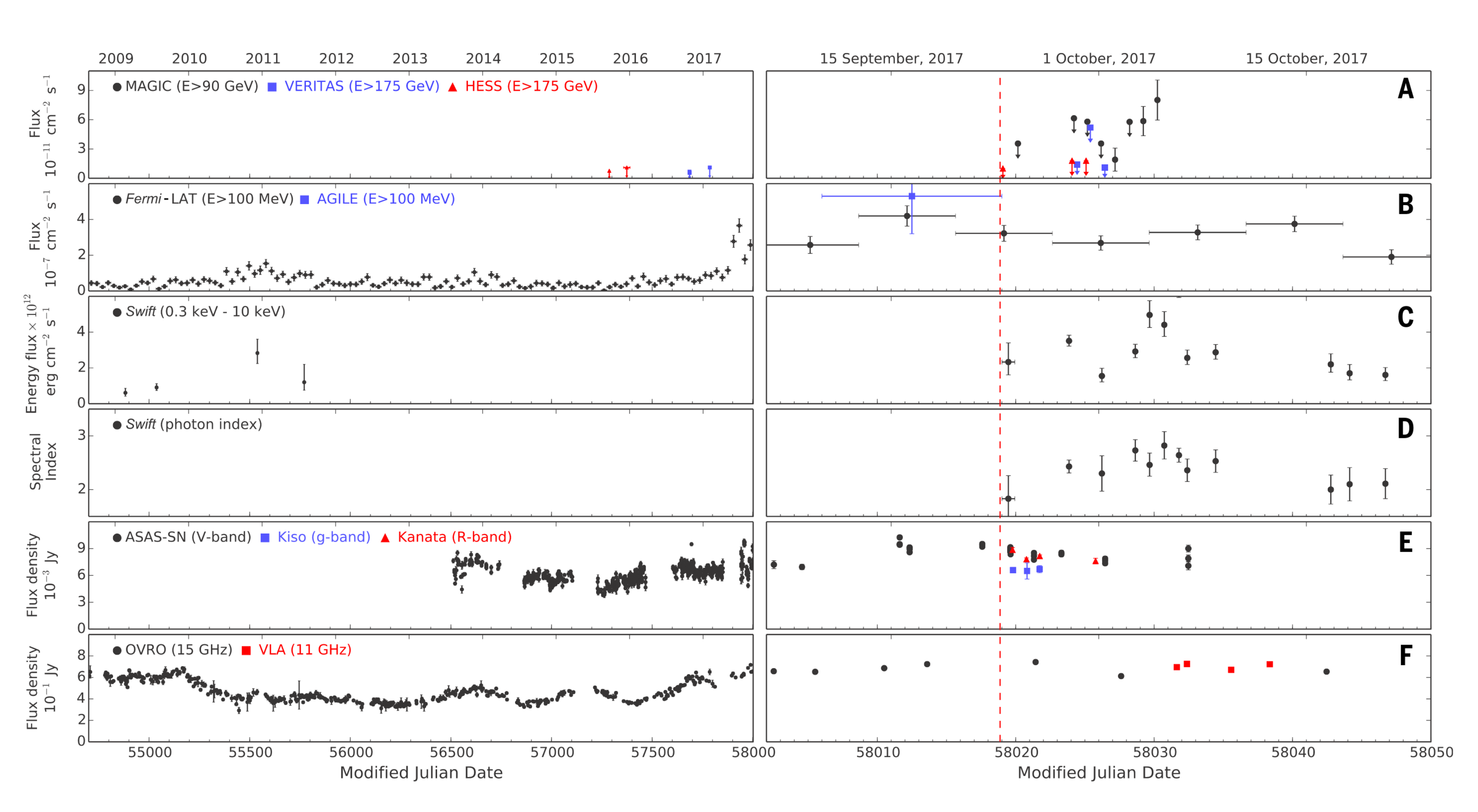}
\caption{Multi-wavelength light curve of TXS 0506+056 before and after the neutrino trigger. From top to bottom: (A) VHE gamma-ray observations by MAGIC,
H.E.S.S. and VERITAS; (B) high-energy gamma-ray observations by Fermi-LAT and AGILE; (C and D) x-ray observations by
Swift XRT; (E) optical light curves from ASAS-SN, Kiso/KWFC, and Kanata/HONIR; and (F) radio observations by OVRO and VLA. The red dashed line marks the detection time of the neutrino IceCube-170922A. The Figure is reproduced/modified from \cite{2018Sci...361.1378I}.}
\label{fig:TXS0506-LC}
\end{figure*}

The  detection of the neutrino IceCube-170922A coincident with a gamma-ray flare from the blazar TXS 0506+056 in 2017 \cite{2018Sci...361.1378I} marked a milestone in multi-messenger astronomy.  By combining gamma-ray and multi-band observations with the neutrino data, the neutrino event IceCube-170922A is thought to be associated with a multi-band flare from the blazar TXS 0506+056 \cite{2018Sci...361.1378I}. Swift and NuSTAR detected X-ray emission from TXS 0506+056\cite{Keivani2018}, finding that an increase in the X-ray flux correlates well with the strong increase at TeV energies over several days after the alert
(see panel C and D of {Figure \ref{fig:TXS0506-LC} }). 
These findings suggest that X-ray activity may correlate with the neutrino event.  The analysis of archival IceCube neutrino data revealed an earlier burst of neutrino events from the same source in 2014/2015 without an accompanying flare of gamma rays \citep{Aartsen2018Sci...361..147I}. Follow-up studies of IceCube-170922A revealed $\sim$7\% optical polarization during flares reported by the Kanata Telescope \citep{Aartsen2018Sci...361.1378I}, but there is no x-ray polarization observation for this event.

Another neutrino candidate source is the nearby obscured AGN NGC 1068, a type-2 Seyfert galaxy, which shows an excess of neutrino events with a significance of 4.2$\sigma$ \cite{2022Sci...378..538I}. In NGC 1068, a SMBH at the center is highly obscured by thick gas and dust \cite{2022Natur.602..403G}. X-ray studies have suggested that NGC 1068 is among the brightest AGNs in intrinsic X-rays \cite{2015ApJ...812..116B}, which are generated through Comptonization of accretion-disk photons in hot plasma above the disk, namely the coronae. Since the SMBH at the center of NGC 1068 is obscured by gas and dust and is also surrounded by strong radiation, efficient neutrino production is expected to occur if cosmic rays are accelerated \cite{2020PhRvL.125a1101M,2020ApJ...891L..33I}.

Additionally, the Chandra observatory has revealed numerous substructures in AGN large-scale jets in the X-ray band \citep{2006ARA&A..44..463H}. For some of these substructures, the observed X-ray spectra are harder than the extrapolation of the radio-to-optical components \citep{2005ApJ...622..797K,2010ApJ...710.1017Z,2018ApJ...858...27Z}, suggesting that they may originate from the inverse Compton (IC) scattering process \citep{2006ARA&A..44..463H} or synchrotron radiation of a distinct high-energy particle component \citep{2015ApJ...805..154M,2020ApJ...903..109T}. The detection of TeV gamma-rays from the large-scale radio lobes of radio galaxies (RGs) Cen A \citep{2010Sci...328..725A,2017AIPC.1792e0030S}, Fornax A \citep{2016ApJ...826....1A}, and NGC 6251 \citep{2024ApJ...965..163Y} further corroborates that these jet substructures serve as sites for the acceleration of high-energy particles. Notably, {the HESS collaboration reported the observations of Cen A at TeV energies that resolve its large-scale jet, providing direct evidence for the acceleration of ultra-relativistic electrons in the jet} \citep{2020Natur.582..356H}. These RGs are also considered potential sources of ultra-high-energy cosmic rays (UHECR) {if the protons are present and being accelerated} \citep{2018MNRAS.479L..76M}.

%Recent optical follow-up observations of neutrino alerts using the Zwicky Transient Facility (ZTF) have identified two optical flares from the centers of galaxies coincident with 100 TeV-scale neutrinos: AT2019dsg associated with the IceCube neutrino event IC191001A \cite{2021NatAs...5..510S} and AT2019fdr associated with IC200530A \cite{2022PhRvL.128v1101R}. The former belongs to the class of tidal disruption events (TDEs) from quiescent black holes, while the latter originates from a probable TDE event in an AGN. TDEs are phenomena in which a star passes close enough to a supermassive black hole to be ripped apart by its tidal forces. In both TDE events, soft thermal X-rays are detected. In AT2019dsg, X-rays were detected beginning 37 days after discovery. The X-ray flux faded extremely rapidly with an unprecedented decline rate in TDEs \cite{2021NatAs...5..510S}. Temporal evolution of X-rays is also evident and in the case of AT2019fdr. The X-rays were detected nearly 300 days after the neutrino detection. 

\subsection {Study the origin of high-energy neutrinos and TeV AGNs with eXTP}

\begin{figure*}
\centering
\includegraphics[width=18cm]{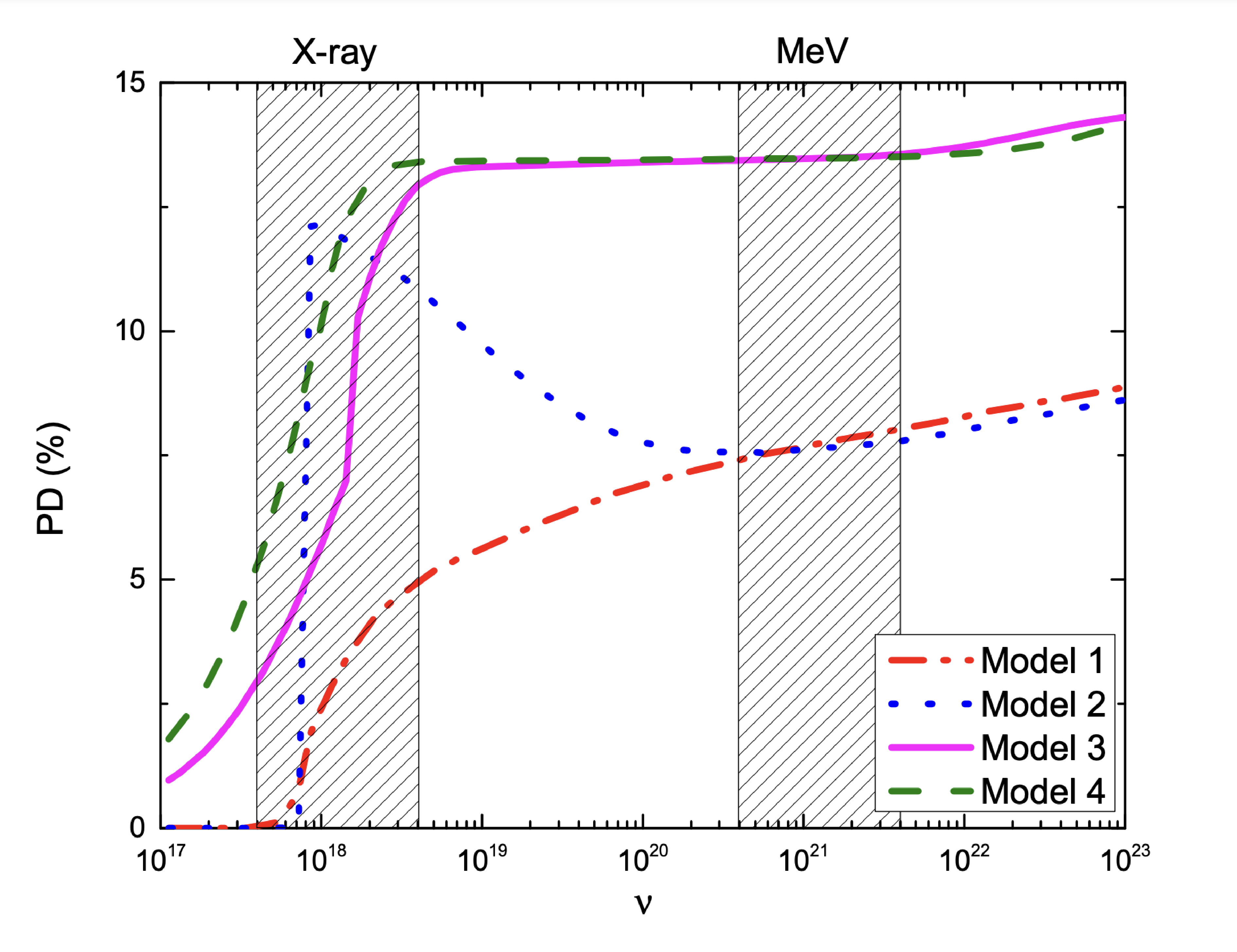}
\caption{ The X-ray to $\gamma$-ray spectral polarization degree based on the spectral fitting of TXS 0506+056 observation.
The shaded regions correspond to the X-ray and MeV $\gamma$-ray bands. The figure is adapted from \cite{2019ApJ...876..109Z} and reproduced by permission of the AAS.}
\label{fig:TXS0506-PD}
\end{figure*}

Several models have been proposed to explain the SED of the TXS 0506+056 flare and the IceCube-170922A event. %(e.g. \citep{Keivani2018,2018ApJ...865..124M,2017A&ARv..25....2P,2018MNRAS.480..192P,2019MNRAS.483L..12C,2019NatAs...3...88G,2019PhRvD..99f3008L,2024arXiv241117632Y}. 
These models can be categorized into three generic groups: 1) a leptonic emission model with a subdominant hadronic component, where the IC emission dominates the high-energy gamma-ray emission but the subdominant hadronic component produces the neutrinos and makes considerable X-ray emission through hadronic cascading synchrotron emission (e.g., \cite{2019NatAs...3...88G};); 2) a hadronic model where  gamma-rays are produced dominantly by 
proton synchrotron emission while neutrinos are produced through $p\gamma$ process.  The X-ray emission consists of both proton synchrotron emission and cascading synchrotron emission (e.g.\citep{2019MNRAS.483L..12C}); and 3) the neutrino is produced by the accretion flow while the high-energy gamma-ray emission is dominated by the electron SSC emission in the relativistic jet (e.g.\citep{2024arXiv241117632Y}). In this model, X-ray emission is generated through Comptonization of accretion-disk photons in a hot plasma above the disk, namely the coronae. The different models predict different properties of the polarization in X-rays \cite{2019ApJ...876..109Z} {(see Figure \ref{fig:TXS0506-PD}). 
Proton synchrotron-dominated models generally show a higher polarization degree than IC-dominated models \cite{2013ApJ...774...18Z,2018ApJ...853L...2P}, which may be studied by eXTP. On the other hand, in the accretion flow model for neutrinos, the X-ray emission is produced by thermal electrons in the coronae around the central SMBH, and the polarization degree in X-rays is lower. It is also found that the spectral polarization degree from X-ray to gamma-ray energies varies differently for different models \cite{2019ApJ...876..109Z}.

X-ray polarization observations would also serve as a powerful tool to elucidate the X-ray  emission  mechanisms of large-scale AGN jet substructures characterized by hard X-ray spectra; either IC scattering \citep{2006ARA&A..44..463H} or synchrotron radiation from a distinct population of high-energy particles \citep{2015ApJ...805..154M,2020ApJ...903..109T}. If the X-ray emission from large-scale jet substructures is indeed produced by synchrotron processes involving such high-energy particles, a high polarization degree is expected, which can be probed by eXTP. 
It then suggests that AGN large-scale jets may serve as sources of gamma-ray emission via IC scatterings of the synchrotron emission, as observed in Cen A } \citep{2020Natur.582..356H}.
%{\bf On the other hand, What is the polarization degree of the x-ray emission for the hadronic TeV emission of  AGN?}
Therefore, the X-ray polarimetry measurements conducted by eXTP for TeV AGNs may reveal intricate details regarding jet composition, particle acceleration, and radiation mechanisms of AGNs.

%\subsubsection {study the TeV source with eXTP}

\section{Summary}\label{sec:sum}
The enhanced X-ray Timing and Polarimetry mission (eXTP) will represent a milestone in high-energy astrophysics, combining advanced polarimetry, high sensitivity, and broad energy coverage (0.5–1000 keV) to tackle fundamental questions in time-domain and multi-messenger astronomy. By systematically exploring gravitational wave (GW) counterparts, gamma-ray bursts (GRBs), magnetars, fast radio bursts (FRBs), tidal disruption events (TDEs), supernovae, and high-energy neutrinos, eXTP will unravel the physics of extreme environments, constrain compact object properties, and test theories of gravity and particle acceleration.

In section \ref{sec:GWn}, we discuss gravitational wave counterparts, emphasizing eXTP’s role in multi-messenger studies of compact binary mergers, extreme mass-ratio inspirals (EMRIs), supermassive binary black holes (SMBBHs), and other GW sources. For stellar-mass mergers, W2C detects prompt gamma-ray emission from BNS/BH-NS mergers for joint GW-electromagnetic parameter estimation, while SFA tracks X-ray afterglows to constrain neutron star physics and magnetar formation. PFA identifies magnetar superflares and merged-magnetar flares, probing dense matter and magnetic fields. For EMRIs, SFA measures Fe K$\alpha$ reflection spectra to calibrate SMBH spins and detect disk-driven quasi-periodic eruptions. For SMBBHs, W2C detects X-ray periodicity and SFA analyzes composite Fe K$\alpha$ lines to identify candidates, complementing PTA data. Additionally, eXTP explores core-collapse supernovae, strange star-merger transients, and millisecond magnetar precession, providing insights into dense matter physics and gravitational wave emission mechanisms.

In section \ref{sec:GRB}, we investigate eXTP’s capabilities to address critical GRB physics across all emission phases. For prompt emission, PFA measures X-ray polarization to distinguish synchrotron (high polarization) vs. photospheric (low polarization) radiation, constraining jet magnetic field configurations. During X-ray plateau and flares, SFA’s high-time resolution captures magnetar-driven energy injection, while PFA tracks polarization evolution to infer jet collimation and central engine activity (black hole vs. magnetar). For X-ray flashes (XRFs), W2C rapidly localizes soft XRFs, and polarization measurements differentiate off-axis jets from low-efficiency explosions. Finally, GRB polarization signatures enable quantum gravity tests by constraining vacuum birefringence effects at Planck-scale energies.

In section \ref{sec:FRB}, we focus on magnetar physics and magnetar related radio transients (including Fast radio bursts and long period radio transients (LPTs)). For magnetar bursts, eXTP’s Spectroscopic Focusing Array (SFA) detects sub-millisecond variability and quasi-periodic oscillations (QPOs), constraining neutron star crustal dynamics and internal magnetic fields. For FRB counterparts, W2C’s all-sky monitoring and SFA’s spectroscopy search for X-ray emission from FRBs, linking them to magnetar activity or compact binary mergers. Additionally, eXTP will study magnetar superflares and LPTs, using polarization measurements (PFA) to characterize emission geometries and distinguish between magnetar-driven and binary interaction models. These observations will advance understanding of extreme magnetar physics, FRB origins, and their connections to compact object mergers.

In section \ref{sec:TDE}, we focus on tidal disruption events (TDEs), where eXTP will investigate supermassive black hole (SMBH) spin and accretion physics. For black hole spin, SFA analyzes X-ray variability from relativistic TDEs to measure SMBH spin via Lense-Thirring precession. For accretion and jets, PFA’s polarization measurements distinguish jet emission from disk processes, while SFA’s spectroscopy resolves outflow dynamics and super-Eddington accretion. Additionally, eXTP will study X-ray polarization in TDEs to characterize emission mechanisms (e.g., stream shocks, inner accretion flow), and X-ray variability to detect quasi-periodic eruptions (QPEs) and accretion disk instabilities. For white dwarf TDEs, eXTP’s sensitivity will detect X-ray signatures of tidal disruption by intermediate-mass black holes (IMBHs), providing multi-messenger insights into EMRI physics and IMBH demographics. These observations will advance understanding of SMBH spin evolution, accretion dynamics, and extreme gravitational environments.

In section \ref{sec:SNe}, we explore the prospects for using eXTP to study supernova-related energetic phenomena, including supernova shock breakout emission, superluminous supernovae (SLSNe), and fast blue optical transients (FBOTs). For shock breakout, SFA detects soft X-ray emission from core-collapse supernovae, probing progenitor structure and shock dynamics. For SLSNe, eXTP’s high-sensitivity spectroscopy tracks magnetar-driven energy injection, while polarization measurements identify asymmetric explosions and jet contributions. For FBOTs, W2C’s rapid response and SFA’s spectroscopy characterize X-ray emission mechanisms, distinguishing between shock-powered and central-engine-driven scenarios. Additionally, eXTP will investigate circumstellar interaction in supernovae, using X-ray spectroscopy to constrain progenitor mass loss and outflow properties. These observations will advance understanding of supernova physics, linking X-ray signatures to explosion mechanisms and progenitor evolution.

In section \ref{sec:Neutrino}, we link high-energy neutrinos to TeV active galactic nuclei (AGNs), leveraging eXTP’s capabilities to characterize emission mechanisms and cosmic-ray acceleration. eXTP’s SFA can be applied to measure X-ray polarization to distinguish between leptonic (inverse Compton-dominated) and hadronic (proton synchrotron-dominated) emission models, clarifying the role of AGN jets in cosmic-ray acceleration and cosmic high energy neutrinos production. 

{\ysx Despite the strong science prospects outlined above, it is important to recognize that the ultimate scientific returns of eXTP will be influenced by astrophysical uncertainties. For certain source classes, such as gravitational-wave events, fast radio bursts, FBOTs, and other rare transients, the dominant uncertainty lies in the intrinsic population statistics—the true event rates, spatial distributions, and luminosity functions remain poorly constrained. For other sources, such as GRBs, TDEs, or high-energy neutrino counterparts, the unknowns involve their intrinsic spectral shapes, polarization degrees, etc.. Nevertheless, even under conservative assumptions, eXTP’s unparalleled combination of sensitivity, spectral-timing-polarimetry coverage, and fast response ensures that it will yield transformative new insights across a wide range of astrophysical phenomena in the era of time-domain and multi-messenger astronomy.}

%%%%%%%%%%%%%%%%%%%%%%%%%%%%%%%%%%%%%%%%%%%%%%%%%%%%%%%
%%% Acknowledgements. ??
%%%%%%%%%%%%%%%%%%%%%%%%%%%%%%%%%%%%%%%%%%%%%%%%%%%%%%%
~

\emph{Acknowledgements.} 
This work is supported by China's  Space Origins Exploration Program. S.-X.Y. acknowledges support from the Chinese Academy of Sciences (grant Nos. E32983U810). WZ is supported by the Strategic Priority Research
Program of the Chinese Academy of Science (grant No. XDB0550300), and the National Natural Science Foundation of
China (grant No. 12325301). 
YH is supported by National Natural Science Foundation of
China (Grant Nos. 12233002 and 12041306), by National SKA Program of China No. 2020SKA0120300, by the National Key R\&D Program of China (2021YFA0718500), and by the support from the Xinjiang Tianchi Program. 
S-NZ is supported by the National Natural Science Foundation of China (No. 12333007), the International Partnership Program of Chinese Academy of Sciences (No.113111KYSB20190020) and the Strategic Priority Research Program of the Chinese Academy of Sciences (No. XDA15020100).
JG is supported by the Beijing Municipal Natural Science Foundation (No. 1242032) and the Youth Innovation Promotion Association of the Chinese Academy of Sciences (No. 2022056).
YL is supported by the National Key Research and Development Program of China (Grant No. 2022YFC2205201 and 2020YFC2201400). AF acknowledges funding by the European Union – NextGenerationEU RFF M4C2 1.1 PRIN 2022 project "2022RJLWHN URKA" and
by INAF 2023 Theory Grant ObFu 1.05.23.06.06 “Understanding R-process \& Kilonovae Aspects (URKA)". \\ Conflict of Interest: The authors declare that they have no conflict of interest.

%%%%%%%%%%%%%%%%%%%%%%%%%%%%%%%%%%%%%%%%%%%%%%%%%%%%%%%
%%% Conflict of interest. ????????????
%%%%%%%%%%%%%%%%%%%%%%%%%%%%%%%%%%%%%%%%%%%%%%%%%%%%%%%
%\InterestConflict{The authors declare that they have no conflict of interest.}

%%%%%%%%%%%%%%%%%%%%%%%%%%%%%%%%%%%%%%%%%%%%%%%%%%%%%%%
%%% Supplements. ????????, ????
%%%%%%%%%%%%%%%%%%%%%%%%%%%%%%%%%%%%%%%%%%%%%%%%%%%%%%%
%\Supplements{}

%%%%%%%%%%%%%%%%%%%%%%%%%%%%%%%%%%%%%%%%%%%%%%%%%%%%%%%
%%% Reference section. ??????
%%% citation in the content using "some words~\cite{1,2}".
%%% ~ is needed to make the reference number is on the same line with the word before it.
%%%%%%%%%%%%%%%%%%%%%%%%%%%%%%%%%%%%%%%%%%%%%%%%%%%%%%%
%\begin{thebibliography}{99}
%\bibitem {ref1} Exmple: M. Aspelmeyer, T. J. Kippenberg, and F. Marquardt, Rev. Mod. Phys. 86, 1391 (2014).

%\bibitem {ref2} Example: R. J. Hunter, \textit{Zeta Potential in Colloid Science} (Academic, New York, 1981), p. 120 %%Books

%\end{thebibliography}

% please donot delete
% \bibliographystyle{ref} % please donot delete
\bibliographystyle{scpma}
% \bibliography{ref} %ref.bib

\bibliography{main} % please donot delete

%%%%%%%%%%%%%%%%%%%%%%%%%%%%%%%%%%%%%%%%%%%%%%%%%%%%%%%
%%% Appendix sections. ??????, ????
%%%%%%%%%%%%%%%%%%%%%%%%%%%%%%%%%%%%%%%%%%%%%%%%%%%%%%%
%\begin{appendix}
%\section{Name}

%\end{appendix}

%\begin{appendices}
%\section{Appendix}
%\end{appendices}
%\appendix

%\appendix

%\renewcommand{\thesection}{Appendix}

%\section{}

%\def\thesubsection{\Arabic.\alph{section}}

%\section{}

%\end{appendices}
%\end{appendix}

\end{multicols}
\end{document}